\documentclass[smallextended]{svjour3}       % onecolumn (second format)
\smartqed  % flush right qed marks, e.g. at end of proof
\usepackage{graphicx}
\usepackage{hyperref} 
 \usepackage{mathptmx}      % use Times fonts if available on your TeX system

\usepackage{bm}
\usepackage{amssymb}
\usepackage{amsmath}
\usepackage{natbib}
\bibpunct[, ]{(}{)}{;}{a}{}{,}
\newcommand{\beq}{\begin{equation}}
\newcommand{\eeq}{\end{equation}}
\newcommand{\bn}{\rule{0pt}{2ex} $ \left\{\begin{array}{l}}
\newcommand{\en}{\end{array}\right. $}

\newenvironment{numlist}{
\setcounter{count}{0}
\begin{list}{\arabic{count}.}{
\setlength{\itemsep}{0ex}
\setlength{\parsep}{0pt}
\setlength{\parskip}{0pt}
\setlength{\topsep}{3pt}
\setlength{\partopsep}{3pt}
\usecounter{count}
}}{\end{list}}
\newcounter{count}

\newcommand{\aion}{a_\mathrm{i}}
\newcommand{\alphaf}{\alpha_\mathrm{f}}

\newcommand{\dd}{\mathrm{d}}
\newcommand{\EF}{\epsilon_\mathrm{F}}
\newcommand{\EFe}{\epsilon_{\mathrm{F},e}}
\newcommand{\EFn}{\epsilon_{\mathrm{F},n}}
\newcommand{\EFx}{\epsilon_\mathrm{F,x}}
\newcommand{\fig}[1]{Fig.~\ref{#1}}
\newcommand{\Gami}{\Gamma_{\mathrm{C}}}
\newcommand{\gammam}{\gamma}
\newcommand{\Gammam}{\Gamma_\mathrm{m}}
\newcommand{\gcc}{\mbox{g~cm$^{-3}$}}
\newcommand{\gfact}{g_\mathrm{i}}

\newcommand{\lambde}{\lambda_e}
\newcommand{\lambdi}{\lambda_\mathrm{i}}
\newcommand{\mel}{m_e}
\newcommand{\mion}{m_\mathrm{i}}

\newcommand{\nb}{n_\mathrm{b}}
\newcommand{\Ne}{\mathcal{N}_B(\epsilon)}
\newcommand{\nel}{n_e}
\newcommand{\nion}{n_\mathrm{i}}
\newcommand{\omc}{\omega_\mathrm{c}}
\newcommand{\omci}{\omega_\mathrm{ci}}
\newcommand{\omg}{\omega_\mathrm{g}}
\newcommand{\omp}{\omega_\mathrm{p}}
\newcommand{\opac}{\varkappa}
\newcommand{\pF}{p_\mathrm{F}}
\newcommand{\pFe}{p_{\mathrm{F}e}}
\newcommand{\pFn}{p_{\mathrm{F}n}}
\newcommand{\pFp}{p_{\mathrm{F}p}}
\newcommand{\req}[1]{Eq.~(\ref{#1})}
\newcommand{\rhob}{\rho_\mathrm{b}}
\newcommand{\rhod}{\rho_\mathrm{drip}}

\newcommand{\singlet}{\mbox{$^1$S$_0$ }}
\newcommand{\sion}{s_\mathrm{i}}
\newcommand{\sSB}{\sigma_\mathrm{SB}}
\newcommand{\Tb}{T_\mathrm{b}}
\newcommand{\TbTs}{$T_\mathrm{b}$\,--\,$T_\mathrm{s}$}
\newcommand{\Tc}{T_\mathrm{c}}
\newcommand{\Tcn}{T_{\mathrm{c}n}}
\newcommand{\Tcp}{T_{\mathrm{c}p}}
\newcommand{\Teff}{T_\mathrm{eff}}
\newcommand{\Tp}{T_\mathrm{p}}
\newcommand{\Tpe}{T_{\mathrm{p},e}}
\newcommand{\Tm}{T_\mathrm{m}}
\newcommand{\triplet}{\mbox{$^3$P$_2$}}
\newcommand{\Ts}{T_\mathrm{s}}
\newcommand{\xr}{x_\mathrm{r}}

\newcommand{\zete}{\zeta_e}
\newcommand{\zeti}{\zeta_\mathrm{i}}

\journalname{Space Science Reviews}
\begin{document}
%%%%%%%%%%%%%%%%%%%%%%%%%%%%%%%%%%%%%%%%%%%%%

\title{Neutron Stars -- Cooling and Transport}

\author{ Alexander~Y.~Potekhin \and
        Jos\'e~A.~Pons \and
        Dany~Page
}

\authorrunning{A.Y. Potekhin, J.A. Pons, D. Page}

\institute{
   A.Y.~Potekhin \at
      Ioffe Institute,
      Politekhnicheskaya 26, 194021 Saint Petersburg,
       Russia;\\
      Central Astronomical Observatory at Pulkovo,
      Pulkovskoe Shosse 65, 196140 Saint Petersburg,
       Russia;\\
      St Petersburg State Polytechnical University,
      Polyteknicheskaya 29, 195251 Saint Petersburg,
       Russia\\
              \email{palex@astro.ioffe.ru}
\and
     J.A.~Pons \at
     Departament de F\'{\i}sica Aplicada, Universitat d'Alacant,
    Ap.\ Correus 99, E-03080 Alacant, Spain\\
      \email{jose.pons@ua.es}
\and
     D.~Page \at
  Instituto de Astronom\'{\i}a, 
  Universidad Nacional Aut\'onoma de M\'exico, M\'exico, D.F. 04510, M\'exico
     \\
    \email{page@astro.unam.mx}              
}

\date{Received: 6 March 2015 / Accepted: 1 July 2015 
~/~ DOI: 10.1007/s11214-015-0180-9
}

\maketitle

%%%%%%%%%%%%%%%%%%%%%%%%%%%%%%%%%%%%%%%%%%%%%
\begin{abstract}

Observations of thermal radiation from neutron stars can
potentially provide information about the states of
supranuclear matter in the interiors of these stars with the
aid of the theory of neutron-star thermal evolution. We
review the basics of this theory for isolated neutron stars
with strong magnetic fields, including most relevant
thermodynamic and kinetic properties in the stellar core,
crust, and blanketing envelopes.

\end{abstract}

% ADDED FOR EASIER NAVIGATION:
\setcounter{tocdepth}{2}
\tableofcontents

%%%%%%%%%%%%%%%%%%%%%%%%%%%%%%%%%%%%%%%%%%%%%
\section{Introduction} 
\label{intro}
The first works on neutron star cooling and thermal
emission 
\citep{Stabler60,Tsuruta64,ChiuSalpeter64,Morton64,BahcallWolf65a,BahcallWolf65b}
appeared at the epoch of the discoveries of X-ray sources
outside the Solar System in the rocket and balloon
experiments \citep{Giacconi-ea62,Bowyer-ea64a,Bowyer-ea64b}.
The authors estimated cooling rates and surface temperatures
in order to answer the question, whether a neutron star can
be detected in this kind of experiments. However, the first
attempts failed to prove the relation between neutron stars
and newly discovered X-ray sources. In particular,
\citet{Bowyer-ea64b} measured the size of the X-ray source
in the Crab Nebula from observations during a lunar
occultation on July 7, 1964. Their result, $\sim 10^{13}$
km, indicated that the source was much larger than a neutron
star should be. Ironically, there was a neutron star there,
the famous Crab pulsar, but it was hidden within a compact
plerion pulsar nebula. \citet{Kardashev64} and later
\citet{Pacini67} conjectured that the Crab Nebula could be
powered by the neutron-star rotational energy, which was
transferred to the nebula via the magnetic field, but this
model remained a hypothesis. Curiously, the Crab pulsar was
observed as a scintillating radio source since 1962
\citep{HewishOkoye65}, but the nature of this source
remained unclear. \citet{Sandage-ea66} identified Sco X-1,
the first detected and the brightest cosmic X-ray source, as
an optical object of 13th magnitude. \citet{Shklovsky67}
analyzed these observations and concluded that the X-ray
radiation of Sco X-1 originated from the accretion of matter
onto a neutron star from its companion. Later this
conjecture was proved to be true \citep{deFreitas77}, but at
the time it was refuted \citep{CameronScoX1}. Because of
these early confusions, the first generally accepted
evidence of neutron stars was provided only by the discovery
of pulsars \citep{Hewish-ea68} after a successful
competition of the theoretical interpretation of pulsars as
rotating neutron stars \citep{Gold68} with numerous
alternative hypotheses (see, e.g., the review by
\citealp{Ginzburg71}).

The foundation of the rigorous cooling theory was laid by
\citet{Tsuruta64} and \citet{TsurutaCameron66}, who
formulated the main elements of the theory: the relation
between the internal and surface temperatures of a neutron
star, the neutrino and photon cooling stages, etc. After the
discovery of neutron stars, a search for their soft X-ray
thermal emission has become a topical challenge, which
stimulated the development of the cooling theory. The first
decade of this development was reviewed by \citet{Tsuruta79}
and \citet{NomotoTsuruta81a}. 

\citet{Thorne77} presented the complete set of equations describing the
mechanical and thermal structure and evolution of a spherically
symmetric star at hydrostatic equilibrium in the framework of General
Relativity (GR). The GR effects on the thermal evolution of neutron
stars were first included into the cooling calculations by
\citet{GlenSutherland80,NomotoTsuruta81b,VanRiperLamb81}. A generally
relativistic cooling code for a spherically symmetric non-barotropic
star was written by \citet{Richardson-ea82}.
\citet{NomotoTsuruta86,NomotoTsuruta87} studied neutron star cooling
using an updated physics input and  discussed the role of different
physical factors for thermal relaxation of different models of neutron
stars. \citet{Tsuruta86} provided a comprehensive review of the neutron
star cooling theory with a comparison of the results of different
research groups obtained by mid-1980s.

The early studies of the neutron-star cooling were mostly focused on the
standard scenario where the neutrino emission from the stellar core was
produced mainly by the modified Urca (Murca) processes, which compete
with neutrino emission via plasmon decay, nucleon bremsstrahlung, etc.
The enhanced (accelerated) cooling due to the direct Urca (Durca)
processes was believed possible only if the core contains a pion
condensate or a quark plasma (e.g.,
\citealp{Tsuruta79,GlenSutherland80,VanRiperLamb81,Richardson-ea82}). 
By the end of 1980s a new cooling agent, kaon condensate, was introduced
\citep{Brown-ea88,PageBaron90}. The studies of the enhanced cooling were
intensified after the discovery by \citet{Lattimer-ea91} that the Durca
process is allowed in the neutron star core with the standard nuclear
composition for some realistic equations of state (EoS) without
``exotic'' models. The standard and  Durca-enhanced neutron star cooling
scenarios were compared in a number of numerical simulations starting
from \citet{PageApplegate92}, who also noticed that nucleon
superfluidity becomes the strongest  cooling regulator in the
Durca-allowed stellar kernels. This result triggered a flow of papers on
the cooling of superfluid neutron stars.

The progress in the theoretical studies of the neutron-star
thermal evolution was influenced in the 1980s and 1990s by
the spectacular progress of the X-ray astronomy, notably due
to the space observatories \textit{Einstein} (1978--1981),
\textit{EXOSAT} (1983--1986), and \textit{ROSAT}
(1990--1998). \textit{ROSAT} was the first to reliably
detect X-ray thermal radiation from isolated neutron stars.
This theoretical and observational progress was reviewed by
\citet{Tsuruta98,YakovlevPethick04,Pageetal2004}.

In the 21st century, the data collected by X-ray
observatories \textit{Chandra} and \textit{XMM-Newton} give
a new impetus to the development of the cooling theory. Some
new theoretical results on the cooling of neutron stars and
relation of the theory to observations were reviewed by
\citet{Yakovlev-ea08,Page09,Tsuruta09}. Recently, 2D
simulations of  the fully coupled thermal and magnetic field
evolution have been possible \citep{PonsMG09,vigano13},
mostly motivated by the increasing number of observed
magnetars and high magnetic field pulsars.  

The theory of thermal evolution of neutron stars has
different aspects associated with rotation, accretion, etc. 
In this review, we will mostly focus on the physics that
determines thermal structure and evolution of  slowly
rotating non-accreting neutron stars, whose thermal emission
can be substantially affected by strong magnetic fields. 
The processes of formation of thermal spectra in the
outermost layers of such stars are explicitly excluded from
this paper but considered in the companion review
(\citealp*{NSemitters}, hereafter Paper~I). We will pay a
special attention to the effects of strong magnetic fields
on the thermal structure and heat conduction in the crust
and heat-blanketing envelopes of neutron stars.

%%%%%%%%%%%%%%%%%%%%%%%%%%%%%%%%%%%%%
\section{The essential physics of neutron star cooling}
\label{sect:phys}
In this section we briefly present the essential physical
ingredients needed to build a model of a cooling neutron
star regardless its magnetic field. The effects of strong
magnetic fields will be discussed in subsequent sections,
starting from Sect.~\ref{sect:magnetic}.

%%%%%%%%%%%%%%%%%%%%%%%%%%%%%%%%%%%%%%%%%%%%%
\subsection{Structure and composition of a neutron star}
\label{sect:struct}
%%%%%%%%%%%%%%%%%%%%%%%%%%%%%%%%%%%%%%%%%%%%%

A neutron star is born hot ($\approx10^{11}$~K) and lepton-rich, but
only a few days after its birth, its temperature drops to a few $\times10^9$~K.
Thus,  the Fermi energy $\EF$ of all particles is much higher than the kinetic
thermal energy in most of the star volume, except in the thin outermost layers 
(a few meters thick), which does not affect the mechanical and
thermal structure of the rest of the star. Therefore,  a good
approximation is to describe the state of matter as cold nuclear matter in
beta equilibrium, resulting in an effectively barotropic EoS.
The mechanical structure of the star is decoupled from its
thermal structure and can be calculated only once and kept fixed during 
the thermal evolution simulations.

To a very good approximation, the  mechanical structure can
be assumed to be spherical. Appreciable deviations from the
spherical symmetry can be caused by ultra-strong magnetic
fields ($B\gtrsim10^{17}$~G) or by rotation with ultra-short
periods (less than a few milliseconds), but we will not
consider such extreme cases.  Then the space-time is
described by the Schwarzschild metric
\citep[e.g.][]{MisnerTW}
\beq
\dd s^2 = - \mathrm{e}^{2\Phi(r)} c^2 \dd t^2 +
\mathrm{e}^{2\Lambda(r)} \dd r^2 + r^2 (\dd \theta^2 +
\sin^2\theta \dd \varphi^2),
\label{Schw}
\eeq
where $(r,\theta,\varphi)$ are the standard spherical
coordinates, $\mathrm{e}^{2\Lambda(r)}=1-2GM_r/c^2r$,
and
$\Phi(r)$ is determined by equation
\beq
   \dd \Phi(r)/\dd P(r) = - \big[P(r) + \rho(r) c^2\big]^{-1}
\label{Phi}
\eeq
with the boundary condition $\mathrm{e}^{2\Phi(R)}=1-r_g/R$
at the stellar radius $R$.  Here,
$r_g=2GM/c^2=2.95(M/M_\odot)$ km is the Schwarzschild
radius, $M\equiv M_R$ is the stellar mass,
$M_r=4\pi\int_0^r\rho(r)r^2\dd r$ is the mass inside a sphere of radius
$r$, 
$G$ is the gravitational constant, $c$ is the speed
of light, $P$ is the pressure, and $\rho$ is the
gravitational mass density.

The mechanical structure of a spherically symmetric star is
described by the Tolman-Oppenheimer-Volkoff equation
\beq
   \frac{\dd P}{\dd r} = 
    - \frac{GM_r\,\rho}{r^2}\,
      \left( 1 + \frac{P}{\rho c^2} \right)\,
      \left( 1 + \frac{4\pi r^3 P}{M_r c^2} \right)\,
      \left( 1 - \frac{2GM_r}{rc^2} \right)^{-1}, % AP: corrected 15.12.15
\label{TOV}
\eeq
where $r$ is the radial coordinate measured from the stellar
center. In order to determine the stellar mechanical
structure, \req{TOV} should be supplemented by an EoS, which
depends on a microscopic physical model
(Sect.~\ref{sect:EOS}).
Several qualitatively different regions can be distinguished
in a neutron star, from the center to the surface: the inner
and outer core, the mantle, the inner and outer crust, the
ocean, and the atmosphere \citep[e.g.,][]{HPY07}.

\emph{The outer core} of a neutron star has mass density
$0.5 \rho_0 \lesssim \rho \lesssim 2\rho_0$, where
$\rho_0=2.8\times10^{14}$~\gcc{} is the nuclear saturation
density (the typical density of a heavy atomic nucleus). It
is usually several kilometers thick and contains most of the stellar mass. The outer core is mostly
composed of neutrons with an admixture of the
protons and leptons -- electrons and
muons ($npe\mu$ matter).

\emph{The inner core}, which can exist in rather
massive neutron stars, $M\gtrsim1.5\,M_\odot$, occupies
the central part with $\rho\gtrsim2\rho_0$. It is defined as the region where the 
composition is uncertain, but probably more rich than simply neutrons and protons.
Its composition and properties are not well known because the results of their calculation strongly
depend on details on the theoretical model of collective
fundamental interactions. Some of the proposed models
envision the following hypothetical options:
\begin{numlist}
\item hyperonization of matter -- the appearance of various
   hyperons (first of all, $\Lambda$- and
$\Sigma^-$-hyperons -- $npe\mu\Lambda\Sigma$ matter);
\item pion condensation -- formation of a Bose condensate
   of collective interactions with the properties of $\pi$-mesons;
\item kaon condensation -- formation of a similar 
     condensate of $K$-mesons;
\item deconfinement -- phase transition to quark matter.
\end{numlist}
The last three options are often called  \emph{exotic}
(\citealp{HPY07}, Chapt.~7). In this paper we will not
consider the exotic matter in any detail.

In the stellar \emph{crust and ocean} the matter is less
extraordinary: it contains electrons, nuclei composed of
protons and neutrons, and, in the inner crust, quasi-free
neutrons. Nevertheless, this region is also under extreme
conditions (density, temperature, magnetic field) that
cannot be reproduced in the laboratory. In the crust, which
is normally $\sim1$ km thick, the nuclei are arranged into a
crystalline lattice, and in the ocean with a typical depth
from a few to $\sim100$ meters (depending on temperature)
they form a liquid (see Sect.~\ref{sect:EOSouter}).

With increasing density, nuclei become progressively
neutron-rich due to the beta-captures that are favored by
the increase of pressure of the degenerate electrons.
Neutrons start to drip out of nuclei at density $\rhod =
4.3\times 10^{11}$ \gcc{}. Thus at $\rho>\rhod$ neutron-rich
nuclei are embedded in the sea of quasi-free neutrons.

At the bottom of the crust, the nuclei may take rodlike and
platelike shapes to compose so called \emph{pasta phases} of
nuclear matter \citep{PethickRavenhall95}. Then they form a
\emph{mantle} with anisotropic kinetic properties
\citep{PethickPotekhin96}. Thermodynamic stability of the
pasta phase state and, therefore, the existence of the
mantle depends on the model of nuclear interactions.
\citet{LorenzRP93} demonstrated stability of the pasta
phases at $\rho\gtrsim10^{14}$~\gcc{} for the FPS EoS model
of \citet{Pandharipande89}, but they were not found to be
stable in modern EoS models SLy \citep{DouchinHaensel01} and
BSk \citep{Pearson-ea12}.

The strong gravity drives the rapid separation of chemical
elements in the crust and the ocean.  Estimates of
characteristic sedimentation time range from seconds to months,
depending on local conditions and composition (see, e.g., Eq.~20
in \citealp{P14}), which is a very short timescale compared to the
stellar age. Therefore the envelopes are thought to be made of chemically
pure layers, which are separated by narrow transition bands
of diffusive mixing \citep{DeBlasio00,ChangBA10}.

%%%%%%%%%%%%%%%%%%%%%%%%%%%%%%%%%%%%%%%%%%%%%
\subsection{Thermal evolution equations}
\label{sect:equations}
%%%%%%%%%%%%%%%%%%%%%%%%%%%%%%%%%%%%%%%%%%%%%

The multidimensional heat transport and thermal evolution
equations in a locally flat reference frame read
\citep[e.g.,][]{AguileraPM08,PonsMG09,vigano13}
\beq
  c_\mathrm{v}\,\mathrm{e}^\Phi\,\frac{\partial T}{\partial t}
      + \nabla\cdot(\mathrm{e}^{2\Phi} \bm{F}) = 
           \mathrm{e}^{2\Phi} (H - Q_\nu),
\qquad
  \bm{F} = - \mathrm{e}^{-\Phi}
 \hat{\kappa}\cdot\nabla(\mathrm{e}^\Phi T),
\label{Tbalance}
\eeq
where $\bm{F}$ is the heat flux density, $H$ is the heating
power per unit volume, $c_\mathrm{v}$ is specific heat
(Sects.~\ref{sect:CV}, \ref{sect:SFc}, and
\ref{sect:EoSmag}), $Q_\nu$ is neutrino emissivity
(Sects.~\ref{sect:nu}, \ref{sect:SFnu},
\ref{sect:numag}), $\hat{\kappa}$ is the thermal
conductivity tensor (Sects.~\ref{sect:conductivity},
\ref{sect:SFcond}, and \ref{sect:condmag}), and
$\nabla=
(\mathrm{e}^{-\Lambda(r)}\partial/\partial r,\,
r^{-1}\partial/\partial \theta,\,
(r\sin\theta)^{-1}\partial/\partial \varphi)$ in
compliance with \req{Schw}.
The inner boundary condition to the system of
equations (\ref{Tbalance}) is $\bm{F}=0$ at
$r=0$. The outer boundary condition is
determined by the properties of a \emph{heat-blanketing
envelope}, which serves as a mediator of the internal heat
into the outgoing thermal radiation. It will be considered in
Sect.~\ref{sect:th_str}. Solutions to the thermal
evolution equations and their implications are briefly
reviewed in Sect.~\ref{sect:cooling}.

For weak magnetic fields, we can
assume that the temperature gradients are essentially
radial, and that in most of the star volume (inner crust and
core) the conductivity tensor is simply a scalar quantity
times the identity matrix. In this limit, corrections for
deviations  from the 1D approximation have little effect on
the total luminosity. However, for strong fields and neutron
stars with locally intense internal heating sources, such as
magnetars,  a more accurate description, beyond the 1D
approximation, must be considered. 

2D calculations of thermal structure and evolution of
strongly magnetized neutron stars have been done by several 
groups 
\citep{GeppertKP04,GeppertKP06,PerezAMP06,AguileraPM08,Kaminker-ea12,Kaminker-ea14}.
In some of these works
\citep{GeppertKP06,PerezAMP06,AguileraPM08}, neutron-star
models with superstrong ($B\sim10^{15}$\,--\,$10^{16}$~G)
toroidal magnetic fields in the crust were considered, in
addition to the less strong
($B\sim10^{12}$\,--\,$10^{14}$~G) poloidal component that
penetrates from the crust into the magnetosphere. The latter
models help to  explain the strongly non-uniform distribution
of the effective temperature over the neutron-star surface
and the possible energy source for magnetars outbursts
\citep{PonsPerna11,PonsRea12}. Only recently
\citep{vigano13}, the fully coupled evolution of temperature
and magnetic field has been studied with detailed numerical
simulations, which allow one to follow the long-term evolution
of magnetars and  their connection with other neutron star
classes. Some results of such calculations will be discussed
in Sect.~\ref{sect:cooling}.

%%%%%%%%%%%%%%%%%%%%%%%%%%%%%%%%%%%%%%%%%%%%%
\subsection{Basic plasma parameters}
\label{sect:param}
%%%%%%%%%%%%%%%%%%%%%%%%%%%%%%%%%%%%%%%%%%%%%

In this section we introduce several basic parameters of
Coulomb plasmas that are used below. To be concrete,
we start with electrons and ions (including bare atomic
nuclei). When other charged particles are present, their
respective parameters are defined analogously, with the
obvious replacements of particle mass, charge, number
density, etc.

Since the major constituents of the neutron-star matter are
mostly degenerate, an important parameter is the Fermi
energy, which (without the rest energy) equals 
\beq
   \EF = c\,\sqrt{(m c)^2 + (\pF)^2}-m c^2,
\label{EF}
\eeq
where $m$ is the
particle mass, and $\pF$ is the Fermi
momentum. For instance, for the Fermi gas in the absence of a quantizing
magnetic field, $\pF = \hbar \,(3 \pi^2 n)^{1/3}$, where
$n$ is the number density, and
$\hbar$ is the reduced Planck constant. It is
convenient to use the dimensionless density parameter
related to the Fermi momentum of electrons, 
$\xr  = \pFe / \mel c$, where
$\mel$ is the electron mass. In
the outer core and the envelopes, as long as the baryons are
non-relativistic, $\xr \approx ( \rho_6 \, Y_e )^{1/3}$,
where $Y_e$ is the number of
electrons per baryon and $\rho_6\equiv\rho/10^6$
g~cm$^{-3}$.

Thermal de Broglie wavelengths of free ions and electrons are
usually defined as
$
   \lambdi = \sqrt{2\pi\hbar^2/\mion T}
$
and
$
   \lambde = \sqrt{2\pi\hbar^2/ \mel
T},
$
where $\mion=Am_\mathrm{u}$ is the ion
mass, and $m_\mathrm{u}$ is the unified atomic mass unit.
Here and hereafter, we use $T$ in energy
units and suppress the Boltzmann constant
(i.e., $10^6\mbox{ K}=86.17$ eV).
The quantum effects on ion motion are important
either at $\lambdi\gtrsim \aion$ or at 
$T \ll \Tp$, where $\Tp = \hbar \omp$
is the ion plasma temperature, and
$
    \omp = \left(  {4 \pi e^2 \,\nion}\,
     Z^2 /\mion \right)^{1/2}
$
is the ion plasma frequency. Debye temperature of a crystal
$\Theta_\mathrm{D}$
is closely related to the plasma temperature. In the
harmonic approximation for the Coulomb crystal, 
$\Theta_\mathrm{D} \approx 0.45\,\Tp$ \citep{Carr61}.

The Coulomb plasmas are called strongly coupled if the
parameter $\Gami=(Z e)^2/\aion T$, which estimates the
electrostatic to thermal energy ratio, is large. Here,
$\aion\equiv(\frac43\pi \nion)^{-1/3}$ is the ion sphere, or Wigner-Seitz cell, radius,
and $\nion$ is the ion number density.
If the plasma only consists
of electrons and non-relativistic ions of one kind, which is
typical for neutron-star envelopes, then
\beq
\Tp = 7.832\, (Z/A)\sqrt{\rho_6}\times10^6\mbox{~K},
\qquad
\Gami = 22.747 \,Z^{5/3}(\rho Y_e)^{1/3}/T_6.
\eeq
Analogously, $\Tpe = \hbar\, (4\pi e^2 n_e/\mel)^{1/2}
= 3.34\,\sqrt{\rho_6\, Z/A}\times10^8$~K is the
electron plasma temperature.
Other plasma parameters, which become important in a strong
magnetic field, will be considered in
Sect.~\ref{sect:mag-par}.

%%%%%%%%%%%%%%%%%%%%%%%%%%%%%%%%%%%%%%%%%%%%%
\subsection{Equation of state}
\label{sect:EOS}
%%%%%%%%%%%%%%%%%%%%%%%%%%%%%%%%%%%%%%%%%%%%%

% --------------------------------------------------
\subsubsection{Equation of state for the outer crust and the
ocean}
\label{sect:EOSouter}

The composition of the outer crust and the ocean of a
neutron star is particularly simple: their basic
constituents are electrons and nuclei, which, to a good
accuracy, can be treated as pointlike. The EoS of such
electron-ion plasmas is well known (see, e.g.,
\citealt{HPY07}, Chapt.~2, and references therein). 

The ions thermodynamic state will go from liquid to solid as
the star cools, and in the solid state from a classical to a
quantum crystal. It is generally assumed that the ions form
a crystalline solid and not an amorphous one. This
assumption is confirmed by molecular dynamics numerical
simulations \citep{Hughto-ea11} and corroborated by the
analysis of observations of neutron-star crust cooling after
an accretion episode (see Sect.~\ref{sect:heating}).

The simplest model of the electron-ion plasmas is  the one
component plasma (OCP) model, which considers Coulomb
interactions of identical pointlike ions and replaces the
degenerate electron gas by a static uniform
charge-compensating background. The OCP has been studied
analytically and numerically in many papers (see
\citealt{HPY07}, Chapt.~2, for references). In the classical
regime ($T\gg\Tp$) its thermodynamic functions depend on a
single parameter $\Gami$. At $\Gami\ll1$ the ions form a
Debye-H\"uckel gas, with increasing $\Gami$ the gas
gradually becomes a liquid, and with further increase of
$\Gami$ the OCP liquid freezes. An analysis of Monte Carlo
simulations of the OCP shows that its ground state is
crystalline when $\Gami>\Gammam=175$ \citep{PC00}. However,
supercooling cannot be excluded up to a value $\Gami\simeq
250$. Indeed, Monte Carlo simulations of freezing of
classical OCP \citep{DeWittSY93} indicate that, as a rule,
the ions do not freeze at the equilibrium melting
temperature $\Tm$ but form a supercooled fluid and freeze at
lower $T$ (depending on initial conditions and other
parameters). This happens because the phase transition is
really tiny.

At $T\lesssim\Tp$, the quantum effects on ion motion become
significant. Then thermodynamic functions depend not only on
$\Gami$, but also on $r_s$. The quantum effects are
especially important for the solid neutron star crust at high
densities, although they can also be significant in the deep layers
of the ocean composed of light elements (for instance, they
prevent solidification of H and He plasmas). The free
energy per unit volume of an OCP crystal can be written as
\beq
   \frac{F_\mathrm{lat}}{V} = \nion C_0
\,\frac{(Ze)^2}{\aion }
    + \frac32\, \nion \, u_1 \Tp + 
     \frac{F_\mathrm{th}}{V} + \frac{F_\mathrm{ah}}{V},
\label{Flat}
\eeq
where the first term is the classical static-lattice energy,
$C_0\approx-0.9$ is the Madelung constant, and the next two
terms describe thermodynamics of the phonon gas in the
harmonic approximation \citep[e.g.,][]{Kittel63}: the second
term accounts for zero-point quantum vibrations, and
$F_\mathrm{th}/V = 3 \nion T \left\langle\ln
[1-\exp(-\hbar\omega_{\bm{k}\alpha} / T)]
\right\rangle_\mathrm{ph}$ is the thermal contribution.
Here, 
$u_1=\langle\omega_{\bm{k}\alpha}\rangle_\mathrm{ph}/\omp\approx0.5$
is the reduced first moment of phonon frequencies 
$\omega_{\bm{k}\alpha}$, and
$\langle\ldots\rangle_\mathrm{ph}$ denotes the averaging
over phonon polarizations $\alpha$ and wave vectors $\bm{k}$
in the first Brillouin zone. The last term in \req{Flat}
arises from anharmonic corrections, which have only been studied
in detail in the classical regime ($T\gg\Tp$; e.g.,
\citealp{FaroukiHamaguchi} and references therein).
An analytical extrapolation of $F_\mathrm{ah}$ for any
$T$ was proposed in \citet{PC10}.

For mixtures of various ion species, the simplest
evaluation of the thermodynamic functions is given by the
average of their values for pure substances, weighed with
their number fractions, which is called the  linear mixing
rule \citep{HTV77}. The linear mixing rule is accurate
within a few percent, if the electrons are strongly
degenerate and $\Gami>1$ for each of the ion species in the
mixture. However, this accuracy may be insufficient for such
subtle phenomena as melting/freezing or phase separation in
the Coulomb plasmas. Corrections to the linear mixing rule
were obtained by \citet{Pmix09}. \citet{MedinCumming10} used
these results to construct a semianalytical model for
prediction of the composition and phase state of
multicomponent mixtures. \citet{Hughto-ea12} confirmed the
qualitative validity of this model by molecular dynamics simulations.

The OCP model is a reasonable first approximation, but in
reality the electrons do not form a uniform background: they
interact with each other and with the ions, which gives rise
to exchange-correlation and polarization corrections,
respectively. The polarization corrections are appreciable
even for strongly degenerate plasmas. For instance, they
can substantially shift the melting transition away from
$\Gami=175$ \citep{PC13}. In the outer envelopes
of a neutron star, the electron degeneracy weakens, and one
should take the $T$-dependence of their EoS into account.
Analytical fits for all above-mentioned contributions to the
EoS of electron-ion plasmas were presented by
\citet{PC10,PC13}.
Their Fortran implementation is publicly available at
\texttt{http://www.ioffe.ru/astro/EIP/}.

An essential input for calculating the EoS is the chemical
composition of the plasma. The ground state of the matter in
the outer crust can be found following the method of
\citet{BPS}. The procedure, based on the minimization of the
Gibbs free energy per nucleon, is described in detail in
\citet{HPY07}. The structure of the crust is completely
determined by the experimental nuclear data up to a density
of the order $\rho\sim6\times10^{10}$ \gcc. At higher
densities the nuclei are so neutron rich that they have not
yet been experimentally studied, and the composition of
these dense layers is model dependent. However, this model
dependence is not very significant in the models based on
modern nuclear physics data
\citep{HP94,RuesterHSB06,PearsonGC11}.

While a newly born neutron star is made of hot matter in
nuclear equilibrium, its  subsequent evolution can lead to
the formation of regions where the matter is out of nuclear
equilibrium. This is the case of an old accreting neutron
star. Burning of the helium layer near the surface is
followed by electron captures and beta decays in deeper
layers. The growing layer of the processed accreted matter
pushes down and eventually replaces the original catalyzed
(ground-state) crust. These processes were studied by
several authors \citep[see][and references therein]{HZ90,HZ08}.

% --------------------------------------------------
\subsubsection{Equation of state for the inner crust and the core}
\label{sect:EOScore}

The pressure in the inner crust of a neutron star is largely
created by degenerate neutrons. However, the electrons and
nuclei may give an important contribution to the heat
capacity (see Sect.~\ref{sect:CV}). In the core, there are
contributions from neutrons, protons, electrons and muons
(and other particles in the exotic models, which we do not
consider here). Different theoretical EoSs of the neutron
fluid and $npe\mu$ matter have been proposed, based on
different methods of theoretical physics: the
Brueckner-Bethe-Goldstone theory, the Green's function
method, variational methods, the relativistic mean field
theory, and the density functional method (see
\citealp{HPY07}, Chapt.~5, for review). The model of
\citet{APR} (APR) has been often cited as the most advanced
one for the core. It was derived from the variational
principle of quantum mechanics, under which an energy
minimum for the trial wave function was sought. The trial
function was constructed by applying the linear combination
of operators describing admissible symmetry transformations
in the coordinate, spin, and isospin spaces to the Slater
determinant consisting of wave functions for free nucleons.
The APR EoS exists in several variants, which differ in the
effective potentials of  nucleon-nucleon interaction and in
relativistic boost corrections. The potentials borrowed from
earlier publications were optimized so as to most accurately
reproduce the results of nuclear physics experiments.

Many theoretical neutron-star EoSs in the literature consist
of crust and core segments obtained using different physical
models. The crust-core interface there has no physical
meaning, and both segments are joined using an ad hoc
matching procedure. This generally leads to thermodynamic
inconsistencies. The EoS models that avoid this problem by
describing the core and the crust in frames of the same
physical model are called \emph{unified}. Examples of the
unified EoSs are the FPS \citep{Pandharipande89,LorenzRP93}, SLy
\citep{DouchinHaensel01}, and BSk
\citep{Goriely-ea10,PearsonGC11,Pearson-ea12} EoS families.
All of them are based on effective Skyrme-like energy
density functionals. In particular, the BSk21 model is based
on a generalized Skyrme functional that most successfully
satisfies various experimental restrictions along with a
number of astrophysical requirements (see the discussion in
\citealp{Potekhin-ea13}). 

%%%%%%%%%%%%%%%%%%%%%%%%%%%%%%%%%%%%%%%%%%%%%
\subsection{Specific heat}
\label{sect:CV}
%%%%%%%%%%%%%%%%%%%%%%%%%%%%%%%%%%%%%%%%%%%%%

% --------------------------------------------------
\subsubsection{Specific heat of electron-ion plasmas}
\label{sect:CVcrust}

The two components that largely dominate the specific heat
$c_\mathrm{v}$ in the crust are the electron gas and the
ions. In
the neutron-star crust and core, the electrons form an
ultra-relativistic highly degenerate Fermi gas, and their
contribution in the heat capacity per unit volume is simply
given by
\beq
c_{\mathrm{v},e} = 
\frac{\pFe^2 }{3 \hbar^3 c} T =
\nel \frac{\pi^2}{\pFe \, c} T.
\label{cve}
\eeq
In the ocean, where the density is lower, approximation
(\ref{cve}) may not work. Then it is advisable to use
accurate approximations, cited in Sect.~\ref{sect:EOSouter}.

%----------------------------------------------------------------------------
\begin{figure}
\begin{center}
  \includegraphics[height=.49\textwidth]{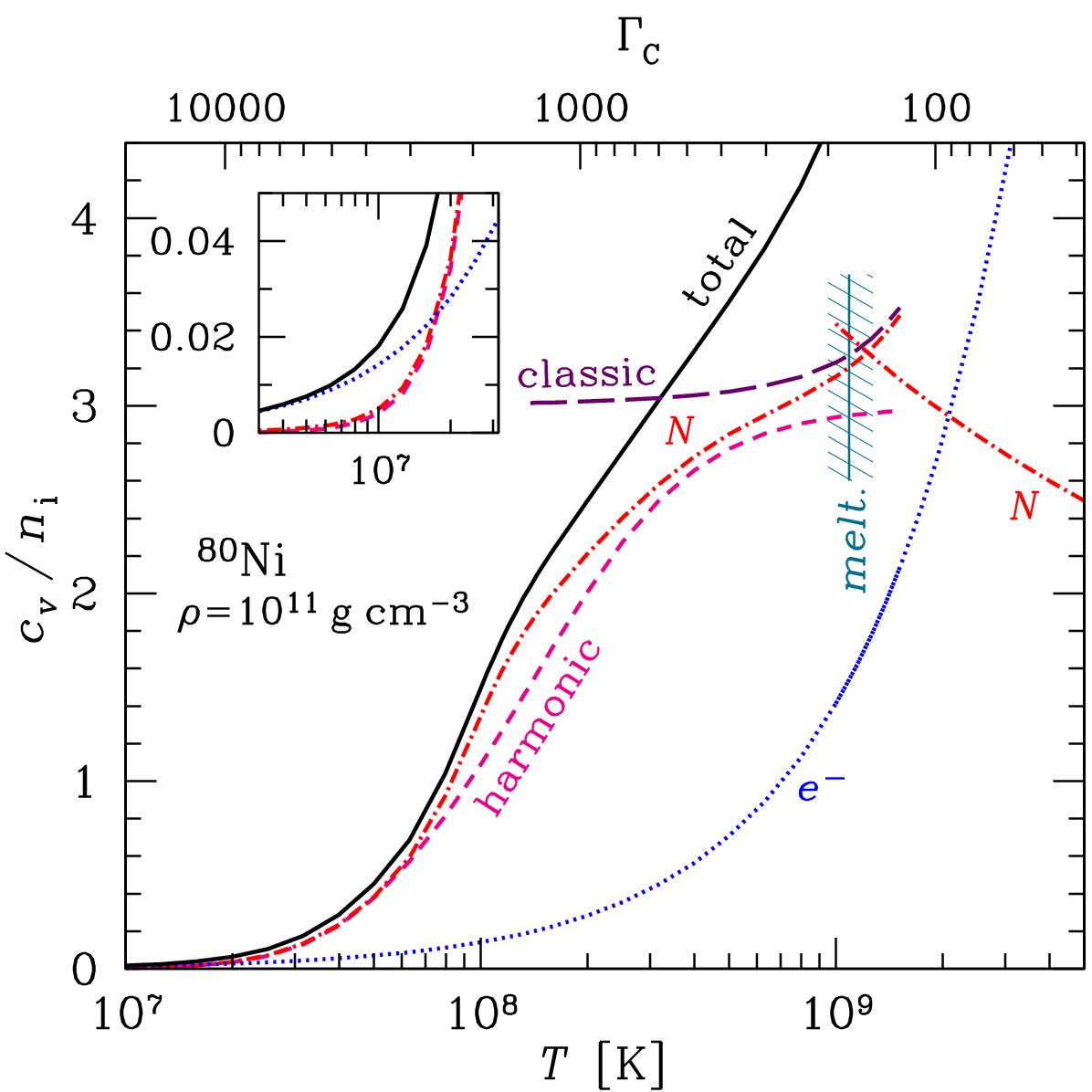}
  \includegraphics[height=.49\textwidth]{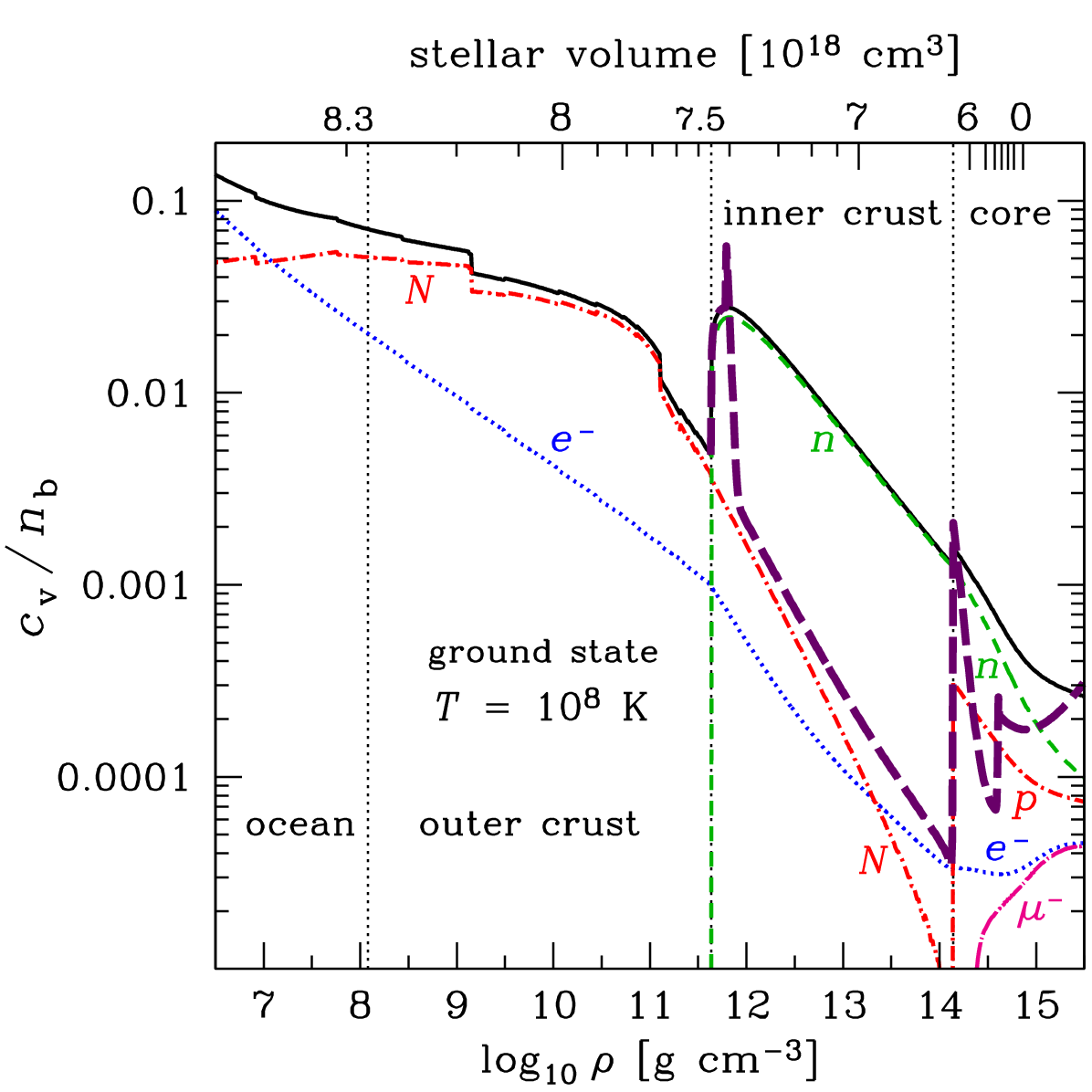}
\caption{
Left panel: 
Heat capacity per ion versus $T$ ({bottom axis}) and $\Gami$
({top axis}) for $^{80}$Ni at $\rho=10^{11}$ \gcc{}. The
solid line displays the total normalized heat capacity
$c_\mathrm{v}/\nion$; the long-dashed line shows this
quantity for a classical Coulomb lattice of ions, including
harmonic and anharmonic terms; the short-dashed line is the
harmonic-lattice approximation in the solid phase; the
dot-dashed line is the same plus anharmonic and electron
polarization corrections in the solid phase. The dotted line
is the electron Fermi gas contribution. The vertical line is
the OCP melting point $\Gami=175$, and the hatched band
shows the range $\Gami=150$\,--\,200, where melting is
expected to occur in realistic conditions. The inset
illustrates the competition between the electron and ion
contributions t low $T$.
Right panel:
Heat capacity per baryon as function of mass density from
the ocean throughout the crust and core of a neutron star at
$T=10^8$~K. The solid line displays the total
$c_\mathrm{v}/n_\mathrm{b}$, and the other lines show its
constituents due to the electrons ($e^-$), neutrons ($n$) in
the inner crust and core, nuclei ($N$), including 
electrostatic terms in the ocean and crust but neglecting
the neutron entrainment effects in the inner crust
(Sect.~\ref{sect:CVn}), protons ($p$) and muons ($\mu^-$) in
the core, assuming that the nucleons are non-superfluid. For
comparison, the thick long dashes display an example of the
total $c_\mathrm{v}/n_\mathrm{b}$ in the inner crust and
core in the case of superfluid nucleons
(Sect.~\ref{sect:SFc}). The top axis shows the volume
contained inside a sphere with given $\rho$ for a 1.4
$M_\odot$ neutron star. The stellar structure and
composition correspond to the BSk21 EoS model.
}
\label{fig:CVcrust}
\end{center}
\end{figure}
%----------------------------------------------------------------------------

In \fig{fig:CVcrust} we show the temperature and density
dependences of the normalized heat capacity of the
ground-state matter in a neutron star. The left panel
illustrates the dependence of $c_\mathrm{v}/n_\mathrm{i}$ on
$T$, and the right panel the dependence of
$c_\mathrm{v}/n_\mathrm{b}$ on $\rho$. Since the electron
polarization effects shift the melting temperature
(Sect.~\ref{sect:EOSouter}), the phase transition may occur
anywhere within the hatched region around the vertical line
$\Gami=175$ in the left panel.

When the temperature of the Coulomb liquid
decreases, the heat capacity per ion increases from the ideal-gas
value $c_\mathrm{v,i}/\nion=\frac32$ at $T\gg\Tm$ to,
approximately, the simple harmonic lattice value
$c_\mathrm{v,i}/\nion=3$ at $T\lesssim\Tm$ (the Dulong-Petit
law for a classical harmonic crystal). This gradual increase
is due to the
Coulomb non-ideality in the liquid phase, which effectively
smears a phase transition between the strongly coupled
Coulomb liquid and OCP crystal \citep[see][]{Baiko-ea98}.
With further cooling, quantum effects suppress the heat
capacity. Once the crystal is deep into the quantum regime
its specific heat is given by the Debye result
\beq
c_\mathrm{v,i}^\mathrm{(D)} =
\nion \frac{12 \pi^4}{5} \left(\frac{T}{\Theta_\mathrm{D}}
\right)^3.
\label{Eq:Cv_Debye}
\eeq
The calculations of \citet{BPY01}
show that the Dulong-Petit law applies at temperatures down
to $T \simeq \Tp$, while the Debye value of
\req{Eq:Cv_Debye} is attained when $T \lesssim 0.1\,\Tp$.
The same authors present a simple analytical approximation
for the heat capacity of a harmonic Coulomb crystal,
accurate to a few parts in $10^5$ at any $T$.

However, the harmonic OCP model is an idealization. The
anharmonic corrections and electron polarization corrections
(Sect.~\ref{sect:EOSouter}) can amount up to
(10\,--\,20)\,\% of $c_\mathrm{v,i}$. Because of the
anharmonic effects, $c_\mathrm{v,i}/\nion$ is not equal to 3
exactly, but is $\sim10$\% larger at $T=\Tm$. If the
above-mentioned supercooling takes place in stellar matter,
various fluid elements solidify at different $T$ below
$\Tm$, and the average heat capacity, as function of
temperature, can contain a bump, associated with latent heat
releases (see Sect.~2.4.6 of \citealp{HPY07} for a
discussion).

The right panel of \fig{fig:CVcrust} shows the density
dependence of the total heat capacity, normalized per one
nucleon, $c_\mathrm{v}/\nb$, throughout the neutron star
from the ocean to the core, and partial contributions to 
$c_\mathrm{v}/\nb$. Different particle fractions are adopted
from the BSk21 model \citep{Goriely-ea10,PearsonGC11,Pearson-ea12}, as
fitted by \citet{Potekhin-ea13}. Here, we have mostly
neglected the effects of nucleon superfluidity to be
discussed in Sect.~\ref{sect:SF}. The importance of these
effects is demonstrated, however, by the heavy long-dashed
line, which displays the total normalized heat capacity
suppressed by nucleon superfluidity (see
Sect.~\ref{sect:SFc}).

% --------------------------------------------------
\subsubsection{Specific heat of neutrons}
\label{sect:CVn}

In the inner crust, besides electrons and nuclei, there are
also neutrons. In a thin layer at densities $\rho$ just
above the neutron drip point $\rhod$, the dripped neutrons
are not paired (non-superfluid) and largely dominate
$c_\mathrm{v}$. Heat capacity of strongly degenerate
non-superfluid neutrons can be accurately evaluated using the
above-referenced analytical fits, but since the neutrons are
strongly degenerate almost everywhere in the neutron star,
the simpler Sommerfeld result for Fermi gases at $T\ll\EF$
is usually applicable,  \beq c_\mathrm{v,x}
\sim\frac{\pi^2}{2}\,\frac{n_\mathrm{x}\,T}{\EFx},
\label{Sommerfeld} \eeq where x stands for the fermion type
($\textrm{x}=n,\,p,\,e,\,\mu$).  For neutrons at $\rho$ only
slightly above $\rhod$, however, the latter formula is
inaccurate because $\EFn$ is not sufficiently large. For
this reason, \citet{PastoreCM15} proposed an interpolation
between  \req{Sommerfeld} and the ideal-gas limit
$c_\mathrm{v} = \frac32n$, \beq c_\mathrm{v,x} \approx
\frac32\,n_\mathrm{x} \left(1 -
\mathrm{e}^{-T/T_\mathrm{cl}} \right), \quad T_\mathrm{cl} =
\frac{3\EFx}{\pi^2}. \label{cvx} \eeq They also showed that
corrections due to the coupling to phonons \citep[e.g.,
Sect.~1.4.4 in][]{BaymPethick} turn out to be unimportant
for $c_{\mathrm{v},n}$. Approximation (\ref{cvx}) is
accurate within $17\%$ for non-relativistic Fermi gases at
any density. For a relativistic Fermi gas, we can preserve
this accuracy by using \req{EF} for $\EFx$ and multiplying
both $T_\mathrm{cl}$ and prefactor $\frac32$ by the ratio
$(mc^2+10T)/(mc^2+5T)$.

%----------------------------------------------------------------------------
\begin{figure}
\begin{center}
\includegraphics[width=\textwidth]{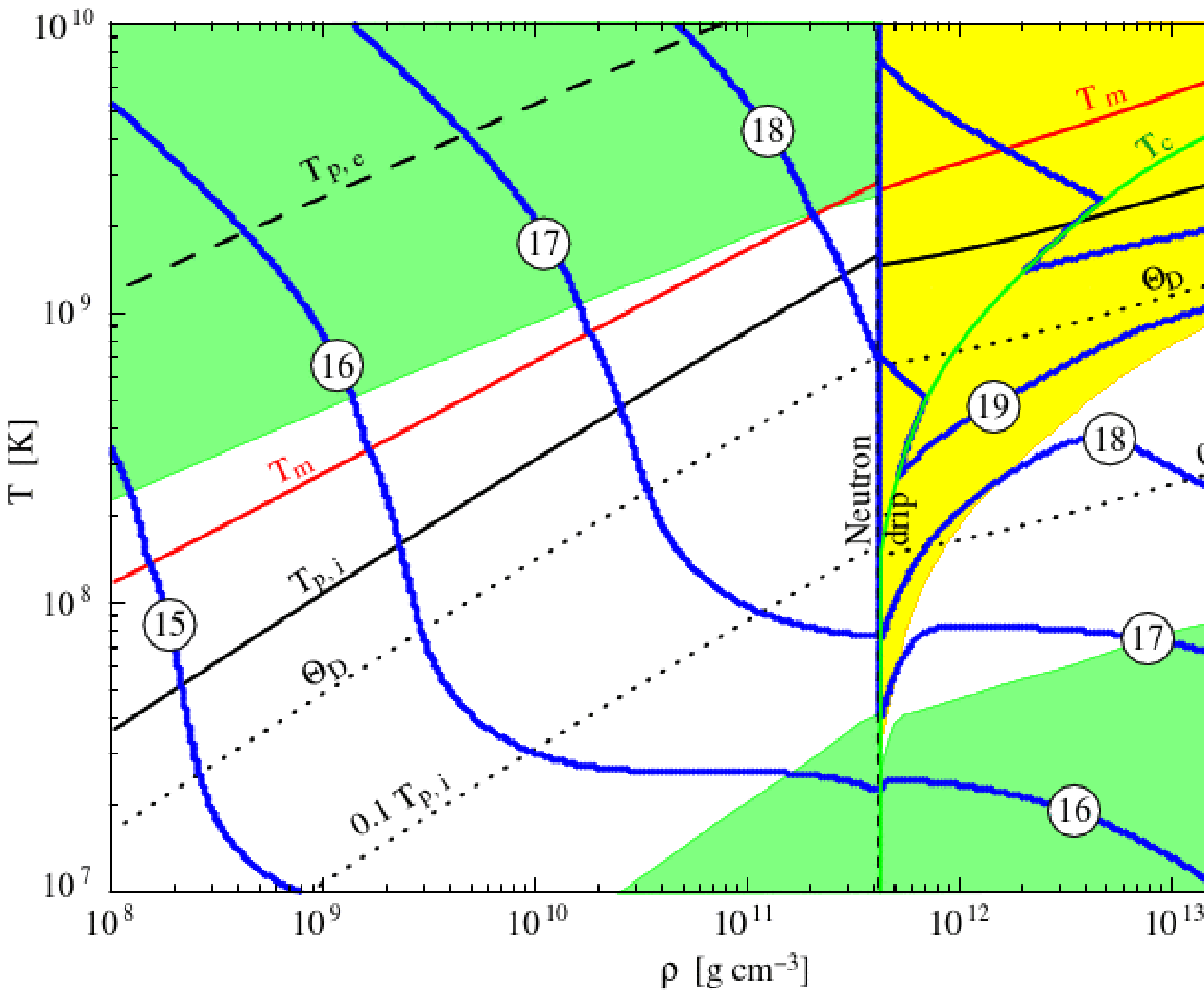}
\caption{
Iso-contour lines of $c_\mathrm{v}$ in the crust, labeled by
the value of $\mathrm{log}_{10} (c_\mathrm{v}/\mathrm{erg \,
cm^{-3} \, K^{-1}})$. Also shown are the melting curve
$\Tm$  and the critical temperature for neutron $^1$S$_0$
superfluidity, $\Tc$, the electron and ion plasma
temperatures, $\Tpe$ and $\Tp$ respectively,
the Debye temperature, $\Theta_\mathrm{D} \simeq 0.45 \Tp$, that marks the transition
from classical to quantum solid and  $0.1 \Tp$ below which the wholly quantum crystal regime is realized.
The outer crust chemical composition is from \citet{HP94} and inner crust from \citet{NV73} with
the neutron drip point at $\rhod = 4.3\times 10^{11}$ g cm $^{-3}$.
The electron contribution dominates in the two dark-shadowed
(green) regions at high $T$ and $\rho$ below $\rhod$
and at low $T$ and high $\rho$, while neutrons dominate
in the light-shadowed (yellow) region at high
$T$ and $\rho$ above $\rhod$,
and ions dominate in the intermediate regime.
The right panel only displays the inner crust but assuming
that about 80\% of the dripped neutrons are entrained,
illustrating the resulting increase in $c_\mathrm{v,i}$,
mainly due to the strong reduction of $\Tp$ and
$\Theta_\mathrm{D}$, significantly extending the regime
where $c_\mathrm{v,i}$ dominates over $c_{\mathrm{v},e}$.
}
\label{fig:CVcrust2}
\end{center}
\end{figure}
%----------------------------------------------------------------------------

With further density increase, the neutrons become
superfluid (Sect.~\ref{sect:SF}), and then their
contribution to $c_\mathrm{v}$ nearly vanishes. However,
even in a superfluid state, the neutrons have a dramatic
effect on $c_\mathrm{v}$. Indeed, \citet{FlowersItoh76}
noticed that since free neutrons move in a periodic
potential created by lattice of atomic nuclei, their energy
spectrum should have a band structure, which can affect
kinetic and neutrino emission phenomena involving the free
neutrons. \citet{Chamel2005} calculated the band structure
of these neutrons, in much the same way as electron band
structure is calculated in solid state physics. The effect
of this band structure is that a large fraction of the
dripped neutrons are ``locked'' to the nuclei, i.e., the
thermal motion of the nuclei entrains a significant part of
the dripped neutrons resulting in a strongly increased ion
effective mass $m_\mathrm{eff,i}$. This increase $\mion
\rightarrow m_\mathrm{eff,i}$ significantly increases 
$c_\mathrm{v,i}$ in the quantum regime since 
$c_\mathrm{v,i}^\mathrm{(D)} \propto \Tp^{-1} \propto
\mion^{3/2}$ \citep{Chamel13}.

The overall ``landscape'' of crustal specific heat is
illustrated in \fig{fig:CVcrust2}. For highly degenerate
electrons $c_{\mathrm{v},e} \propto T$, while for ions
$c_\mathrm{v,i}$ decreases as $T^3$ according to
\req{Eq:Cv_Debye}, therefore the electron contribution
dominates at $T \ll \Theta_\mathrm{D}$, and the ion contribution
prevails at $T\gtrsim\Theta_\mathrm{D}$ (cf.~the inset in
the left panel of \fig{fig:CVcrust}). On the other hand, in the
non-degenerate regime $c_{\mathrm{v},e}/c_\mathrm{v,i}\sim Z$,
therefore the contribution of the electrons dominates again
for $Z>1$  at $T \gtrsim \EF$ in the liquid phase (also cf.~the
left panel of \fig{fig:CVcrust}). The effect of dripped
neutron band structure on low-level collective excitations
in the inner crust and the resulting increase of
$c_\mathrm{v,i}$ is illustrated in the right panel of
\fig{fig:CVcrust2}.

%%%%%%%%%%%%%%%%%%%%%%%%%%%%%%%%%%%%%%%%%%%%%
\subsubsection{Specific heat of the core}
\label{sect:CVcore}

The specific heat is simpler to evaluate in the core than in
the crust but it has larger uncertainties. The core is a
homogeneous quantum liquid of strongly degenerate fermions,
and its specific heat is simply taken as the sum of its
components contribution: $c_\mathrm{v} = \sum_\mathrm{x}
c_\mathrm{v,x}$ where x stands for neutrons ($n$), protons
($p$), electrons ($e$), muons ($\mu$), and any other
component as hyperons or quarks that may appear at high
densities. For each fermionic component, one can use
\req{Sommerfeld}, but for baryons one should replace the bare
fermion mass $m_\mathrm{x}$ by an effective mass
$m_\mathrm{x}^*$, which encapsulates most effects of
interactions. In principle, $m_\mathrm{x}^*$ should be
calculated from the same microphysical interaction as
employed for the EoS; cf.{} Sect.~\ref{sect:medium}. For
leptons ($e$ and $\mu$), interactions have a negligible
effect on $m_\mathrm{x}^*$ and the bare fermion mass value
can be used. The nucleon heat capacity in the core is
strongly affected by pairing (superfluidity) effects, 
as discussed in Sect.~\ref{sect:SFc}.

%%%%%%%%%%%%%%%%%%%%%%%%%%%%%%%%%%%%%%%%%%%%%
\subsection{Neutrino emissivity}
\label{sect:nu}
%%%%%%%%%%%%%%%%%%%%%%%%%%%%%%%%%%%%%%%%%%%%%

The neutrino luminosity of a neutron star is, in most cases,
strongly dominated by the core contribution, simply because
the core comprises a lion's share of the total mass. The
crust contribution can, however, prevail in the case of
strong superfluidity in the core, which suppresses the
neutrino emissivities. Crust neutrino emission is also
essential during the early thermal relaxation phase (the
first few decades of the life of the star), or the crust
relaxation after energetic transient events (e.g., strong
bursts of accretion in X-ray binaries and flares in
magnetars).

\citet{YKGH01} reviewed the main neutrino emission
mechanisms in neutron star crusts and cores and collected
fitting formulae for the neutrino emissivity in each reaction as 
a function of density and temperature. The
summary of the most important processes is given in
Table~\ref{tab:nu}. The last column of this table contains
references to the analytical fitting formulae that can be
directly employed to calculate the relevant emission rates.
These processes are briefly described below.

%%%%%%%%%%%%%%%%%%%%%%%%%%%%%%%%%%%%%%%%%%%%%%%%%%%%%%%%%%%%%%%%%%%%%%%%%%%%
\begin{table}[t]
\caption{Main neutrino emission processes$^\mathrm{a}$}
\setlength{\tabcolsep}{4pt}
\begin{center}
  \begin{tabular}{llll}
  \hline \hline
   & Process / Control function  & Symbolic
   notation$^\mathrm{b}$ & Formulae for $Q_\nu$ and/or $R$\rule[-1.5ex]{0pt}{4ex} \\
\hline
\multicolumn{4}{c}{\emph{In the
crust}\rule{0pt}{2.5ex}}\\[.5ex]
 1 & Plasmon decay                   & $ \Gamma \to \nu + \bar{\nu}$
                                     & Eqs.~(15)\,--\,(32) of [1]  \\
 2 & Electron-nucleus bremsstrahlung
   & $ e^- + N \to e^- + N + \nu + \bar{\nu}$
                                     & Eqs.~(6), (16)\,--\,(21) of [2] \\[.5ex]
 3 & Electron-positron annihilation      & $e^- + e^+ \to \nu + \bar{\nu}$
                                     & Eq.~(22) of [3]  \\
 4$^\mathrm{c}$ & Electron synchrotron
   & $e^-\stackrel{B}{\to} e^- + \nu + \bar\nu$
                          & Eq.~(48)\,--\,(57) of [3]  \\[1ex]
  \hline
\multicolumn{4}{c}{\emph{In the core}\rule{0pt}{2.5ex}}\\[.5ex]
 \parbox[c]{2ex}{1$^\mathrm{d}$\\} & \parbox[c]{10em}{Direct Urca\\
              (Durca)}
   & $\begin{array}{l}
      n \rightarrow p + e^- + \bar\nu_e\,, \\
      \,\,\, p + e^- \rightarrow n+ \nu_e
      \end{array}$
   & Eq.~(120) of [3] \\[2ex]
   & Magnetic modification$^\mathrm{c}$ & $R_B^\mathrm{(D)}$
                                   & Eqs.~(247)\,--\,(250) of [3] \\[.5ex]
   & Reduction factors$^\mathrm{e}$ & $R_x^\mathrm{(D)}$ 
                                   & Eqs.~(199), (202)--(206) of [3] \\[1ex]
 \parbox[c]{1ex}{2\\} & \parbox[c]{3.4cm}{Modified Urca
           (Murca)\\ (neutron branch)}
   & $\begin{array}{l}
       n+n \rightarrow n+p+e^-+\bar\nu_e\,, \\
      \,\,\, n+p+e^- \rightarrow n+n+\nu_e
      \end{array}$
   & Eq.~(140) of [3] \\[2ex]
   & Reduction factors$^\mathrm{e}$ & $R_{x}^{(\mathrm{M}n)}$ 
   & \parbox[c]{20ex}{Appendix\,of\,[4]}\\[1.ex]
 \parbox[c]{1ex}{3\\} & \parbox[c]{3.4cm}{Murca\\ (proton branch)}
   & $\begin{array}{l}
      p+n \rightarrow p+p+e^-+\bar\nu_e\,, \\
         \,\,\, p+p+e^- \rightarrow p+n+\nu_e
      \end{array}$
   & \parbox[c]{28ex}{\mbox{Eq.\,(142) of [3], corrected at}
       \mbox{\,\,\,$3\pFp>\pFn+\pFe$ as per [4]}}\\[1.8ex]
   & Reduction factors$^\mathrm{e}$ & $R_{x}^{(\mathrm{M}p)}$ 
                & Appendix (and Eq.~(25)) of [4] \\[1ex]
 \parbox[c]{1ex}{4\\} & \parbox[c]{3.4cm}{Baryon-baryon \\ bremsstrahlung}
   & \bn n+n \rightarrow n+n+\nu+\bar\nu \\
         n+p \rightarrow n+p+\nu+\bar\nu \\
         p+p \rightarrow p+p+\nu+\bar\nu \en
   & \parbox[c]{14ex}{Eq.~(165) of [3] \\
        Eq.~(166) of [3] \\
        Eq.~(167) of [3]} \\[4ex]
   & Reduction factors$^\mathrm{e}$
   &  \bn R_{x}^{(nn)} \\[1.5ex] R_{x}^{(np)} \\[1.8ex] R_{x}^{(pp)} \en
   & \parbox[c]{28ex}{\mbox{Eqs.~(221),~(222),~(228)~of~[3]}
          \mbox{\,\,\,and Eq.~(60) of [4]}\\[.2ex]
        \mbox{Eq.~(220), (229) of [3]}
          \mbox{\,\,\,and Eq.~(54) of [4]}\\[.2ex]
        Eq.~(221) of [3]} \\[6ex]
 \parbox[c]{2ex}{5$^\mathrm{e}$}
   & \parbox[c]{3.4cm}{Cooper pairing of baryons}
  & \bn n+n \to [nn] + \nu + \bar\nu \\
       p+p \to [pp] + \nu + \bar\nu \en 
  & \parbox[c]{28ex}{\mbox{Eqs.~(236), (241) of [3],}
    \mbox{\,\,\,corrected as per [5] (Sect.~\ref{sect:SFnu})}} \\[2ex]
 \parbox[c]{3ex}{6$^\mathrm{c,e}$\\}
 & \parbox[c]{3.4cm}{Electron-fluxoid \\ bremsstrahlung}
  & $e^- + f \to e^- + f +\nu+\bar\nu$
          & \parbox[c]{28ex}{\mbox{Eqs.~(253), (263), (266)\,--\,(268)} \\
                \mbox{ \,\,\,of [3]}} \\ [2ex]
  \hline \hline
  \end{tabular}
\end{center}
\textit{Notes.}
{}$^\mathrm{a}$\,References:
[1]~\citet{KantorGusakov07};
[2]~\citet{OfengeimKY14};  
[3]~\citet{YKGH01};
[4]~\citet{Gusakov02};
[5]~\citet{Leinson09,Leinson10}.\hspace*{1ex}
{}$^\mathrm{b}$\,$\Gamma$ means a plasmon, $e^-$ an electron,
$e^+$ a positron, $\nu$ a neutrino, $\bar{\nu}$ an
antineutrino (in general, of any flavor, but
$\nu_e$ or $\bar\nu_e$ stands
for the electron neutrino or antineutrino,
respectively), $p$ a proton, $n$ a neutron, $[pp]$ and $[nn]$
their paired states, $N$ stands
for an atomic nucleus, and $f$ for a proton fluxoid.
At densities where muons are present, they participate in
the Urca and bremsstrahlung processes fully analogous to the
processes 1, 2, 3, 6 in the core (see details in Ref.~[1]).
$R$ with subscripts/superscripts signifies a control function
(correction factor) due to superfluidity or magnetic field.
Subscript $x$ in $R_x$ substitutes for different 
 superfluidity types (proton or neutron, singlet or
triplet); $B$ indicates magnetic field.\hspace*{1ex}
{}$^\mathrm{c}$\,The effect of strong magnetic field
(see Sect.~\ref{sect:numag}).\hspace*{1ex}
{}$^\mathrm{d}$\,At densities beyond the Durca threshold 
(see Sect.~\ref{sect:nu-core}).\hspace*{1ex}
{}$^\mathrm{e}$\,The effect of
superfluidity (see Sect.~\ref{sect:SFnu}).
\label{tab:nu}
\end{table}
%%%%%%%%%%%%%%%%%%%%%%%%%%%%%%%%%%%%%%%%%%%%%%%%%%%%%%%%%%%%%%%%%%%%%%%%%%%%

%%%%%%%%%%%%%%%%%%%%%%%%%%%%%%%%%%%%%%%%%%%%%
\subsubsection{Neutrino emission in the crust}
\label{sect:nu-crust}

There is a variety of neutrino processes acting in the
crust. In a non-magnetized crust the most important ones are
the \emph{plasmon decay} process and the \emph{electron-ion
bremsstrahlung} process (see Table~\ref{tab:nu}). The
\emph{pair annihilation} process can be also important if
the crust is sufficiently hot.

\begin{figure}
\begin{center}
\includegraphics[width=.5\textwidth]{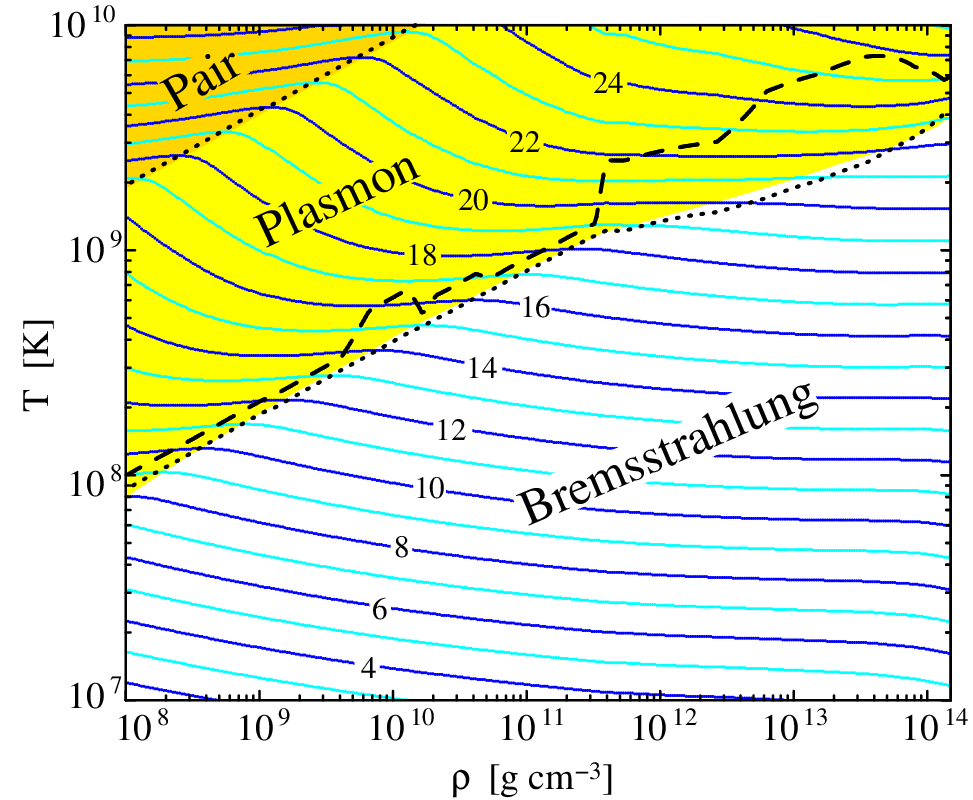}
\hspace*{0.3ex}
\includegraphics[width=.485\textwidth]{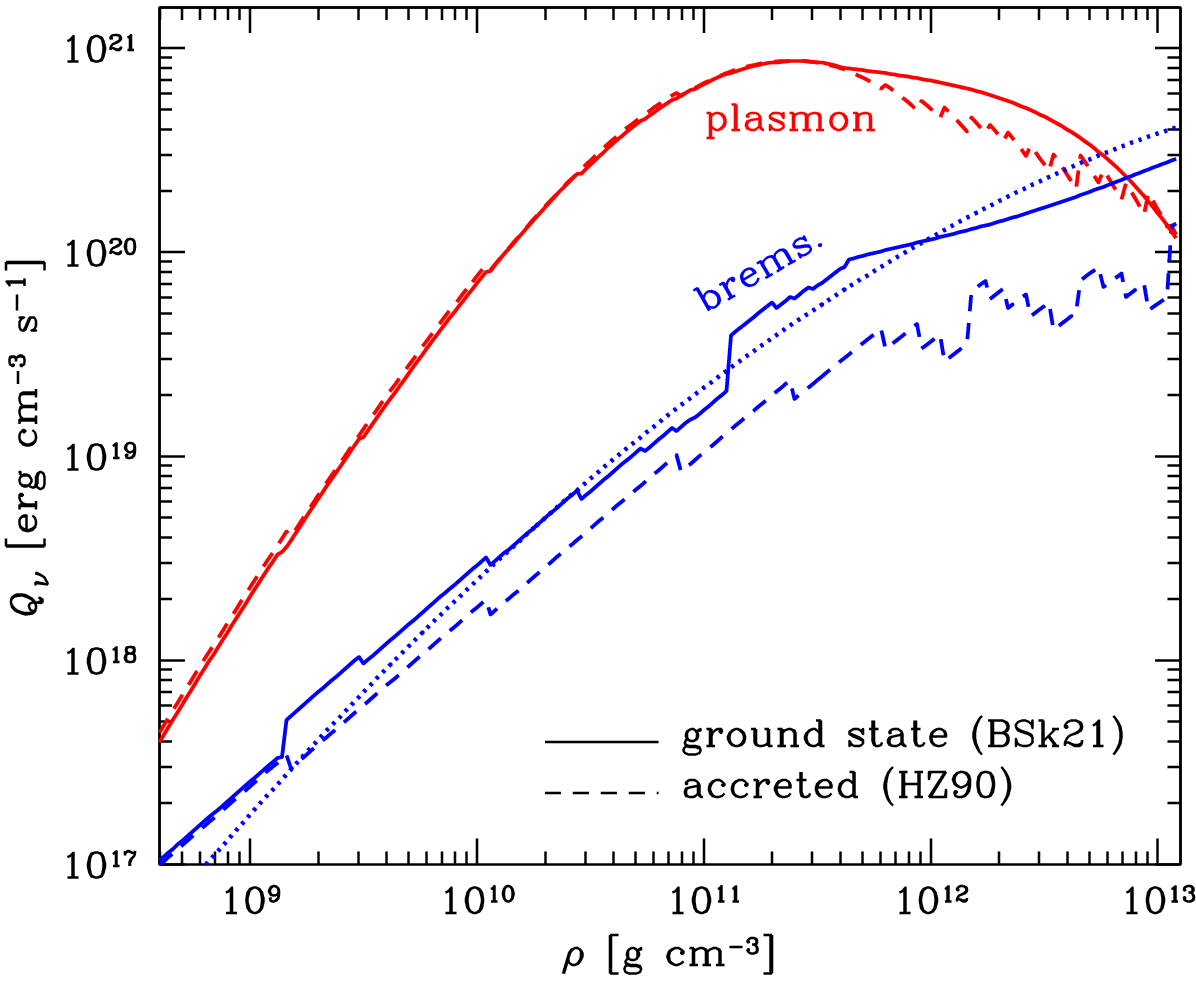}
\caption{Neutrino emissivity $Q_\nu$ in a non-magnetized crust from 
the pair annihilation, plasmon decay,
and electron-ion bremsstrahlung processes.
Left panel: The contour lines are labeled by the value of 
$\mathrm{log}_{10} [Q_\nu/ (\mathrm{erg \, cm}^{-3} \, \mathrm{s}^{-1})]$.
Regions where the pair, plasma, and bremsstrahlung processes dominate are indicated:
the boundaries happen to be quite well described by the two dotted lines that show
$\frac{5}{3} \Tpe$ and $\frac{1}{13} \Tpe$.
(Also indicated is the ion melting curve, dashed line.)
Right panel: Density dependences of $Q_\nu$ for the
ground-state nuclear matter (solid lines) and
for the accreted crust (dashed lines) at $T=10^9$~K. The
dotted line represents an older fit to the bremsstrahlung process
(see text for detail).
\label{Fig:Nu_crust}
}
\end{center}
\end{figure}

The total emissivity from the sum of these processes is
illustrated in the left panel of \fig{Fig:Nu_crust}. The
first thing to notice is the enormous range of values of
$Q_\nu$ covered in the $\rho-T$ range displayed in this
figure, which spans 26 orders of magnitude. This is a direct
consequence of the strong $T$ dependence of the neutrino
processes. The pair annihilation process is efficient only
at low densities and very high temperatures, but when $T \ll
T_\mathrm{F,e}$ very few positrons are present and the
process is strongly suppressed. In the whole range of this
plot, $T_\mathrm{F,e} \gg 10^{10}$ K but pair annihilation
still dominates  at low $\rho$ and high $T$. In the opposite
high-$\rho$ and low-$T$ regime the dominant process is
electron-ion bremsstrahlung, for which
$Q_\nu^\mathrm{(brems)}\propto T^8$. At intermediate $T$ and
$\rho$ the plasmon decay process is most important and, when
it strongly dominates, its emissivity behaves as
$Q_\nu^\mathrm{(pl)}\propto T^4$.

The right panel of \fig{Fig:Nu_crust} illustrates the
density dependence of $Q_\nu^{(\mathrm{pl})}$ and 
$Q_\nu^{(\mathrm{brems})}$ in either ground-state or
accreted crust of a  neutron star with $T=10^9$~K. Pair
annihilation is negligible in this case.
$Q_\nu^{(\mathrm{pl})}$ is calculated according to
\citet{YKGH01} and $Q_\nu^{(\mathrm{brems})}$ according to
\citet{OfengeimKY14}. For comparison, an older fit to
$Q_\nu^{(\mathrm{brems})}$ \citep{Kaminker-ea99} is plotted
by the dotted line. The ground-state composition and the
nuclear size are described by the BSk21 model
(\citealp{Goriely-ea10,Pearson-ea12}, as fitted by
\citealp{Potekhin-ea13}). The accreted composition is taken
from \citet{HZ90}; in this case the approximation by
\citet{ItohKohyama83} is used for the nuclear size.

%---------------------------------------------------------------------------------------
%---------------------------------------------------------------------------------------

The band structure of the energy spectrum of neutrons in the
inner crust, which was mentioned in Sect.~\ref{sect:CVcrust},
should reduce the neutrino reactions of the bremsstrahlung
type and initiate an additional neutrino emission due to
direct inter-band transitions of the neutrons, in analogy
with Cooper pairing of neutrons discussed in
Sect.~\ref{sect:SFnu}. These effects have been mentioned by
\citet{YKGH01}, but remain unexplored.

Electron and positron captures and decays by atomic nuclei
(beta processes), which accompany cooling of matter and
non-equilibrium nuclear reactions, produce neutrino
emission. A pair of consecutive beta capture and decay
reactions is a nuclear Urca process. Urca processes
involving electrons were put forward by
\citet{GamowSchoenberg41}, while those involving positrons
were introduced by \citet{Pinaev}. In the neutron star
crust, the appropriate neutrino luminosity depends on
cooling rate and should be especially strong at $T
\sim$(2--4) $\times 10^9$ K when the main fraction of free
neutrons is captured by nuclei. However, there are other
efficient neutrino reactions open at such temperatures,
which make the neutrino emission due to beta processes
insignificant \citep{YKGH01}. On the other hand, heating
produced by non-equilibrium nuclear reactions (the deep
crustal heating, \citealp{HZ90}) that accompany accretion
mentioned in Sect.~\ref{sect:EOSouter}, may be more
important than non-equilibrium neutrino cooling.

There are a number of other neutrino-emission processes
\citep{YKGH01}, which are less efficient than those listed
in Table~\ref{tab:nu}.  In the inner crust with dripped
neutrons, $n-n$ bremsstrahlung is very efficient but it is
suppressed by pairing and, hence, only acts in the layers
where $T>\Tcn$, where $\Tcn$ is the neutron pairing critical
temperature (Sect.~\ref{sect:pairing}). This process
operates in a wide range of densities and temperatures, and
the density dependence of its emissivity is generally
smooth. Neutrino emission from the formation and breaking of
Cooper pairs makes a significant contribution, much stronger
than the bremsstrahlung, but is confined to a restricted
region of $\rho$ and $T$ (Sect.~\ref{sect:SFnu}). In the
presence of a very strong magnetic field, some of the
above-mentioned processes are modified, and new channels for
neutrino emission may open (Sect.~\ref{sect:magnetic}).

%%%%%%%%%%%%%%%%%%%%%%%%%%%%%%%%%%%%%%%%%%%%%
\subsubsection{Neutrino emission in the core}
\label{sect:nu-core}

\citet{YKGH01} discussed a wealth of neutrino reactions
which may be important in the $npe\mu\Lambda\Sigma^-$ matter
in a neutron star core, which include 
\begin{numlist}
\item 8 direct Urca (Durca)
processes of the electron or muon production and capture by
baryons (baryon direct Urca processes), 

\item 32  modified
Urca (Murca) processes, also associated with the electron or muon
production and capture by baryons (baryon Murca
processes),

\item 12 processes of neutrino-pair emission in
strong baryon-baryon collisions (baryon bremsstrahlung),

\item 4 Murca processes associated with muon decay
and production by electrons (lepton Murca process),

\item 7 processes of neutrino pair emission in Coulomb
collisions (Coulomb bremsstrahlung).
\end{numlist}
In this paper we basically restrict ourselves to the
$npe\mu$ matter. We refer the reader to the review by
\citet{YKGH01} for the more general case, as well as for a
discussion of other exotic models (such as the pion or kaon
condensates). It appears that the reactions that proceed in
the $npe\mu$ matter are often sufficient for the
neutron-star cooling, even when the appearance
of the $\Sigma^-$ and $\Lambda$ hyperons is allowed. The
reason is that these hyperons can appear at high densities
only, where the baryon Durca processes are likely to be
allowed and dominate, for realistic EoSs.

The Durca cycle consists of the beta decay and electron
capture processes (see Table \ref{tab:nu}). They are
threshold reactions open at sufficiently high densities, and
not for every EoS model. For the degenerate nucleons they
are only possible if the proton fraction exceeds a certain
threshold. In the $npe$ matter (without muons) this
threshold is $\approx11\%$, which follows readily from the
energy and momentum conservation combined with the condition
of electric charge neutrality of  matter. Indeed, for
strongly degenerate fermions the Pauli blocking implies that
the reaction is possible only if the energies of the
reacting particles are close to their respective Fermi
energies. Then the momentum conservation assumes the
inequality $p_\mathrm{Fn} \leqslant{p}_\mathrm{Fe}
+{p}_\mathrm{Fp}$, that is $n_n^{1/3} \leqslant n_e^{1/3} +
n_p^{1/3}$. For the $npe$ matter $n_e=n_p$ because of the
charge neutrality, therefore $n_n \leqslant 8n_p$, or
$n_p\geqslant \nb/9$, where $\nb$ is the total baryon number
density. The presence of muons can increase this threshold
by several percent. If 
${p}_\mathrm{F\mu}\geqslant{p}_\mathrm{Fn}-{p}_\mathrm{Fp}$,
then the muon Durca process adds to the
electron Durca.

If allowed, the Durca processes produce a rapid (enhanced)
cooling of neutron stars. If they are forbidden, the main
reactions are those of the baryon Murca and bremsstrahlung
processes which produce a slow (standard) cooling. The Murca
process is a second order process, in which a bystander
neutron or proton participates to allow momentum
conservation (see Table \ref{tab:nu}). Since this process
involves five degenerate fermions, instead of three for the
Durca process, its efficiency is reduced, simply by phase
space limitation, by a factor of order $(T/\EF)^2$, which
gives an overall temperature-dependence $T^8$ instead of
$T^6$. This reduction, for typical conditions in the
neutron-star core, amounts to 6 orders of magnitude. It is
certainly the dominant process for not too high densities in
absence of pairing, and is the essence of the ``standard
cooling scenario''. However, in presence of superfluidity, 
neutrino emission by the formation of Cooper pairs
(Sect.~\ref{sect:SFnu}) can dominate over the Murca process.

Other neutrino reactions in the core involve neutrino-pair
bremsstrahlung in Coulomb collisions lepton modified Urca
processes, electron-positron annihilation, etc. All of them
are not significant under the typical conditions in the
non-exotic core. For instance, the plasmon decay process
that is efficient in the neutron star crust
(Sect.~\ref{sect:nu-crust}) is exponentially suppressed in
the core, because the electron plasmon energy in the core
($\sim\hbar\omp\sim10$~MeV) is much larger than the thermal
energy.

In a strong magnetic field penetrating into the core, some
of the above-mentioned processes can be modified, and new
channels for neutrino emission may open (see
Sect.~\ref{sect:magnetic}).

%%%%%%%%%%%%%%%%%%%%%%%%%%%%%%%%%%%%%%%%%%%%%
\subsubsection{Remarks on in-medium effects}
\label{sect:medium}

Neutrino emissivity $Q_\nu$ may be strongly modified by
in-medium (collective) effects at the high densities of
neutron stars \citep[see][for a review]{Voskresensky01}. For
instance, these effects may result in renormalization of
electroweak interaction parameters.  Moreover, the in-medium
effects may open new channels for neutrino emission.
\citet{VoskresenskySenatorov86} found that the direct and
modified Urca processes appreciably exceed the estimates
obtained neglecting the collective effects, provided the
density is sufficiently large. On the other hand, the
in-medium effects suppress the $nn$ bremsstrahlung in the
neutron-star core by a factor of 10\,--\,20
\citep{Blaschke-ea95}. According to the study by 
\citet{Schaab-ea97}, the medium effects on the emissivity of
the Murca process cause a more rapid cooling than obtained
for the standard scenario and result in a strong density
dependence, which gives a smooth crossover from the standard
to the enhanced cooling scenario (see
Sect.~\ref{sect:scenarios}) for increasing star masses.

The problem of calculation of the in-medium effects in the
neutron star matter is complicated. Various theoretical
approaches were used to solve it, results of different
techniques being different typically by a factor of a few
(see, e.g., \citealp{Blaschke-ea95}, and references
therein). The renormalization of the electroweak coupling is
usually taken into account in an approximate manner by
replacing the bare baryon masses $m_B$ with effective ones,
$m_B^*$ \citep[e.g.,][]{YKGH01}. The values of these
effective masses should be taken from microscopic theories.

The in-medium effects are also important for the
Cooper-pairing neutrino emission mechanism related to baryon
superfluidity, as discussed in Sect.~\ref{sect:SFnu} below,
for heat capacity (Sect.~\ref{sect:CVcore}), and for baryon
heat conduction (Sect.~\ref{sect:corecond}) in the core of a
neutron star.

%%%%%%%%%%%%%%%%%%%%%%%%%%%%%%%%%%%%%%%%%%%%%
\subsection{Thermal conductivity}
\label{sect:conductivity}
%%%%%%%%%%%%%%%%%%%%%%%%%%%%%%%%%%%%%%%%%%%%%

The most important heat carriers in the crust and ocean of
the star  are the electrons. In the atmosphere, the heat is
carried mainly by photons. In general, the two mechanisms
work in parallel, hence
$
   \kappa=\kappa_\mathrm{r}+\kappa_e,
$
where $\kappa_\mathrm{r}$ and $\kappa_e$  denote
the radiative (r) and electron (e) components of the thermal
conductivity $\kappa$. The radiative transfer is considered
in Paper~I. In this paper we will pay most attention to the
electron heat conduction mechanism. Both the electron and
photon heat conduction are affected by strong
magnetic fields. We will consider these effects in
Sect.~\ref{sect:magnetic}.

The elementary theory in which  the effective collision rate
$\nu$ of heat carriers with effective mass $m^*$ and number
density $n$ does not depend on their velocity, gives
\citep{Ziman}
\beq
 \kappa = a \,{n T}/{m^\ast\nu},
\label{elementary}
\eeq
where $a$ is a numerical coefficient: $a=3/2$ for a
non-degenerate gas, and $a=\pi^2/3$ for strongly degenerate
particles. (We remind that we use energy units for $T$;
otherwise $a$ should be multiplied by the squared
Boltzmann constant.)

\begin{figure}
\begin{center}
\includegraphics[width=\textwidth]{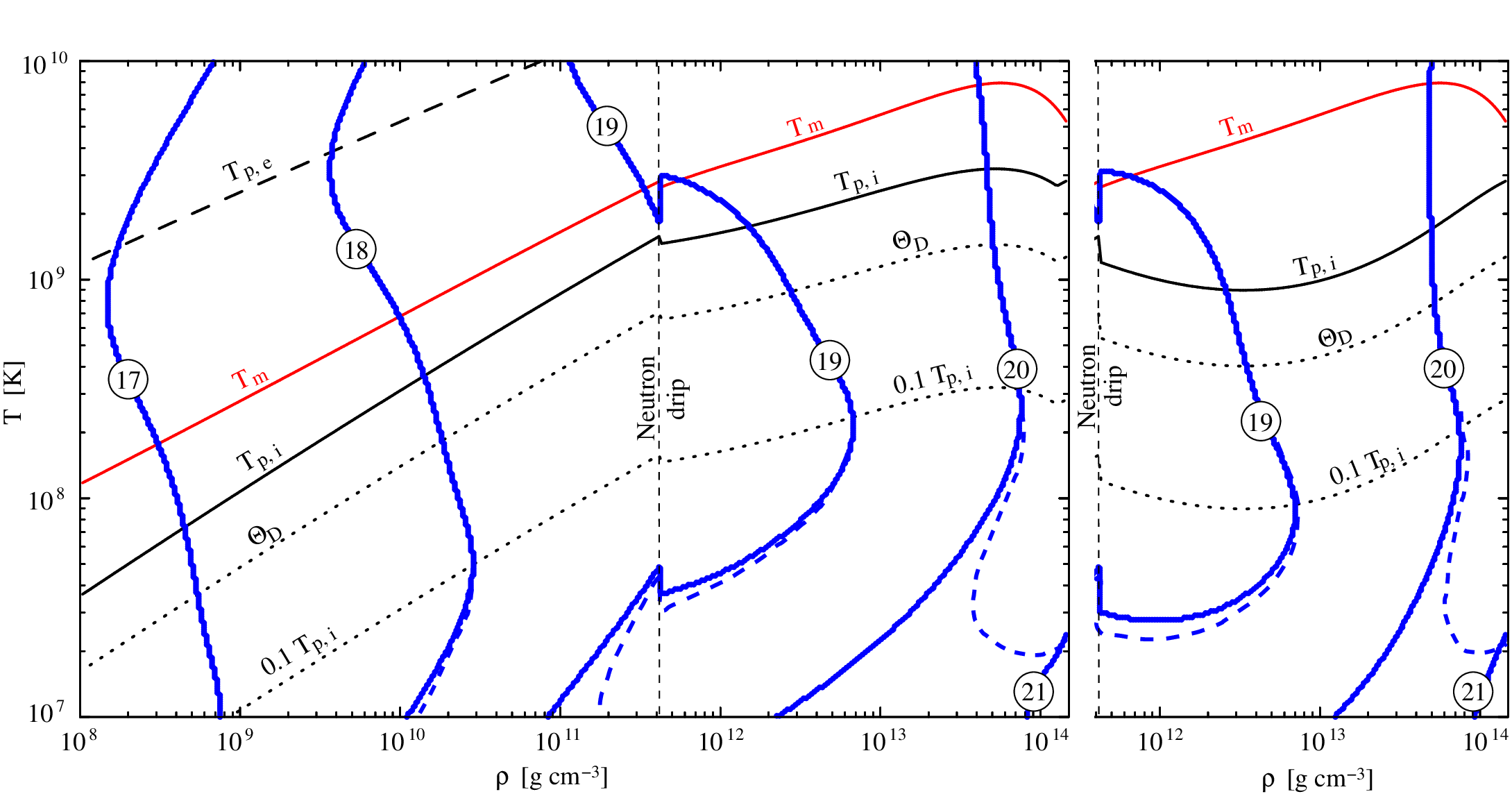}
\caption{
Iso-contour lines of the electron thermal conductivity $\kappa_e$ in the crust, labeled by the value of
$\mathrm{log}_{10} (\kappa_e/\mathrm{erg \, s^{-1} \, cm^{-1} \, K^{-1}})$, using the results of \citet{GYP01}.
Also shown are the melting curve $\Tm$, 
the electron and ion plasma temperatures, $\Tpe$ and $\Tp$
respectively,
the Debye temperature, $\Theta_\mathrm{D} \simeq 0.45 \Tp$, that marks the transition
from classical to quantum solid and  $0.1 \Tp$ below which the wholly quantum crystal regime is realized.
The crust composition is the same as in \fig{fig:CVcrust2}.
The right panel only displays the inner crust but assuming that about 80\% of the dripped neutrons are entrained:
the strong reduction of $\Tp$ and $\Theta_\mathrm{D}$ pushes the onset of the wholly quantum regime to lower $T$.
The dashed contour lines illustrate the reduction of $\kappa_e$ from impurity scattering, assuming an impurity
parameter $Q_\mathrm{imp} =1$.
}
\label{fig:cond2}
\end{center}
\end{figure}

The most important heat carriers and respective scattering
processes that control the thermal conductivity $\kappa$ are
listed in Table~\ref{tab:cond}, and briefly discussed below. The last column of
the table contains references to either analytical fitting
formulae or publicly available computer codes for the evaluation
of $\kappa$.
Figure~\ref{fig:cond2} illustrates the magnitude of
$\kappa_e$ and characteristic
temperatures in the crust.

%%%%%%%%%%%%%%%%%%%%%%%%%%%%%%%%%%%%%%%%%%%%%%%%%%%%%%%%%%%%%%%%%%%%%%%%%%%%
\begin{table}[t]
\caption{Main contributions to thermal conductivity}
\setlength{\tabcolsep}{4pt}
\begin{center}
  \begin{tabular}{lll}
  \hline \hline
   & Conduction type and regime    & References$^\mathrm{a}$\rule[-1.5ex]{0pt}{4ex} \\
\hline
 1$^\mathrm{b}$ & Photon conduction
      & Eqs.~(14)\,--\,(20) of [1]\rule{0pt}{2.5ex}  \\
   & -- plasma cutoff correction &  Sect.~3.3 of [2] \\
   & -- magnetic field modifications$^\mathrm{c}$ & Eqs.~(21)\,--\,(23) of [1]  \\
 2 & Electron conduction in the ocean and the crust: &
                                see Appendix~\ref{sect:conduct}\\
   & -- Electron-ion / electron-phonon scattering & [3] (theory),
                                              [4] (public code) \\
   & \phantom{--} -- the effects of magnetic fields$^\mathrm{d}$ &
             [5] (theory), [4] (public code) \\
   & \phantom{--} -- the effects of finite nuclear sizes in the inner
                         crust & [6] (theory), [4] (public code) \\
   & -- Electron scattering on impurities in the crust &
                                     see Appendix~\ref{sect:imp} \\
   & -- Electron-electron scattering: \\
   & \phantom{--} -- strongly degenerate electrons
        & Eqs. (10), (21)\,--\,(23) of [7] \\
   & \phantom{--} -- arbitrary degeneracy
        & see Appendix~\ref{sect:interpolate} \\
 3 & Baryon conduction in the core& Eqs.~(7), (12), (21), (28)\,--\,(30) of
[8] \\
  & -- Effects of superfluidity$^\mathrm{e}$ &  Eqs.~(45)\,--\,(48),
(50)\,--\,(53) of [8] \\
 4 & Lepton conduction in the core & Eqs.~(4)\,--\,(6), (16), (17),
(33)\,--\,(37) of
[9] \\
  & -- Effects of superfluidity$^\mathrm{e}$
   & Eqs.~(45), (54)\,--\,(61), (84)\,--\,(92)$^\mathrm{f}$ of
[9] \\[1ex]
  \hline \hline
  \end{tabular}
\end{center}
\textit{Notes.}
{}$^\mathrm{a}$\,References:
[1]~\citet{PY01};
[2]~\citet{Potekhin-ea03};
[3]~\citet{Potekhin-ea99};
[4]~http://www.ioffe.ru/astro/conduct/;
[5]~\citet{P96,P99};
[6]~\citet{GYP01};
[7]~\citet{ShterninYakovlev06};
[8]~\citet{BaikoHY01};
[9]~\citet{ShterninYakovlev07}.\hspace*{1ex}
{}$^\mathrm{b}$\,For fully ionized atmospheres only. For
partially ionized atmospheres, see references in \citet{P14}.\hspace*{1ex}
{}$^\mathrm{c}$\,See Sect.~\ref{sect:radiopa}.\hspace*{1ex}
{}$^\mathrm{d}$\,See Sect.~\ref{sect:condmag-e}.\hspace*{1ex}
{}$^\mathrm{e}$\,See Sect.~\ref{sect:SFcond}.\hspace*{1ex}
{}$^\mathrm{f}$\,The power index 2
should be suppressed at its first occurrences in the third
and fourth lines of Eq.~(92) of Ref.~[9].\vspace*{-1ex}
\label{tab:cond}
\end{table}
%%%%%%%%%%%%%%%%%%%%%%%%%%%%%%%%%%%%%%%%%%%%%%%%%%%%%%%%%%%%%%%%%%%%%%%%%%%%

% -----------------------------------
\subsubsection{Heat conduction in the outer envelopes}

Electron heat conduction is the most important process in
the neutron star envelopes that determines thermal
luminosity of neutron stars. In this case, $m^\ast = \mel \,
\sqrt{1+\xr^2}$ in \req{elementary}, and $\nu=\nu_e$ is mostly determined by
electron-ion ($e$i) and electron-electron ($ee$) Coulomb
collisions. In the crystalline phase, the electron-ion
scattering takes the form of scattering on phonons
(collective ion excitations). The Matthiessen rule
\citep[e.g.,][]{Ziman} assumes that effective frequencies of
different collisions simply add up, i.e.,
$\nu_e=\nu_{e\mathrm{i}}+\nu_{ee}$. This is strictly valid
for extremely degenerate electrons \citep{HubbardLampe}. In
general case it remains a good estimate, because
$\nu_{e\mathrm{i}}+\nu_{ee}\leq\nu_e\leq\nu_{e\mathrm{i}}+\nu_{ee}+\delta\nu$,
where $\delta\nu\ll\min(\nu_{e\mathrm{i}},\nu_{ee})$
\citep{Ziman}. The relative importance of the different
types of collisions and practical formulae for evaluation of
$\nu_e$ can be different, depending on the composition and
phase state of the plasma (see Appendix \ref{sect:conduct}).

\citet{ChugunovHaensel07} considered an alternative heat
transport by the plasma ions (phonons in the solid OCP),
which works in parallel with the transport by the electrons.
The ion (phonon) heat conduction is usually unimportant in neutron stars.
Although the ion thermal conductivity can be larger than the
electron conductivity across the strong magnetic field, the
multidimensional modeling shows
that in such cases the heat is mainly transported by the
electrons non-radially (i.e., not straight across the field
lines; see Sect.~\ref{sect:cooling}).

% -----------------------------------
\subsubsection{Heat conduction in the inner crust}

The inner crust of a neutron star is characterized by the
presence of free neutrons. This has two important
consequences. First, heat transport by neutrons can compete
with the transport by the electrons and phonons. Second,
electron-neutron scattering adds to the other electron
scattering mechanisms considered above and in
Appendix~\ref{sect:conduct}.

The thermal conductivity by neutrons, $\kappa_n$, was
studied in several papers
\citep[e.g.,][]{FlowersItoh76,BisnoRomanova82}. A general
expression for $\kappa_n$ in non-superfluid matter is given
by \req{elementary} with $n=n_n$, the number density of
neutrons, $m^*=m_n^\ast$, the neutron effective mass
modified by medium effects, and $\nu_n=\nu_{n\mathrm{i}}+\nu_{nn}$.
The neutron-neutron collision frequency, $\nu_{nn}$, can be
calculated in the same manner as in uniform matter of a
neutron-star core (Sect.~\ref{sect:corecond}). However, for
strongly degenerate neutrons these collisions are much less
efficient than the neutron-ion ones. Therefore, one can set
$\nu_n \approx \nu_{n\mathrm{i}}$, at least for order-of-magnitude
estimates. For the scattering of the neutrons by
uncorrelated nuclei,  $\nu_{n\mathrm{i}} = n_\mathrm{i} \,
\mathrm{v}_{\mathrm{F}n} \, S_{n\mathrm{i}}$, where
$\mathrm{v}_{\mathrm{F}n}=\pFn/
\sqrt{c^2+(\pFn/m_n^\ast)^2}$  is the neutron Fermi velocity
and $S_{n\mathrm{i}}$ is the transport cross section. For a
crude estimate at sufficiently low neutron energies in the
neutron star crust one can set
\citep[e.g.,][]{BisnoRomanova82} $S_{n\mathrm{i}}= \pi \,
R_n^2$, where $R_n$ is the neutron radius of an atomic
nucleus (fitted, e.g., in \citealp{Potekhin-ea13}).
Estimated in this way, $\kappa_n$ is negligible, being at
least two orders of magnitude smaller than $\kappa_e$ in the
entire inner crust at $T \lesssim 10^9$ K. However,
$\nu_{n\mathrm{i}}$ can be strongly affected by ion-ion
correlations and by superfluidity (Sect.~\ref{sect:SFcond}).

In addition, the electron conduction in the inner crust is
affected by the size of a nucleus, which becomes
non-negligible compared to the mean distance between the
nuclei, so that the approximation of pointlike scatterers is
not applicable anymore. Then one should take into account
the form factor, which depends on the size and shape of the
charge distribution in a nucleus. A finite charge
distribution reduces $\nu_{e\mathrm{i}}$ with respect to the
model of a pointlike charge, thereby increasing the
conductivity \citep{GYP01}. The effect mainly depends on the
ratio of the root mean square charge radius of a nucleus
$R_\mathrm{ch}$ to the Wigner-Seitz cell radius $\aion$.
\citet{GYP01} presented fitting formulae for the dependences
of the thermal and electrical conductivities on the
parameter $x_\mathrm{nuc}=\sqrt{5/3}\,R_\mathrm{ch}/\aion$.
The latter parameter has been fitted as function of density
for modern BSk models of nuclear matter
\citep{Potekhin-ea13} and for some other models (Appendix~B
in \citealt{HPY07}).

%----------------------------------------------------------------------------
%----------------------------------------------------------------------------

%%%%%%%%%%%%%%%%%%%%%%%%%%%%%%%%%%%%%%%%%%%%%
\subsubsection{Heat conduction in the core}
\label{sect:corecond}
%%%%%%%%%%%%%%%%%%%%%%%%%%%%%%%%%%%%%%%%%%%%%

The first detailed studies of the kinetic coefficients in
neutron star cores were performed by \citet{FlowersItoh79},
who constructed the exact solution of the multicomponent
system of transport equations in the $npe$ matter. But since
the proton fraction is small and the electron-neutron
interaction is weak, the kinetic coefficients can be split
in two almost independent parts -- the neutron kinetic
coefficients mediated by nucleon-nucleon collisions and
electron kinetic coefficients mediated by the collisions
between charged particles; the proton kinetic coefficients
are small. In the non-superfluid $npe\mu$ matter, the
neutrons are the main heat carriers at $T\lesssim10^8$~K,
while the heat transport by leptons $e^-$ and $\mu^-$ is
competitive at $T\gtrsim10^9$~K
\citep{ShterninYakovlev07,ShterninBH13}.

% -----------------------------------
\paragraph{Baryon heat conduction.}

\citet{FlowersItoh79} based their calculations on the free
nucleon scattering amplitudes, neglecting the Fermi-liquid
effects and nucleon many-body effects. Their results were
later reconsidered by \citet{BaikoHY01}.

The thermal conductivity is written in the form analogous to
\req{elementary}:
\beq
    \kappa_n= \frac{\pi^2 T n_e \tau_n }{ 3m_n^*},
 \quad
     \kappa_p=\frac{\pi^2 T n_p \tau_p}{3m_p^*},
\eeq
where the effective relaxation times $\tau_n$ and $\tau_p$
are provided by solution of the system of algebraic
equations \citep[e.g.,][]{ShterninBH13}
\beq
  \sum_{j=n,p}\nu_{ij} \tau_j=1,
\qquad
     \nu_{ij}= \frac{64 m_i^* m_j^{*2} T^2
        }{ 5 m_N^2 \hbar^3} \,S_{ij}
\qquad (i,j=n,p),
\eeq
where $\nu_{ij}$ are effective collision frequencies, $m_N$
is the bare nucleon mass in vacuo, and $S_{ij}$ are the effective
cross-sections.

Many-body effects in the context of transport coefficients
of pure neutron matter were first addressed by
\citet{WambachAP93} and later reconsidered in many papers.
There are two kinds of these effects: the three-body part of
the effective potential for the nucleon-nucleon interactions
and the in-medium effects (cf.{} Sect.~\ref{sect:medium})
that affect nucleon-nucleon
scattering cross-sections. \citet{BaikoHY01} calculated
$S_{ij}$ in the approximation of pairwise interactions
between nucleons with appropriate effective masses, using
the Bonn potential model for the elastic nucleon-nucleon
scattering \citep{Machleidt-ea87} with and without the
in-medium effects. They presented the results in the form $
S_{ij} = S_{ij}^{(0)}K_{ij}, $ where $S_{ij}^{(0)}$
corresponds to scattering of bare particles, and $K_{ij}$
describes the in-medium effects. They also constructed a simple
analytical fits to their results for $S_{ij}^{(0)}$ and
$K_{ij}$ (referenced in Table~\ref{tab:cond}).

\citet{ShterninBH13} studied the many-body
effects on the kinetic coefficients of nucleons in the
$npe\mu$ matter in beta equilibrium using the
Brueckner-Hartree-Fock (BHF) method. According to this
study, the three-body forces suppress the thermal
conductivity. This suppression is small at low densities but
increases to a factor of $\sim4$ at the baryon number
density of $\nb=0.6$ fm$^{-3}$. However, the use of the
effective masses partly grasps this difference. For this
reason it proves to be sufficient to multiply the
conductivities obtained in the effective-mass approximation
\citep{BaikoHY01} by a factor of 0.6 to reproduce the BHF
thermal conductivity \citep{ShterninBH13} with an accuracy
of several percent in the entire density range of interest.

% -----------------------------------
\paragraph{Lepton heat conduction.}

The up-to-date electron and muon contributions to thermal
conductivities of neutron star cores were calculated by
\citet{ShterninYakovlev07}. Their treatment included the
Landau damping of electromagnetic interactions owing to the
exchange of transverse plasmons. This effect was studied by
\citet{HeiselbergPethick93} for a degenerate quark plasma,
but was neglected in the previous studies of the lepton heat
conductivities in the $npe\mu$ matter
\citep[e.g.,][]{FlowersItoh81,GnedinYakovlev95}.

The electron and muon thermal conductivities are additive,
$\kappa_{e\mu}=\kappa_e+\kappa_\mu$, and can be written 
in the familiar form of \req{elementary}:
\beq
    \kappa_e=\frac{\pi^2 T n_e \tau_e }{ 3m_e^*}, \quad
     \kappa_\mu=\frac{\pi^2 T n_\mu \tau_\mu }{ 3m_\mu^*},
\label{kappa-emu}
\eeq
where $\kappa_e$ and $\kappa_\mu$ are the partial thermal
conductivities of electrons and muons, respectively; $n_e$
and $n_\mu$ are number densities of these particles, $m_e^*$
and $m_\mu^*$ are their dynamical masses at the Fermi
surfaces, determined by their chemical potentials. In
neutron star cores at beta equilibrium these chemical
potentials are equal, therefore $m_e^*=m_\mu^*$. The
effective electron and muon relaxation times can be written
as \citep{GnedinYakovlev95}
\beq
   \tau_e =
   \frac{\nu_\mu -\nu'_{e\mu} }{ \nu_e \nu_\mu - \nu'_{e\mu}\nu'_{\mu e}},
   \quad
   \tau_\mu =
   \frac{\nu_e -\nu'_{\mu e} }{ \nu_e \nu_\mu - \nu'_{e\mu}\nu'_{\mu e}},
\label{taus}
\eeq
where
$
     \nu_e = \sum_i \nu_{ei}=\nu_{ee}+\nu_{e \mu}+\nu_{ep}
$
and
$  
     \nu_\mu = \sum_i \nu_{\mu i}=\nu_{\mu \mu}+ \nu_{\mu e}+ \nu_{\mu p}
$
are the total effective collision frequencies of electrons
and muons with all charged particles $i=e,\mu,p$; 
$\nu_{ei}$ and $\nu_{\mu i}$ are partial effective collision
frequencies, while $\nu'_{e \mu}$ and $\nu'_{\mu e}$ are
additional effective collision frequencies, which couple
heat transport of the electrons and muons. All these 
collision frequencies can be expressed as multidimensional
integrals over momenta of colliding particles.
\citet{ShterninYakovlev07} calculated these integrals in the
weak-screening approximation and described the results by
simple analytical formulae (referenced in
Table~\ref{tab:cond}). In the case of strongly degenerate
ultra-relativistic leptons, which is typical for neutron
star cores, the latter authors obtained a much simpler
expression, which can be written as
\beq
    \kappa_{e,\mu} \approx 20.8 \,c\, ({\pF}_{e,\mu}/\hbar)^2.
\eeq
The latter simplification, however, does not hold if the
protons are superfluid.

%%%%%%%%%%%%%%%%%%%%%%%%%%%%%%%%%%%%%%%%%%%%%
\section{Superfluidity and superconductivity}
\label{sect:SF}
Soon after the development of the BCS theory \citep{BCS57},
which explains superconductivity by Cooper pairing of
fermions \citep{Cooper56}, \citet{BohrMP} argued that the
same phenomenon of pairing is occurring inside nuclei (later
this suggestion was confirmed  experimentally).
\citet{Migdal59} extended the idea to the interior of
neutron stars.  \citet{GinzburgKirzhnits} formulated a
number of important propositions concerning neutron
superfluidity in the interior of neutron stars, the
formation of Feynman-Onsager vortices, a critical
superfluidity temperature ($\Tc\lesssim10^{10}$~K) and its
dependence on the density ($\rho\sim10^{13}$\,--\,$10^{15}$
\gcc), and discussed the influence of neutron superfluidity
on heat capacity and therefore on the thermal evolution of a
neutron star. \citet{BPP69a} and \citet{Ginzburg70} analyzed
the consequences of neutron superfluidity and proton
superconductivity: rotation of the superfluid component in
the form of quantized vortices and splitting of the internal
stellar magnetic field into fluxoids
(Sect.~\ref{sect:fluxoid}). Later many different authors
considered various types of pairing of nucleons, hyperons,
or quarks using different model potentials.

Although we will not consider exotic models of neutron star
cores, let us mention that superfluidity is possible in
these models as well. For instance,
\citet{TakatsukaTamagaki95} reviewed calculations of neutron
and proton superfluid gaps in pion condensed matter. Some
authors have discussed superfluidity in quark matter
\citep[e.g.,][]{StejnerWM09}. If hyperons are present, they can 
also be in a superfluid state \citep{BalbergBarnea98}. For a 
detailed recent review of superfluidity in the interiors of neutron 
stars, see \citet{Page-ea14}.

% -----------------------------------
\subsection{Pairing types and critical temperatures}
\label{sect:pairing}

The Cooper pairing appears as a result of the attraction of
particles with the anti-parallel momenta,which is expected
to occur, at low enough temperature, in any degenerate
system of fermions in which there is an attractive
interaction between particles whose momenta $\bm{p}$ lie
close to the Fermi surface \citep{Cooper56}. The strength of
the interaction determines the critical temperature $\Tc$ at
which the pairing phase transition will occur. In a normal
system the particle energy $\epsilon$ varies smoothly when
the momentum crosses the Fermi surface, while in the
presence of pairing a discontinuity develops, with a
forbidden energy zone having a minimum width of
$2\Delta_\mathrm{pair}$ at $p=\pF$, which can be regarded as
the binding energy of a Cooper pair.

The BCS equations that describe symmetric nuclear matter in
atomic nuclei and asymmetric neutron-rich
matter in neutron stars have much in common but have also some differences.
For instance, pairing in atomic nuclei takes place in the
singlet state of a nucleon pair. In this case, the energy
gap is isotropic, that is independent of the orientation of
nucleon momenta. On the other hand, one can expect
triplet-state pairing in the neutron-star matter, which
leads to anisotropic gap. Singlet-state neutron
superfluidity develops in the inner neutron star crust and
disappears in the core, where an effective neutron-neutron
singlet-state attraction transforms into repulsion.
Triplet-state neutron superfluidity appears in the
neutron-star core. Protons in the core can undergo the
singlet-state pairing.

The triplet pair states may have different projections $m_J$
of the total pair momentum onto the quantization axis:
$|m_J| = 0$, 1, and 2. The actual (energetically favorable)
state may be a superposition of states with different
$m_J$.  Owing to uncertainties of microscopic theories this
state is still unknown; it depends possibly on density and
temperature. In simulations of neutron star cooling, one
usually considers the triplet-state pairing with $|m_J| = 0$
and 2, since their effects on the heat capacity and neutrino
luminosity are qualitatively different
\citep[e.g.,][]{YLS99,YKGH01}.

The critical temperature $\Tc$ is very sensitive to the
strength of the repulsive core of the nucleon-nucleon
interaction. It is related to the superfluid energy gap by
$\Tc = 0.5669\Delta_\mathrm{pair}$ for the singlet gap
(e.g., \citealp{LaLi-SP2}, Sect.~40). For the triplet gap,
the situation is more complicated, because the gap is
anisotropic \citep[e.g.,][]{AmundsenOstgaard85b,Baldo-ea92,YLS99}. Examples of the
dependence of $\Tc$ on gravitational mass density in the
crust and core of a neutron star are shown in
Fig.~\ref{fig:Tc}. Here, we employed the gap parametrization
of \citet{KaminkerHY01} with the parameter values and
notations for different models of superfluidity according to
\citet{Ho-ea15} together with the $\rho$-dependences of
free-nucleon number densities $n_n$ and $n_p$ from the fits
\citep{Potekhin-ea13} for the BSk21 model of crust and core
composition. Figure~\ref{fig:Tc} demonstrates a large
scatter of theoretical predictions, but also general
features. We see that the \singlet{} superfluidity of
neutrons occurs mostly in the inner crust and the \triplet{}
superfluidity mostly in the core.  The critical temperatures
of neutrons in the triplet states, $\Tcn(\triplet)$, and
protons, $\Tcp(\singlet)$, have usually a maximum at a
supranuclear density $\rho>\rho_0$. Typical magnitudes of
$\Tc$ vary from one model to another within a factor of a
few.  Neutron \triplet{} superfluidity has, in general, much
lower $\Tc$ than \singlet{} pairing of neutrons in the inner
crust and protons in the core.

% -----------------------------------
\subsection{Superfluid effects on heat capacity}
\label{sect:SFc}

%----------------------------------------------------------------------------
\begin{figure}
\begin{center}
  \includegraphics[height=.44\textwidth]{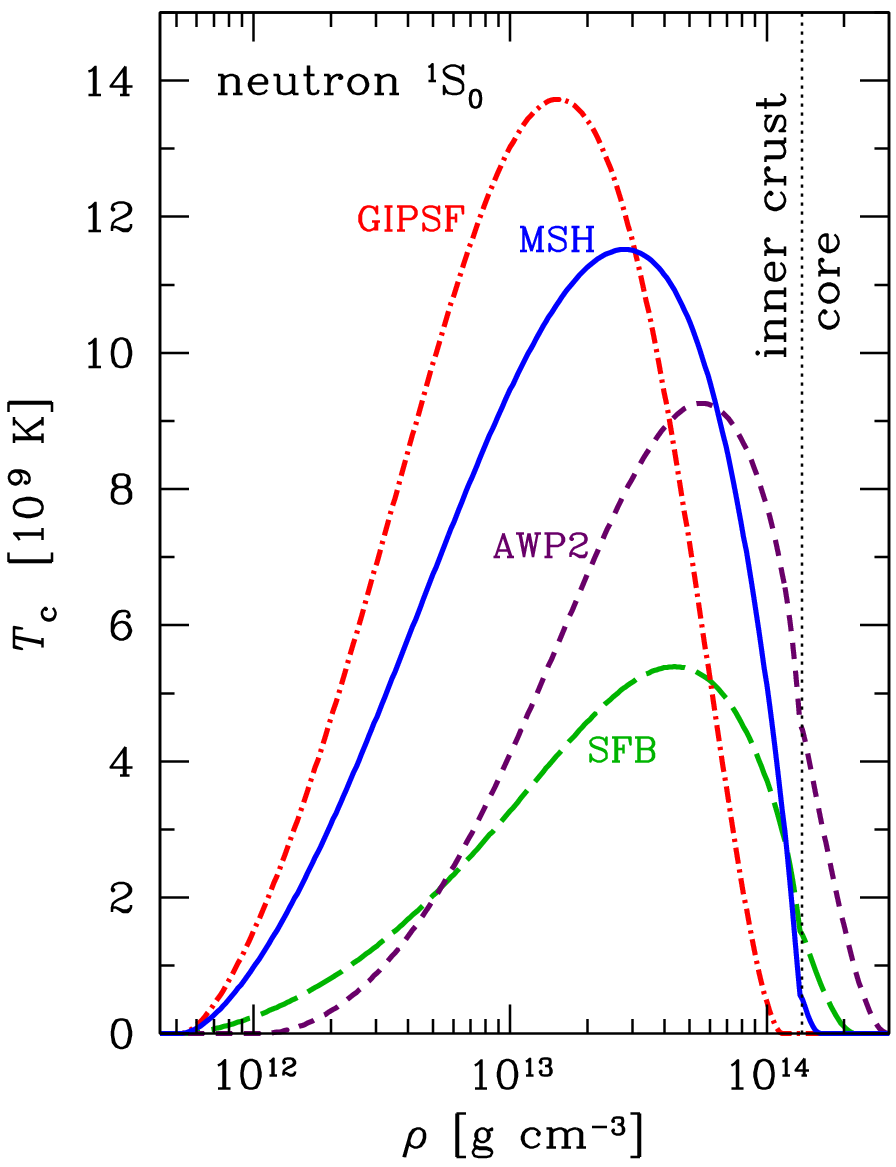}
  \includegraphics[height=.44\textwidth]{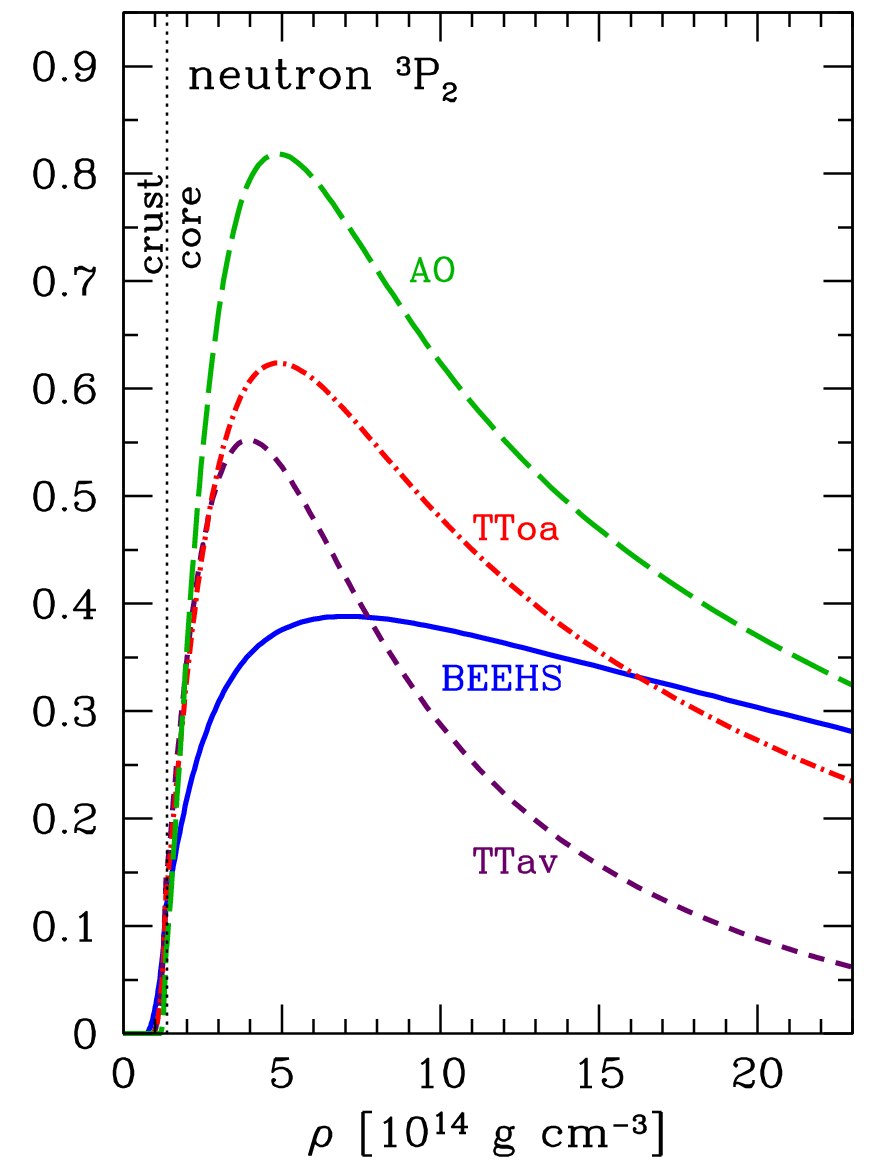}
  \includegraphics[height=.44\textwidth]{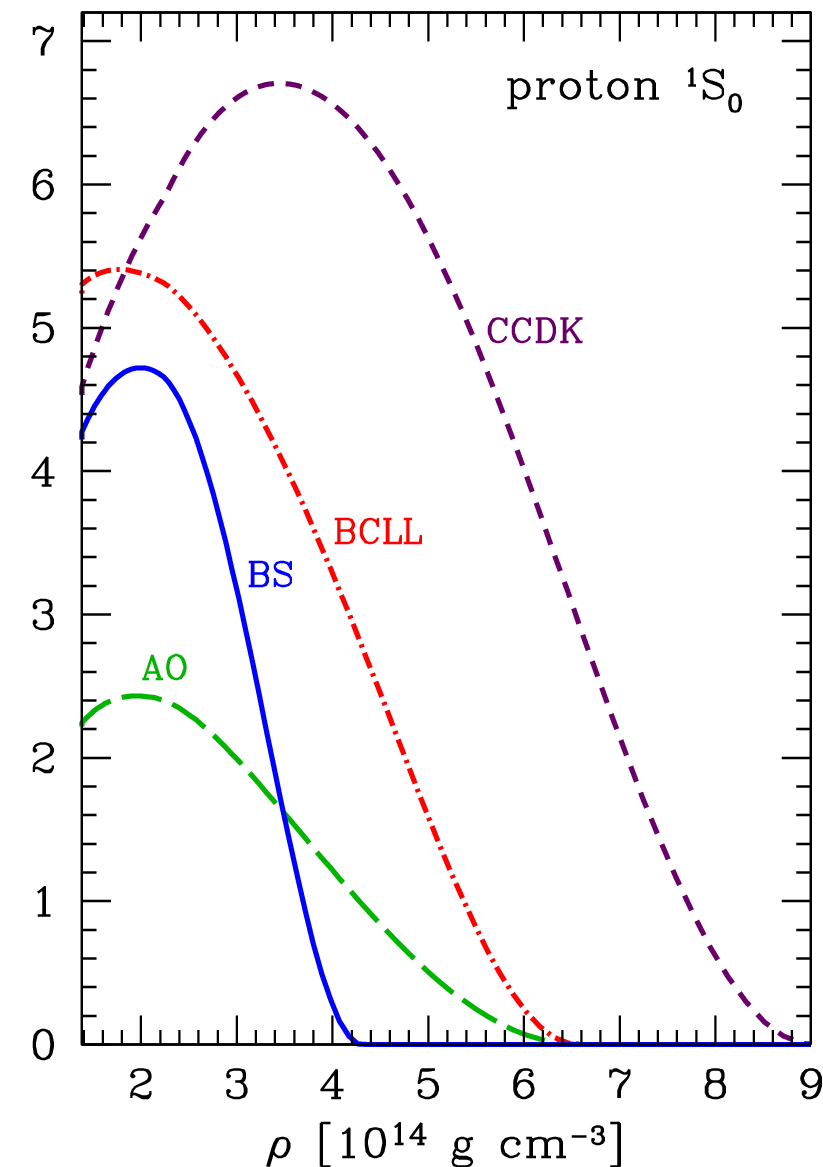}
\caption{Critical temperatures of singlet neutron (left
panel), triplet neutron (middle panel), and singlet proton
(right panel) superfluidities in the inner crust and core of
a neutron star, as functions of gravitational mass density
$\rho$, for different
superfluidity models, as marked near the curves
\citep[see][]{Ho-ea15}: 
AO \citep{AmundsenOstgaard85a,AmundsenOstgaard85b}, 
AWP2 \citep{AinsworthWP89},
BCLL \citep{Baldo-ea92}, 
BEEHS \citep{Baldo-ea98}, 
BS \citep{BaldoSchulze07}, 
CCDK \citep{Chen-ea93}, 
GIPSF \citep{Gandolfi-ea08}, 
MSH \citep{MargueronSH08}, 
SFB \citep{SchwenkFB03}, 
TTav and TToa \citep{TakatsukaTamagaki04}.
}
\label{fig:Tc}
\end{center}
\end{figure}
%----------------------------------------------------------------------------

Once a component x of the neutron star matter becomes superfluid, 
its specific heat $c_\mathrm{v,x}$ is strongly altered. When $T$ reaches
$T_\mathrm{c,x}$, the critical temperature for the pairing
phase transition, $c_\mathrm{v,x}$ jumps by a factor
$\gtrsim 2$. However, as $T$ continues to decrease, the heat
capacity becomes progressively suppressed. At $T\ll\Tc$ the
energy gap in the nucleon spectrum strongly reduces the heat
capacity even compared to its value in the absence of
pairing. These effects are
implemented in numerical calculations through ``control functions''
$R_\mathrm{c}(T/T_\mathrm{c,x})$ as
\beq
 c_\mathrm{v,x}
    = R_\mathrm{c} \, c_\mathrm{v,x}^{(0)},
\eeq
where $c_\mathrm{v,x}^{(0)}$ denotes the value in the normal
phase, \req{Sommerfeld}. The control function depends on the
type of pairing. This dependence was studied by
\citet{LY94}. Analytical fitting formulae for $R_\mathrm{c}$
in the $npe\mu$ matter for the main types of superfluidity
listed above are given by Eq.~(18) of
\citet{YLS99}.\footnote{In the latter paper, an accidental
minus sign in front of the term $(0.2846\mathrm{v})^2$ in
the denominator of the fitting formula for $R_\mathrm{c}$ in
the case of ``type C'' (\triplet, $|m_J|=2$) superfluidity
must be replaced by the plus sign (D.G.~Yakovlev, personal
communication).\label{RC-corr}}

\begin{figure}
\begin{center}
\includegraphics[width=.453\textwidth]{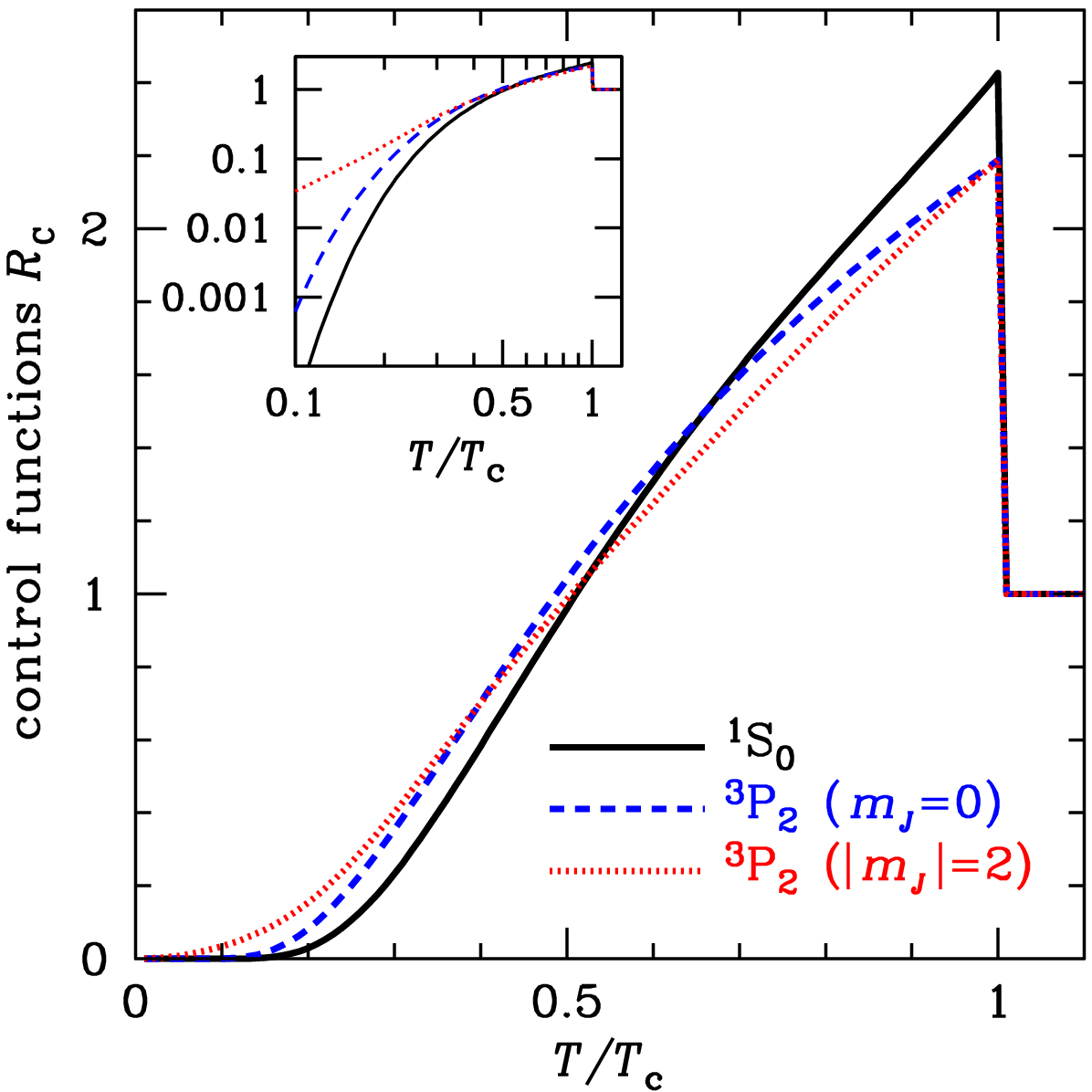}
\hspace*{1ex}
\includegraphics[width=.52\textwidth]{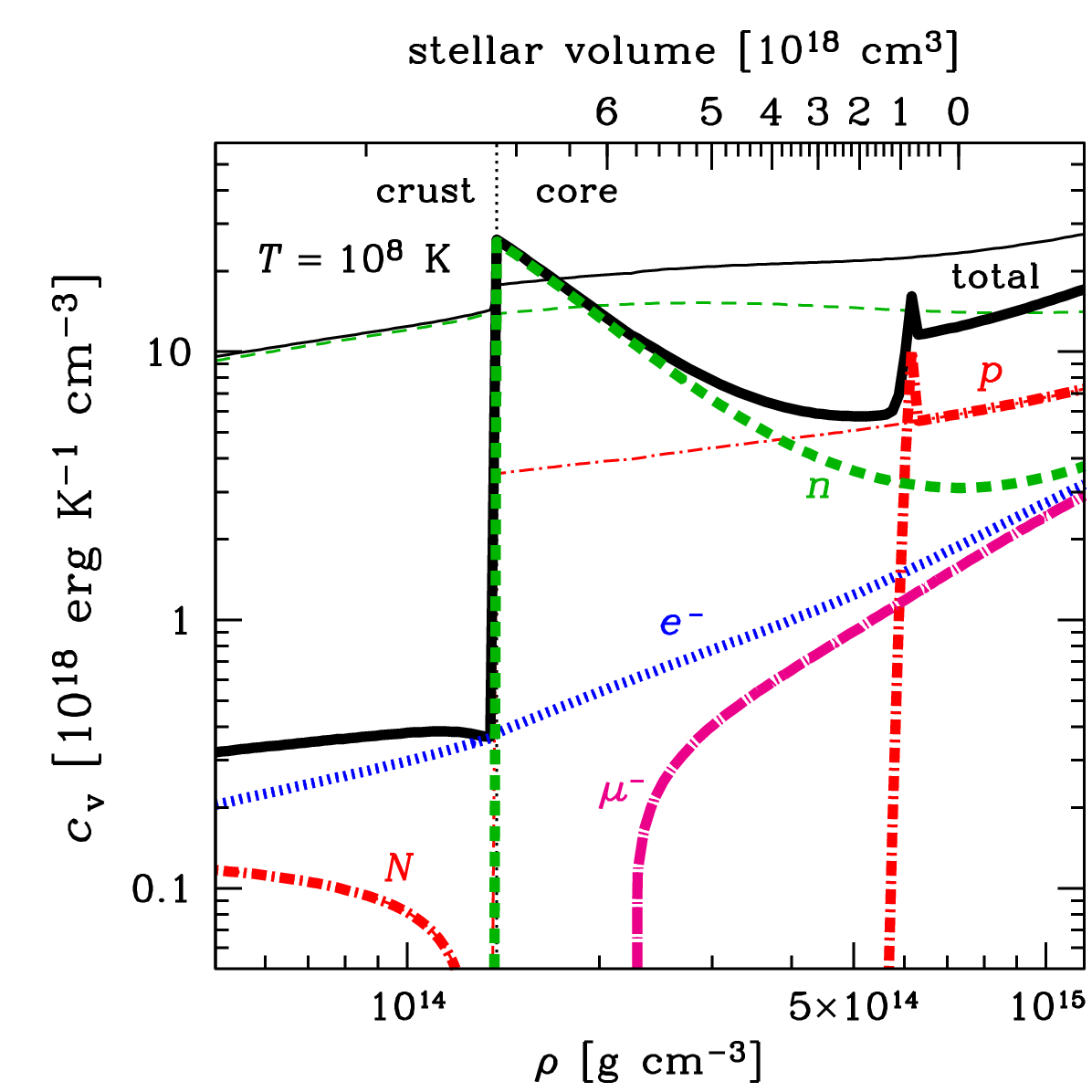}
\caption{
Left panel:
Specific heat control functions for the \singlet,
\triplet{} ($m_J=0$), and \triplet{} ($|m_J|=2$) 
types of pairing listed in Sect.~\ref{sect:pairing}.
The inset displays the same functions on a logarithmic
scale.
Right panel: Total and partial specific heats near the
bottom of the crust and in the core of a neutron star at 
$T=10^8$~K as functions of density. The solid lines
show the total $c_\mathrm{v}$, and the other lines show the
contributions of electrons ($e^-$), neutrons ($n$), nuclei
($N$, in the crust), muons and protons ($\mu^-$ and $p$, in
the core). Thin lines show results of a calculation with
nucleons assumed to be unpaired, and thick lines take
pairing into account. The top axis shows the volume
contained inside a sphere with given $\rho$ for a 1.4
$M_\odot$ neutron star. The stellar structure and
composition are adopted from the BSk21 model.
\label{Fig:Cv_core}}
\end{center}
\end{figure}

Three examples of the control functions, calculated
according to \citet{YLS99} (with the correction mentioned in
footnote \ref{RC-corr}), are shown in the left panel of
\fig{Fig:Cv_core}. One can notice that $c_\mathrm{v,x}$
nearly vanishes when $T$ drops below $\sim 0.1
T_\mathrm{c,x}$. Therefore, in the case of extensive pairing
of baryons, the heat capacity of the core can be reduced to
its leptonic part. This would result in a drastic reduction
of the total specific heat, as already demonstrated by the
heavy long-dashed line in \fig{fig:CVcrust}, where we
adopted MSH, TToa (assuming $m_J=0$), and BS superfluidity
models for neutrons in the crust and core, and protons in
the core, respectively, according to the notations in the
caption to \fig{fig:Tc}. 

Another example of the distribution of $c_\mathrm{v}$ among
the various core constituents is shown in the right panel of
\fig{Fig:Cv_core}. Here, we have adopted SFB, BEEHS (with
$m_J=0$), and BCLL pairing gaps. 
The behavior of $c_\mathrm{v}$ as function of
$\rho$ proves to be qualitatively similar for different sets of
superfluid gap models. In all cases this behavior strongly
differs from that for unpaired
nucleons, which is shown by thin lines for comparison.

%----------------------------------------------------------------------------
%---------------------------------------------------------------------------------------

% -----------------------------------
\subsection{Superfluid effects on neutrino emission}
\label{sect:SFnu}

The enormous impact of pairing on the cooling comes directly
from the appearance of the energy gap $\Delta_\mathrm{pair}$
at the Fermi surface which leads to a suppression of all
processes involving single particle excitations of the
paired species. When $T\ll\Tc$ the suppression is of the
order of $\mathrm{e}^{-\Delta_\mathrm{pair}/T}$ and hence
dramatic. Its exact value depends on the details of the
phase space involved in each specific process. In numerical
calculations it is introduced as a control function. As well
as for the heat capacity, for the neutrino emissivity one
writes
\beq
   Q_\nu = R_\textrm{(pairing type)}^\textrm{(process type)}
   Q_\nu^{(0)},
\eeq
where $Q_\nu^{(0)}$ relates to the same process
in the absence of pairing. These control functions
(reduction factors) are available in the form of analytical
fits, referenced in Table~\ref{tab:nu}.

The superfluidity not only reduces the emissivity of the
usual neutrino reactions but also initiates a specific
``pair breaking and formation'' (PBF) neutrino emission
mechanism. The superfluid or superconducting condensate is
in thermal equilibrium with the single particle (``broken
pairs'') excitations and there is continuous formation and
breaking of Cooper pairs. The formation of a Cooper pair
liberates energy which can be taken away by a $\nu-\bar\nu$
pair \citep{FlowersRS76,VoskresenskySenatorov87}. This
effect is most pronounced near the Fermi surface. When $T$
falls below $\Tc$, the neutrino emissivity produced by the
Cooper pairing sharply increases. The PBF mechanism is
sensitive to the model adopted for calculating the
superfluid gaps in the baryon spectra: it is more important
for lower gaps (weaker superfluid). Its emissivity is a
sharp function of density and temperature. The main neutrino
energy release takes place in the temperature interval
between $\sim\Tc/5$ and $\Tc$. The control functions and the
intensity of the Cooper-pair neutrino emission are available
as analytical fits collected by \citet{YKGH01} (see
references therein for the original derivations), as
indicated in Table~\ref{tab:nu} above.

\citet{VoskresenskySenatorov87} noticed that the PBF
mechanism is sensitive to the in-medium renormalization of
the nucleon weak-interaction vertex due to strong
interactions (cf.{} Sect.~\ref{sect:medium}). Later this
effect has been reexamined in many papers for different
types of baryon pairing -- see \citet{Leinson09,Leinson10}
for modern results and a critical analysis of previous
works. The net result is that the collective effects
virtually kill down the PBF emission for the singlet pairing
of nucleons, but leave this mechanism viable for the triplet
pairing. Quantitatively, PBF emissivity estimated without
in-medium effects \citep{YKL99} has to be multiplied by a
small factor of $(\pF/m^* c)^2$ in the \singlet{} case, but
by a moderate factor of $\approx0.19$ in the \triplet{}
case. This result lies at the basis of the ``minimal cooling
scenario'' and the explanation of the observed fast cooling
of the neutron star in the Cassiopeia A supernova remnant
(see Sect.~\ref{sect:cooling}).

Superconductivity of protons may also induce another type of
neutrino emission, electron-fluxoid scattering, in the
presence of a strong magnetic field. It will be addressed in
Sect.~\ref{sect:numag}.

% -----------------------------------
\subsection{Superfluid effects on heat conduction}
\label{sect:SFcond}

The effects of nucleon superfluidity on the heat transport
in neutron stars were discussed qualitatively by
\citet{FlowersItoh76,FlowersItoh81}. The thermal
conductivity of electrons and muons was reconsidered by
\citet{GnedinYakovlev95} and later by
\citet{ShterninYakovlev07}, who obtained accurate analytical
expressions valid for a wide class of models of superfluid
and non-superfluid matter. \citet{BaikoHY01} reanalyzed the
thermal conduction by neutrons, utilizing some new
developments in the  nucleon--nucleon interaction theory.
The latter authors showed that the low-temperature behavior
of the nucleon thermal conductivity is very sensitive to the
relation between critical temperatures of neutrons and
protons.

The lepton heat conduction in the core can also be affected
by proton superconductivity, because superconductivity modifies
the transverse polarization function and screening
functions in neutron-star matter. These effects were studied by
\citet{ShterninYakovlev07}. These authors, as well as
\citet{BaikoHY01}, managed to describe the effects of
superfluidity by analytical functions, which facilitate
their inclusion in simulations of neutron-star thermal
evolution (see Table~\ref{tab:cond}).

In the presence of neutron superfluidity, there may be
another channel of heat transport, the so-called convective
counterflow of the normal component of matter with respect to
the superfluid one. This mechanism is known to be quite
effective in superfluid helium \citep[e.g.,][]{Tilley90},
but in the context of neutron stars the situation is unclear and 
has not been studied in sufficient detail.

Heat can also be transported through the neutron star crust by
collective modes of superfluid neutron matter, called
superfluid phonons \citep{aguilera09}. At $\rho \approx
10^{12}-10^{14}$ \gcc{} the conductivity due to superfluid
phonons was estimated to be significantly larger than that
due to lattice phonons and comparable to electron
conductivity when $T \approx 10^8$~K. The authors found that
this mode of heat conduction could limit the
anisotropy of temperature distribution at the surface of
highly magnetized neutron stars. However, new
studies of  the low-energy collective excitations in the
inner crust of the neutron star \citep{Chamel12,Chamel13},
including neutron band structure effects, show that there is
a strong mixing between  the Bogoliubov-Anderson bosons of
the neutron superfluid and the longitudinal crystal lattice
phonons. In addition, the speed of the transverse shear mode
is  greatly reduced as a large fraction of superfluid
neutrons  are entrained by nuclei. This results in an
increased specific heat of the inner crust, but also in a
decrease of the electron thermal conductivity. On the other
hand, the entrainment of the unbound neutrons decreases the
density of conduction neutrons, i.e., neutrons that are
effectively free. The density of the conduction neutrons can
be much smaller than the total density of unbound neutrons
\citep{Chamel12}, which results in a decrease of the neutron
thermal conductivity.

%%%%%%%%%%%%%%%%%%%%%%%%%%%%%%%%%%%%%
\section{The effects of strong magnetic fields}
\label{sect:magnetic}
%%   ---------------------------
\subsection{Magnetic-field parameters}
\label{sect:mag-par}

Convenient dimensionless parameters that characterize the
magnetic field in a plasma are the ratios of the electron
cyclotron energy $\hbar\omc$ to the atomic unit of energy,
electron rest energy, and temperature:
\beq
   \gammam=
         \frac{\hbar^3 B }{ \mel^2 c e^3} =
         \frac{B}{B_0},
\qquad
   b=\frac{\hbar eB}{\mel^2 c^3} = 
           \frac{B}{B_\mathrm{QED}}\,,
\qquad
   \zete = \frac{\hbar\omc}{T} = 134.34\,\frac{
B_{12}}{T_6}.
\label{zeta_e}
\eeq
Here, $\omc=eB/\mel c$ is the electron cyclotron frequency,
$B_0=2.3505\times10^9$~G is the atomic unit of magnetic
field, $B_\mathrm{QED}=4.414\times10^{13}$~G is the critical
field in Quantum Electrodynamics \citep{Schwinger}, and
$B_{12}\equiv B/10^{12}$~G.

Motion of charged particles in a magnetic field is quantized
in discrete Landau levels. In the non-relativistic theory,
the energy of an electron in a magnetic field equals
$N\hbar\omc+\mel p_z^2/2$, where $p_z$ is the momentum
component along $\bm{B}$, $N=n_\mathrm{L}+\frac12\mp\frac12$
characterizes a Landau level, the term $\mp\frac12$ is the
spin projection on the field, and $n_\mathrm{L}$ is the
non-negative integer Landau number related to the
quantization of the kinetic motion transverse to the field.
In  the relativistic theory \citep[e.g.,][]{SokTer}, the
kinetic energy $\epsilon$ of an electron at the Landau level
$N$ depends on its longitudinal momentum $p_z$ as
\beq
 \epsilon_N(p_z) = c\,\left(\mel^2 c^2 + 2\hbar\omc
   \mel N+p_z^2\right)^{1/2} - \mel c^2.
\label{magnenergy}
\eeq
The levels $N\geqslant 1$ are double-degenerate with respect
to the spin projection $s$. Their splitting $\delta\epsilon$
due to the anomalous magnetic moment of the electron is
negligible, because it is much smaller than $\hbar\omc$
\citep[e.g.,][]{Schwinger,SuhMathews}: 
\beq
   \delta\epsilon\approx\frac{\alphaf}{2\pi}\times
\left\{\begin{array}{l}
   \hbar\omc \mbox{~at~}b\ll1, \\
    \mel c^2\,[\ln b-1.584]^2 \mbox{~at~}b\gg1,
\end{array}\right.
\eeq
where $\alphaf$ is the fine structure constant.

The Landau quantization becomes important when the electron
cyclotron energy $\hbar\omc$ is at least comparable to both
the electron Fermi energy $\EF$ and temperature $T$. If
$\hbar\omc$ is appreciably larger than both $\EF$ and $T$,
then the electrons reside on the ground Landau level, and the
field is called \emph{strongly quantizing}. The condition
$\hbar\omc>T$ is equivalent to $\zete>1$. The condition
$\hbar\omc>\EFe$ translates into $\rho < \rho_B$, where
\beq
  \rho_B 
  \approx 7045 \,Y_e^{-1}\,B_{12}^{3/2}\text{ \gcc}.
\label{rho_B}
\eeq
In the opposite limit, where either $\zete\ll1$ or
$\rho\gg\rho_B$, the field can be considered as
\emph{nonquantizing}.

For the ions, the cyclotron energy is $\hbar\omci =
Z\,(\mel/\mion)\, \hbar\omc$, and the Landau quantization is
important when the parameter
\beq
   \zeti = \hbar\omci/T = 0.0737\,(Z/A)\,
           B_{12}/T_6
\label{omci}
\eeq
is not small. The energy spectrum of an ion essentially differs from
\req{magnenergy} because of the non-negligible anomalous
magnetic moments. In the non-relativistic theory, the energy
of an ion equals 
$   \epsilon= (n_\mathrm{L}+\frac12)\hbar\omci+\mion p_z^2/2
   + \frac14\, \gfact \zeti \sion,
$
where $n_\mathrm{L}$ is the ion Landau number, $p_z$ is the
longitudinal momentum, $\gfact$ is the $g$-factor
($\gfact=2$ in the Dirac theory, but, e.g.,
$\gfact=5.5857$ for the protons), and $\sion$ is the integer
quantum number corresponding to the spin
projection on $\bm{B}$ in units of $\hbar/2$. If the ions are
relativistic, the situation is much more complicated. For
baryons with spin $\frac12$ (e.g., protons)
the energy spectrum was derived by \citet{Broderick00}.

% -----------------------------------
\subsection{Magnetic field effects on the equation of state
and heat capacity}
\label{sect:EoSmag}

% -----------------------------------
\subsubsection{Magnetized core}

A magnetic field can affect the thermodynamics of the
Coulomb plasmas, if the Landau quantization is important,
i.e., under the conditions that are quantified in Sect.~\ref{sect:mag-par}.
In particular, \req{rho_B} can be recast into
\begin{equation}
   B \gtrsim (3.8\times 10^{19}\mbox{ G})\,( Y_e
   n_\mathrm{b}/\mbox{fm}^{-3})^{2/3}.
\label{Blarge}
\end{equation}
We have $n_\mathrm{b} \sim 0.1$ fm$^{-3}$ near the crust-core
interface,
and $Y_e$ is typically several percent throughout the core.
Therefore, the electron component of pressure in the
core might be affected by the fields $B\gtrsim 10^{18}$~G.

One can easily generalize \req{Blarge} for other fermions
($\mu$-mesons, nucleons) in the ideal-gas model. In this
case, $Y_e$ should be replaced by the number of given
particles per baryon, and the right-hand side should be
multiplied by $m_\mu/\mel=206.77$ for muons and $\sim10^3$
(of the order of nucleon-to-electron mass and
electron-to-nucleon magnetic moment ratios) for protons and
neutrons. Accordingly, the partial pressures of muons and
nucleons in the core cannot be affected by any realistic
($B\lesssim\mbox{a few}\times10^{18}$~G) magnetic field. 

\citet{Broderick00} developed elaborated models of matter in
ultra-magnetized cores of neutron stars. They considered not
only the ideal $npe\mu$ gas, but also interacting matter in
the framework of the relativistic mean field (RMF) model.
The magnetic field affects their EoS at $B\gtrsim10^{18}$~G.
As follows both from the estimates based on the virial
theorem \citep{LaiShapiro91} and from numerical hydrodynamic
simulations (e.g., \citealp{FriebenRezzolla12}, and
references therein), this field is close to the upper limit
on $B$ for dynamically stable stellar configurations. The effect is
even smaller when the magnetization of matter is included consistently in
the EoS \citep{Chatt15}.
Therefore, it is unlikely that a magnetic modification of
the EoS could be important in the cores of neutron stars.

% -----------------------------------
\subsubsection{Magnetized crust and ocean}

At  $B\gtrsim10^{16}$~G,  nuclear shell energies become
comparable with the  proton cyclotron energy. Thus the
interaction of nucleon magnetic moments and proton orbital
moments with magnetic field may cause appreciable modifications
of nuclear shell energies. These modifications and their
consequences for magnetars were studied by
\citet{KondratyevMC01}, who found large changes in the
nuclear magic numbers under the influence of such magnetic
fields. This effect may alter significantly the equilibrium 
chemical composition of a magnetar crust.

\citet{MuzikarSS80} calculated the triplet-state neutron
pairing in magnetized neutron-star cores. According to these
calculations, magnetic fields $B\gtrsim10^{16}$~G make
the superfluidity with nodes at the Fermi surface
energetically preferable to the usual superfluidity without
nodes. Accordingly, the superfluid reduction factors for the
heat capacity and neutrino emissivity (the control
functions) may be different in ultra-strong fields.

\citet{Chamel-ea12} studied the impact of superstrong
magnetic fields on the composition and EoS of the neutron
star crust. In particular, they found that the neutron-drip
pressure increases almost linearly by 40\% from its
zero-field value in the interval $10^{16}\mbox{
G}<B<5\times10^{16}\mbox{ G}$. With further increase of the
field strength, the drip pressure becomes directly
proportional to $B$.

%%%%%%%%%%%%%%%%%%%%%%%%%%%%%%%%%%%%%%%%%%%%%%%%
\begin{figure}
\includegraphics[width=.49\textwidth]{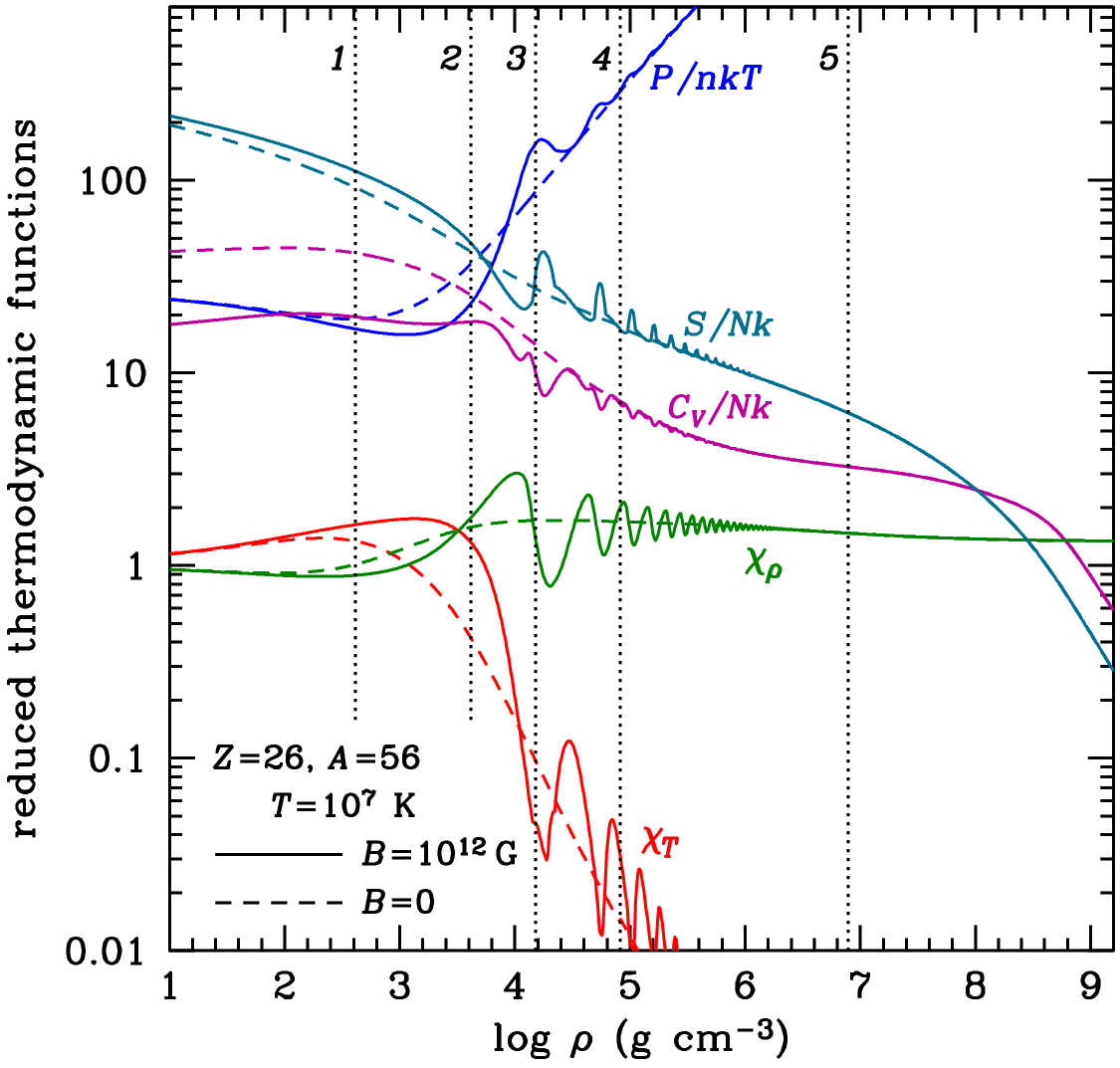}
\includegraphics[width=.49\textwidth]{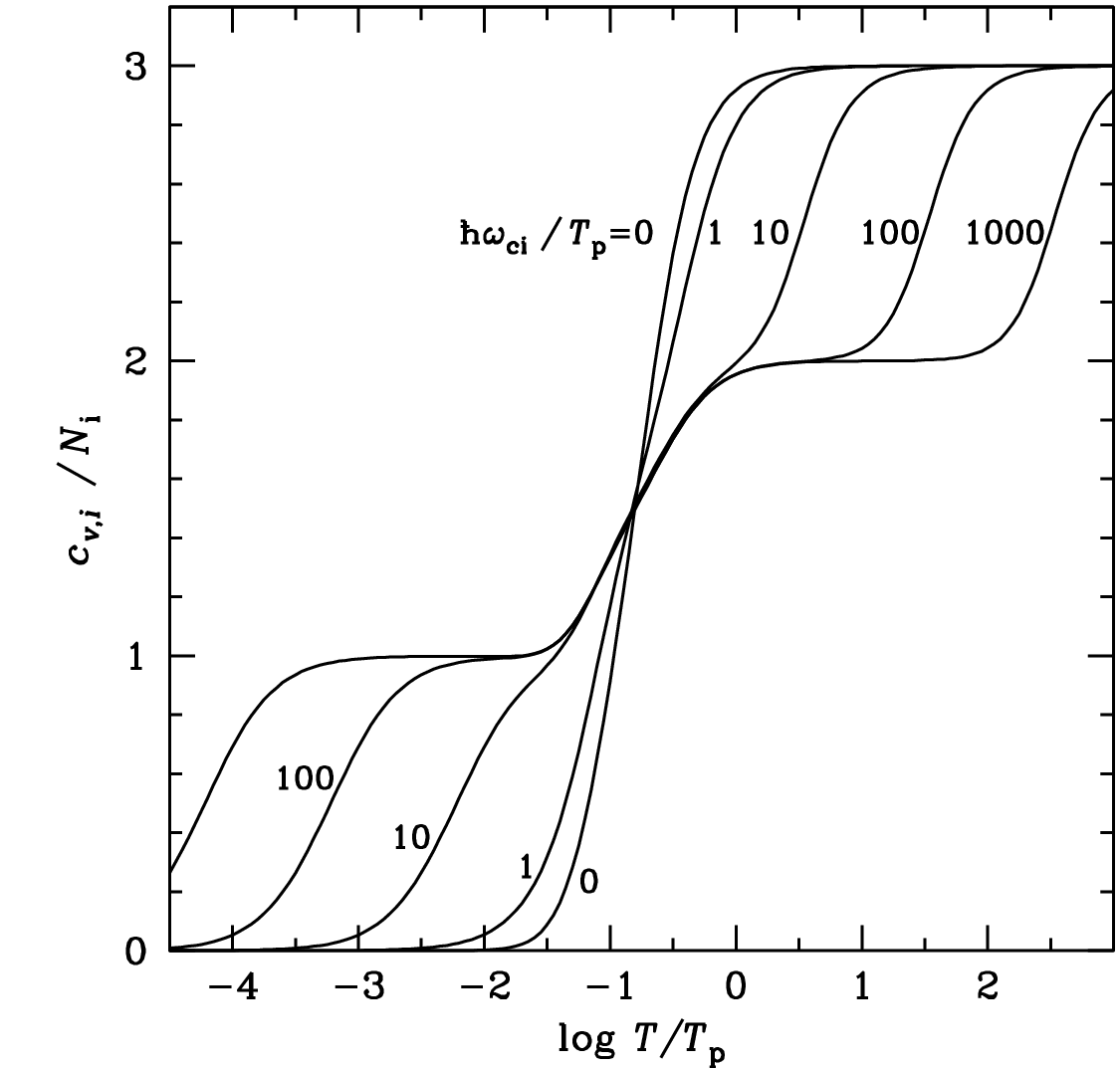}
\caption{Left panel:
Normalized pressure $P/\nion T$; entropy $S$ and heat
capacity $C_V$ per one ion, and
logarithmic derivatives of pressure over density and
temperature, $\chi_\rho$ and $\chi_T$, for a fully-ionized
non-magnetic (dashed lines) and magnetized ($B=10^{12}$~G,
solid lines) iron plasma at $T=10^7$~K. The vertical dotted
lines mark the densities at which the electron Fermi
temperature equals $T$ without (\emph{1}) or with (\emph{2})
the magnetic field, $\rho=\rho_B$ (\emph{3}),
$\Gami=\Gammam$ (\emph{4}), and $\Tp=T$ (\emph{5}).
(Figure~6 from \citealp{PC13}, reproduced with the
permission of \copyright ESO.)
Right panel: 
Normalized  thermal phonon contribution to the reduced heat
capacity as a function of $\log_{10}(T/\Tp)$ at different values
of the ratio $\hbar\omci/\Tp$, marked near the curves.
}
\label{fig:eosmag}
\end{figure}
%%%%%%%%%%%%%%%%%%%%%%%%%%%%%%%%%%%%%%%%%%%%%%%%

Thus the ultra-strong fields $B\gtrsim10^{16}$~G can affect
various aspects of the physics of the inner crust
in quite non-trivial way. Hereafter
we will consider only  fields $B\lesssim10^{16}$~G. They can
be quantizing  in the outer crust of a neutron star, but not
in the inner crust or the core. Analytical fitting formulae
for the thermodynamic functions of the electron-ion plasmas
in such fields, as well as a computer code that implements
these fits\footnote{Also available at
\texttt{http://www.ioffe.ru/astro/EIP/}.}, were published by \citet{PC13}. Such fields affect
the electron part of thermodynamic functions in the outer
envelopes only, as illustrated in the left panel of
Fig.~\ref{fig:eosmag} in the case of fully-ionized iron 
at $T=10^7$~K and $B=10^{12}$~G  (for illustration, the
density range is extended to $\rho\lesssim10^5$ neglecting
the bound states that can be important in this
$\rho$\,--\,$T$ domain).  We plot the principal
thermodynamic quantities normalized per one ion as functions
of density. For comparison we also show them in the absence
of quantizing magnetic field. The vertical dotted lines
marked by numbers separate different characteristic domains,
consecutively entered with increasing density: onset of
electron degeneracy at $B=0$ and at $B=10^{12}$~G,
population of excited Landau levels ($\rho=\rho_B$), melting
point with formation of a classical Coulomb crystal
($T_\mathrm{m}=T$), and onset of the quantum effects in the
crystal ($\Tp=T$). The gradually decreasing oscillations
correspond to consecutive filling of the electron Landau
levels. The magnetic field $B=10^{12}$~G does not affect the
ion contributions at this $T$.

The contributions of ions to the thermodynamic functions are
affected by the magnetic field if the parameter $\zeti$,
defined by \req{omci}, is large. This may occur in a
superstrong field of a magnetar. The right panel of
Fig.~\ref{fig:eosmag} illustrates the effects of a
superstrong field on $c_\mathrm{v,i}$ for the model of a
harmonic Coulomb crystal \citep{Baiko09}. Here we plot the
thermal phonon contribution to the heat capacity of the bcc
Coulomb lattice calculated as the derivative
$c_\mathrm{v}=T\partial S/\partial T$ of the fit to the
phonon entropy $S$ given by Eq.~(77) of \citet{PC13}. This
approximation is more accurate for the heat capacity than
the alternative approximation that provides exact
fulfillment of the Maxwell relations (Eq.~80 of the same
paper). The three steps on the curves in the right panel of
Fig.~\ref{fig:eosmag} correspond to contributions of three
branches of the phonon spectrum, which are affected
differently by the quantizing magnetic field.

% -----------------------------------
\subsection{Magnetic field effects on neutrino emission}
\label{sect:numag}

% -----------------------------------
\subsubsection{Magnetic Durca process} 

We have mentioned in Sect.~\ref{sect:nu-core} that the Durca
reaction is the most efficient neutrino emission process,
but it can only operate above a certain threshold density in
the central parts of the cores of sufficiently massive
neutron stars.  \citet{LeinsonPerez98} noted that a
superstrong magnetic field can substantially weaken this
requirement. An accurate study of this effect was performed
by \citet{BaikoYakovlev99}. They showed that the border
between the open and closed Durca regimes is smeared out
over some $B$-dependent scale and described this smearing by
simple formulae. In practice this effect should be very
important for neutron stars with $B\gtrsim10^{16}$~G. At
less extreme fields ($10^{14}\mbox{~G}\lesssim B \lesssim
10^{16}$~G) it is important for neutron stars whose mass
happens to be close (within a few percent) to the Durca
threshold mass. \citet{BaikoYakovlev99} also showed that a
strong magnetic field has a non-trivial effect (oscillations
of the reaction rate) in the permitted domain of the Durca
reaction, but the latter effect, albeit interesting, appears
to be unimportant.

% ----------------------------------- 
\subsubsection{Pair annihilation}

 The $e^-e^+$ pair annihilation process in strong magnetic fields 
was studied by \citet{Kaminker-ea92} and \citet{KaminkerYakovlev94}. 
In a hot, non-degenerate plasma ($T \gtrsim 10^{10}$ K) only
ultra-strong magnetic fields $B\gtrsim10^{16}$~G can
significantly affect the neutrino emissivity. Such fields
can be quantizing in the $\rho-T$ domain where the pair
emission dominates (see Fig.~\ref{Fig:Nu_crust}). They
amplify $Q_\mathrm{pair}$ by increasing the number densities
of electrons and positrons via very strong quantization of
their motion. Lower fields may also influence
$Q_\mathrm{pair}$ but less significantly. A field $B \sim
10^{14}$~G may quantize the motion of positrons at $T
\lesssim 10^9$~K and increase the positron number density.
In this way the presence of a strong magnetic fields greatly enhances
$Q_\mathrm{pair}$ in a not too hot plasma. However, this
enhancement usually takes place where the pair annihilation
emissivity is much lower than the contribution from other neutrino
reactions, and therefore it is unimportant for studies of
neutron-star thermal evolution.

% -----------------------------------
\subsubsection{Synchrotron radiation} 

\begin{figure}
\begin{center}
\includegraphics[width=.49\textwidth]{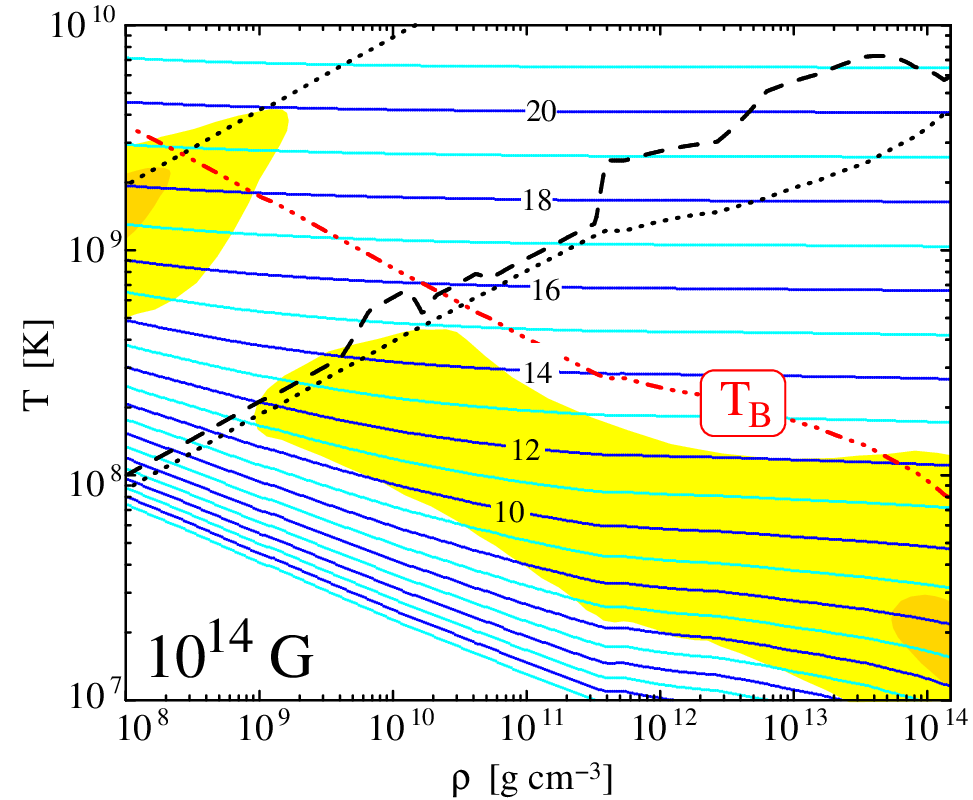}
\includegraphics[width=.49\textwidth]{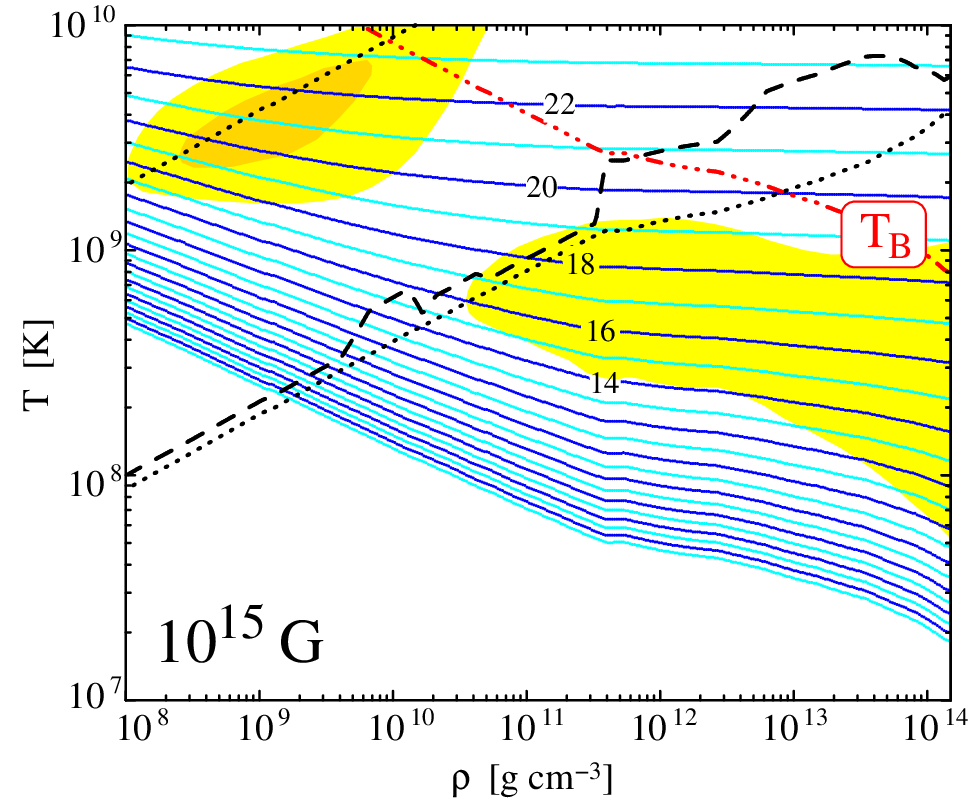}
\caption{Neutrino emissivity in a magnetized crust from the synchrotron processes for two,
uniform, magnetic field strengths of $10^{14}$ G (left panel) and $10^{15}$ G (right panel).
The contour lines are labeled by the value of 
$\mathrm{log}_{10} [Q_\nu/ (\mathrm{erg \, cm}^{-3} \, \mathrm{s}^{-1})]$.
Regions where this process dominates over the ones shown in \fig{Fig:Nu_crust} are
lightly shadowed (in yellow) and regions where it dominates by more than a factor of 10 are 
darkly shadowed (in orange).
The two dotted lines show the dominance transitions between the three processes
presented in \fig{Fig:Nu_crust}.
(Also indicated is the ion melting curve, dashed line.)
\label{Fig:Nu_synch_crust}
}
\end{center}
\end{figure}

A relativistic electron propagating in the magnetic field
can emit neutrinos because of its rotation around the
magnetic field lines. This process is quite analogous to the
usual synchrotron emission of photons. The calculation of
the corresponding neutrino emissivity, $Q_\mathrm{syn}$, is
similar to that of the pair annihilation process. It was
studied, e.g., by \citet{Kaminker-ea92}, \citet{Vidaurre95}, and
\citet{Bezchastnov-ea97}. In Fig.~\ref{Fig:Nu_synch_crust}
we show the plot of $Q_\mathrm{syn}$ on the $\rho-T$ plane
for two field strengths typical for magnetars, $B=10^{14}$~G
and $10^{15}$~G. It is clear from this plot that the
the synchrotron process can be dominant in the crust of 
magnetars in a large temperature range.

%---------------------------------------------------------------------------------------
%---------------------------------------------------------------------------------------

% ----------------------------------- 
\subsubsection{Electron-fluxoid scattering}
\label{sect:fluxoid}

The internal stellar magnetic field can be confined in the crust or be
distributed over the entire star. In the latter case, a
transition to a superconducting state in the course of
stellar cooling is accompanied by a dramatic change in the
spatial structure of the magnetic field. Initially
homogeneous field splits into an ensemble of Abrikosov
fluxoids -- quantized magnetic flux tubes, which contain a
superstrong magnetic field, embedded in the field-free
superconducting medium. Neutrino synchrotron radiation is
then modified and may be treated as neutrino pair emission
due to scattering of electrons on the fluxoids. This
mechanism was studied by \citet{KaminkerYH97}, who obtained
an analytical fit to the corresponding neutrino emissivity
(referenced in Table~\ref{tab:nu}). The concentration of the
field within the fluxoids  amplifies the neutrino
emissivity, compared to the usual synchrotron regime, when
temperatures drops below the critical one for the protons,
$\Tcp$. As long as $T$ is not much lower than $\Tcp$, the
Cooper pairing mechanism remains much more powerful (unless
$B\gg10^{16}$~G, in which case the electron-fluxoid
scattering may be more powerful at any $T$). At $T\ll \Tcp$,
the electron-fluxoid scattering becomes the dominant
neutrino emission mechanism for the neutron stars with
strong and superstrong fields ($B\gtrsim10^{12}$~G).

% -----------------------------------
\subsection{Magnetic field effects on heat conduction}
\label{sect:condmag}

% ------------------------------------------
\subsubsection{Photon heat conduction}
\label{sect:radiopa}

The thermal conductivity $\kappa$ is related to the opacity
$\opac$ by the equation 
\begin{equation} 
  \kappa =
  \frac{16\sSB T^3}{3\rho\opac},
\label{opacity}
\end{equation} 
where $\sSB$ is the Stefan-Boltzmann constant.  The spectral
radiative opacities for two normal polarization modes in
strongly magnetized neutron-star photospheres are reviewed
in Paper~I. These opacities, $\opac_{\omega,j}(\theta_B)$,
where $j=1,2$ marks the extraordinary and ordinary
polarization modes, depend on the angle $\theta_B$ between
the wave vector and magnetic field.  In the diffusion
approximation,  they combine into the effective opacities
for the transport along ($\opac_{\omega,j}^\|$) and across
($\opac_{\omega,j}^\perp$) magnetic field according to 
\begin{equation}
\left\{
   \begin{array}{c}
 (\opac_{\omega,j}^\|)^{-1}
\\
   (\opac_{\omega,j}^{\perp})^{-1\rule{0pt}{2ex}}
   \end{array}
  \right\}
 = \frac34 \int_0^\pi
\left\{
   \begin{array}{c}
2\cos^2\theta_B \\
\sin^2\theta_B
\end{array}
  \right\}
\frac{\sin\theta_B\,\dd\theta_B}{\opac_{\omega,j}(\theta_B)}\,.
\label{kappa-eff}
\end{equation}      
The effective opacity for energy transport
at angle $\theta$
to $\bm{B}$ in each
polarization mode is given by
$
1/\bar\opac_j =
\cos^2\theta/\bar\opac_j^\|
 + \sin^2\theta/\bar\opac_j^\perp,
$
where $\bar\opac$ is the
Rosseland mean of $\opac_\omega$,
\begin{equation}
\frac{1}{\bar\opac_j} \equiv
    \int_0^{\infty}\frac{u(z)}{\opac_{\omega,j}}
 \dd z ,
\quad
 u(z) = \frac{15}{4\pi^4}
\frac{z^4\mathrm{e}^z}{(\mathrm{e}^z-1)^2},
\quad 
  z =  \frac{\hbar\omega}{T}.
\end{equation}

For fully ionized plasmas, the radiative opacities are
contributed from the free-free absorption and Thomson
scattering.
\citet{SilYak} studied the Rosseland
opacities for a non-polarized radiation in magnetized fully
ionized plasmas
\begin{equation}
\opac_\mathrm{r}^{\|,\perp}=\left[ 
    1/\bar\opac_1^{\|,\perp} + 1/\bar\opac_2^{\|,\perp}
\right]^{-1}
\end{equation}
using the Born approximation for the
free-free  contribution.
\citet{PY01} obtained simple analytical fits for
$\opac_\mathrm{r}^{\|,\perp}$, including a correction to the
Born approximation, as functions of $\rho$, $T$, $Z$, $A$,
and the magnetic-field parameter $\zete$ defined by \req{zeta_e}.
Asymptotically, $\opac_\mathrm{r}\propto \zete^{-2}$ at
$\zete\to \infty$.

At finite but large $\zete$, the radiative opacities of
fully ionized matter are strongly reduced. The reduction is
$\sim10$ times stronger for the Thomson scattering than for
the free-free absorption. In deep, strongly magnetized
photospheric layers the Thomson scattering dominates only if
$T_6\gtrsim10\,\rho^{2/7} \gtrsim 10\,B_{12}^{2/7}$;
otherwise the free-free absorption prevails \citep{PY01}.

The outermost envelopes of neutron stars can be incompletely
ionized in the cases of large $Z$ or $B$. The presence of
bound species can strongly affect the radiative opacities
and the spectrum of emitted radiation, as discussed in
Paper~I. However, the layers that are responsible for the
heat flux from the interior of the neutron star to the
surface, as a rule, lie at sufficiently large densities,
where the plasma is fully ionized by pressure. Therefore the
bound species are usually unimportant for the effective
surface temperature of a neutron star.

% ------------------------------------------
\subsubsection{Electron heat conduction}
\label{sect:condmag-e}

A non-quantizing magnetic field does not affect 
thermodynamic functions of the plasma. However,
it does affect the electron heat conduction, if
the \emph{Hall magnetization parameter} 
\begin{equation}
   \omg\tau\approx 1760\,\frac{B_{12}}{\sqrt{1+x_\mathrm{r}^2}}
   \,\frac{\tau}{10^{-16}\mathrm{~s}}
\label{omegatau}
\end{equation}
 is not small.
Here, $\omg=\omc/\sqrt{1+x_\mathrm{r}^2}$ 
is the electron gyrofrequency, and
$\tau$ 
is the effective relaxation
time. In a degenerate Coulomb plasma with a non-quantizing
magnetic field, the main contribution is given by the
electron-ion scattering according to \req{tau}.
This regime has been studied by \citet{YakovlevUrpin80}.

Electric and thermal currents
induced in a magnetized plasma
under the effect of an electric field $\bm{E}$, a
weak gradient $\nabla\mu$ of the electron 
chemical potential, and a weak temperature gradient $\nabla T$
 can be decomposed into conduction
and magnetization
 components \citep[e.g.,][]{Hernquist84}.
The latter ones relate to surface effects and
must be subtracted.
Let $\bm{j}_e$ and $\bm{j}_T$ be the conduction components of
the electric and thermal current densities. They can be written as
\begin{equation}
   \bm{j}_e = \hat{\sigma}\cdot\bm{E}^\ast - \hat{\alpha}\cdot\nabla T,
\quad
\bm{j}_T = T\hat{\alpha}\cdot\bm{E}^\ast-\hat{\tilde\kappa}\cdot\nabla T,
\label{J}
\end{equation}
where $\bm{E}^\ast = \bm{E}+\nabla\mu/e$
is the electrochemical field.
The symbols $\hat{\sigma}$, $\hat{\alpha}$, 
and $\hat{\tilde\kappa}$ denote second-rank tensors
($\hat{\sigma}$ is the conductivity tensor)
which reduce to scalars at $B=0$.
Equations (\ref{J}) can be rewritten as
\begin{equation}
   \bm{E}^\ast = \hat{R}\cdot\bm{j}_e - \hat{Q}\cdot\nabla T,
\quad
   \bm{j}_T = T \hat{\alpha}\cdot\hat{R}\cdot\bm{j}_e - \hat{\kappa}\cdot\nabla T,
\label{J1}
\end{equation}
where 
$\hat{R}=\hat{\sigma}^{-1}$,
$\hat{Q}=-\hat{R}\cdot\hat{\alpha}$, and
$\hat{\kappa} = \hat{\tilde\kappa} + T\hat{\alpha}\cdot \hat{Q}$
are the tensors of 
specific resistance, thermopower, and thermal conductivity,
 respectively.

Electron heat and charge transport controlled by electron-ion collisions in quantizing
magnetic fields of neutron stars was studied by
\citet{KaminkerYakovlev81,Yakovlev84,Hernquist84,P96,P99}.
The components of tensors $\hat{\sigma}$, $\hat{\alpha}$, and
$\hat{\tilde\kappa}$ can
be expressed as \citep{P99}
\begin{equation}
\left\{
   \begin{array}{c}
       \sigma_{ij} \\ \alpha_{ij} \\ \tilde\kappa_{ij}
   \end{array}
   \right\}
  =
     \int_{0}^\infty \left\{
   \begin{array}{c}
       e^2 \\ e {(\mu - \epsilon)/ T} \\ {(\mu - \epsilon)^2/ T}
   \end{array}
   \right\}
   \frac{\Ne}{m_e+\epsilon/c^2} \,\tau_{ij}(\epsilon)
   \left[ -\frac{\partial}{\partial\epsilon}
        \frac{1}{\mathrm{e}^{(\epsilon-\mu)/T}+1} \right]
      \mathrm{d}\epsilon,
\label{sigma-tau}
\end{equation}
where
\begin{equation}
    \Ne = \frac{1}{ 2 \pi^2 a_\mathrm{m}^2 \hbar}
            \sum_{N=0}^{N_\mathrm{max}} (2-\delta_{N,0})|p_z|,
\label{N_e}
\end{equation}
$N_\mathrm{max}$ is the maximum Landau number for a
given electron energy $\epsilon$, 
% $m_e^*(\epsilon) = m_e + \epsilon/c^2$ is the effective dynamical electron mass,
and $|p_z|$ depends
on $\epsilon$ and $N$ according to \req{magnenergy}.
In a non-quantizing magnetic field, i.e., at
$N_\mathrm{max}\gg1$, the sum can be replaced by the
integral, which gives $\Ne=(p/\hbar)^3/3\pi^2$, where $p$ is
the momentum that corresponds to the energy $\epsilon$.
The functions $\tau_{ij}(\epsilon)$
play role of relaxation times for the components of 
tensors $\hat{\sigma}$, $\hat{\alpha}$, and $\hat{\tilde\kappa}$,
determined by electron scattering.
In general, they differ from the mean free time
$\tau_{e\mathrm{i}}(\epsilon)=1/\nu_{e\mathrm{i}}(\epsilon)$
between scattering events for an electron with energy
$\epsilon$.
Because of the symmetry properties of
the tensors $\hat{\sigma}$, $\hat{\alpha}$, and
$\hat{\tilde\kappa}$, in the
coordinate frame with $z$ axis directed along $\bm{B}$,
there are only three different
non-zero components of $\tau_{ij}$:
$\tau_{zz}$ related to longitudinal currents,
$\tau_{xx}=\tau_{yy}$
 related to transverse currents,
and $\tau_{xy}=-\tau_{yx}$ related to the Hall
currents.

%%%%%%%%%%%%%%%%%%%%%%%%%%%%%%%%%%%%%%%%%%%%%%%%%%%%%%%%%%%%%%
\begin{figure}
\begin{center}
\includegraphics[height=.49\textwidth]{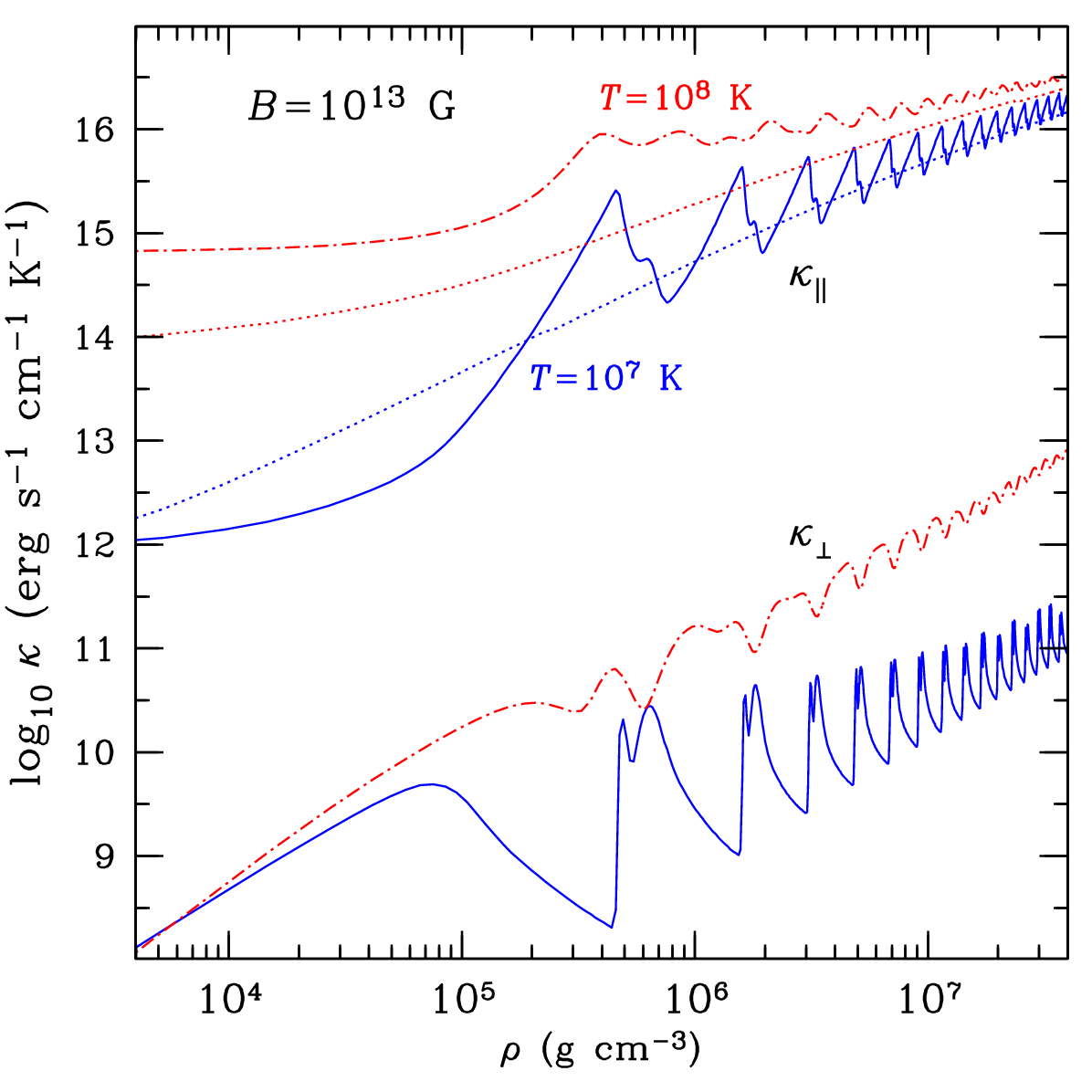}
\,
\includegraphics[height=.49\textwidth]{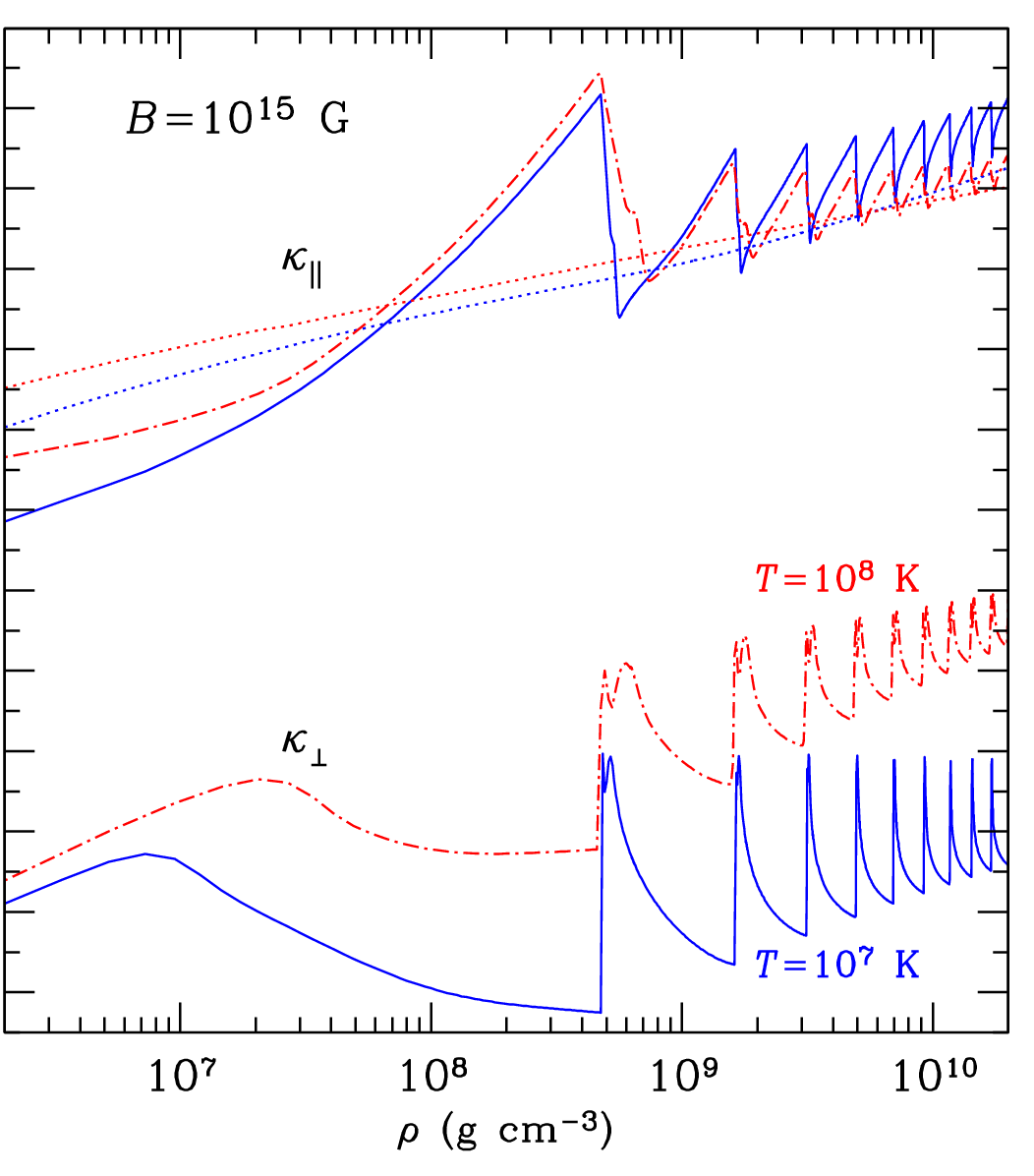}
\caption{Electron thermal conductivities along (upper
curves) and across (lower curves) magnetic field
$B=10^{13}$~G (left panel) and $10^{15}$~G (right panel) as
functions of mass density at temperatures $T=10^7$~K (solid
lines) and $10^8$~K (dot-dashed lines). For comparison, the
non-magnetic thermal conductivities are shown by dotted
lines.
}
\label{fig:cond}
\end{center}
\end{figure}
%%%%%%%%%%%%%%%%%%%%%%%%%%%%%%%%%%%%%%%%%%%%%%%%%%%%%%%%%%%%%%%%

In a quantizing magnetic field, there are
two different effective mean-free times
$\tau_\|(\epsilon)$ and $\tau_\perp(\epsilon)$, 
corresponding to electron
transport parallel and perpendicular to  $\bm{B}$.
In this case, the classical expressions
\citep[e.g.,][]{YakovlevUrpin80} are recovered:
\begin{equation}
      \tau_{zz}=\tau_\|,
\quad
   \tau_{xx}=\frac{\tau_\perp}{ 1+(\omg\tau_\perp)^2},
\quad
   \tau_{yx}=\frac{\omg\tau_\perp^2}{ 1+(\omg\tau_\perp)^2}.
\label{tau-quant}
\end{equation}
It is convenient to keep using \req{tau} for
$\tau_\|$ and $\tau_\perp$, but with
different Coulomb logarithms $\Lambda_\|(\epsilon)$ and
$\Lambda_\perp(\epsilon)$. \citet{P99} calculated these
Coulomb logarithms and fitted them by analytic
expressions. Their Fortran 
implementation is available at
\texttt{http://www.ioffe.ru/astro/conduct/}.
In the limit of non-quantizing magnetic field,
$\tau_\|=\tau_\perp=\tau_{e\mathrm{i}}(\epsilon)$ is
given by \req{tau} with $\epsilon=\mu$. 

When the electrons are strongly degenerate,
the derivative in the square brackets in \req{sigma-tau}
is sharply peaked.
Then \req{sigma-tau} gives
\begin{equation}
   \sigma_{ij} \approx \frac{e^2 c^2 n_e}{ \mu}\,\tau_{ij}(\mu),
\qquad
   \kappa_{ij} \approx \tilde\kappa_{ij}
   \approx \frac{\pi^2 T }{ 3e^2}\,\sigma_{ij}.
\label{Wiedemann-mag}
\end{equation}
The latter relation is the Wiedemann-Franz law generalized
to the magnetic case. On the
other hand, \req{sigma-tau} satisfactorily describes the
conductivities in general, including the opposite case of weakly
degenerate electrons.

Figure~\ref{fig:cond} illustrates the $\rho$-dependence of the
thermal conductivities along 
($\kappa_\|$) and across ($\kappa_\perp$) the magnetic field. The first, most significant
peak at $\kappa_\|$ is related to the filling of the first
Landau level by the electrons at $\rho\sim\rho_B$. The other
peaks correspond to consecutive filling of higher Landau levels.

%%%%%%%%%%%%%%%%%%%%%%%%%%%%%%%%%%%%%
\section{Thermal structure of neutron stars}
\label{sect:th_str}
% -----------------------------------
\subsection{Blanketing envelopes}
\label{sect:blanket}

The very different thermal relaxation timescales of the
envelope and the crust of a neutron stars makes computationally expensive
to perform cooling simulations in a numerical grid that
comprises both regions. Radiative equilibrium is established in the low-density
region much faster than the crust evolves, so that the envelope reaches a
stationary state on shorter timescales. Thus, the usual
approach is to use results of stationary envelope models to
obtain a relation between the photon flux 
$F_\mathrm{ph}$ radiated from the surface and the flux
$F_\mathrm{b}$ and temperature $T_\mathrm{b}$ at the crust/envelope
boundary, $\rho=\rhob$. This relation supplements the
evolution equations for the interior [\req{Tbalance}]
as an outer boundary condition.

The boundary density $\rhob$ is chosen as a
trade-off between two requirements: first, that the thermal
relaxation time of the layer with $\rho<\rhob$ is  short compared to the
characteristic variability time of the studied thermal
radiation, which favors smaller $\rhob$, and
second, that $T$ does not strongly vary at $\rho >
\rhob$, which favors larger $\rhob$. For
weakly magnetized, isolated cooling neutron stars,
$\rhob$ is usually set at $10^{10}$~\gcc{}
\citep{GPE83}, but in general it varies from $10^8$ \gcc{} for neutron
stars with rapid variations of thermal emission
\citep{Shternin-ea07} to
$\rhod$ for relatively hot and strongly
magnetized neutron stars \citep{Potekhin-ea03}.

At every $T_\mathrm{b}$, $F_\mathrm{ph}$ or,
equivalently, the effective surface temperature $\Ts$,
depends on the properties of the heat-blanketing envelopes. 
In the absence of neutrino energy losses in the envelope
(that is the case for most cooling neutron stars,
except for the hottest ones), the flux $F_\mathrm{ph}$ at
the surface is equal to the flux $F_\mathrm{b}$ at the inner
boundary of the blanketing envelope. Then it is sufficient
to know the \TbTs{} relation for cooling simulations.

\citet{GPE83} carried out a
comprehensive study of the thermal structure of the
non-magnetized blanketing envelopes composed of iron, using
the best physics input available at that time. They
considered the envelopes with $\log_{10} \Ts$[K]$ \geq 5.25$ (there
were no reliable calculation of the thermal conductivities for lower temperatures) 
and fitted the numerical solutions by a remarkably simple formula
\beq
     T_\mathrm{b} = 1.288 \times 10^8 \,
     (T_\mathrm{s6}^4/g_{14})^{0.455}~~\mathrm{K},
\label{therm-GPE}
\eeq
where $T_\mathrm{s6}=\Ts/10^6$~K.
An analytical derivation of a similar expression
was given by \citet{VP01}. A more accurate but less simple
fit was constructed by \citet{PCY97}.

The \TbTs{} relation is mainly regulated by the thermal
conductivity in the ``sensitivity strip'' \citep{GPE83} that
plays the role of a ``bottleneck'' for the heat leakage. Its
position lies around the line where
$\kappa_\mathrm{r}=\kappa_e$ (as a rule, around
$\rho\sim10^5$\,--\,$10^7$ \gcc{} for $B=0$) and depends on
the stellar structure, the boundary temperature
$T_\mathrm{b}$, the magnetic field $\bm{B}$ in the vicinity of
the given surface point, and the chemical composition of the
envelope. Since the magnetic field hampers heat transport
across $\bm{B}$, the depth of the sensitivity strip can be
different at different places of a star with a strong
magnetic field: it lies deeper at the places where the
magnetic field lines are parallel to the surface
\citep{VP01}.

The blanketing envelopes are more transparent to the heat
flux, if they are composed of light chemical elements. This
effect was studied in detail by \citet{PCY97} for
non-magnetic envelopes and by \citet{Potekhin-ea03} for strongly
magnetized envelopes. The effect is related to  the $Z$-dependence
of the collision frequencies  $\nu_{e\mathrm{i}}$. The higher
is $Z$, the larger is $\nu_{e\mathrm{i}}$ and the lower is the
conductivity. A temperature variation by a factor of 30 can
change the thermal conductivity of iron plasma less 
than altering the chemical composition from  Fe to He
at a fixed $T$. This effect has important consequences  for
the relationship between the surface and internal
temperatures  of neutron stars.
For example, combined effects of strong magnetic fields and
light-element composition simplify the interpretation of
magnetars: these effects allow one to interpret
observations assuming less extreme (therefore, more
realistic) heating in the crust \citep{Kaminker-ea09,PonsMG09,vigano13}.

The envelope is thin (its depth $z_\mathrm{b}\sim
100$~m, if $\rhob=10^{10}$ \gcc) and contains a tiny
fraction of the neutron-star mass ($\sim 10^{-7}$, if
$\rhob=10^{10}$ \gcc). Therefore one can neglect the
variation of the gravitational acceleration in this layer.
Neglecting also the non-uniformity of the energy flux through
the envelope due to the neutrino emission (which is small,
if the neutron star is not too hot, as we discuss below) and the 
variation of the temperature $\Ts$ over the surface (which
varies on larger length scales than  $z_\mathrm{b}$), 
one can obtain, instead of \req{Tbalance},
the much simpler \emph{thermal structure
equation} \citep{GPE83,VanRiper88}
\beq
   \frac{{\rm d}\ln T}{{\rm d}\ln P} =
   \frac{3}{ 16}\,\frac{P\opac}{ g}\,\frac{\Ts^4}{ T^4},
\label{th-str}
\eeq
where$\opac$ is the total opacity,
related to the conductivity $\kappa$ via \req{opacity}.

The assumption of a constant flux, however, breaks down for
magnetars, most of which have atypically high effective
temperatures. In this case one should solve the complete set
of equations, taking neutrino emission and heat sources into
account. The neutrino emission from the crust limits the
effective surface temperature of a cooling neutron star
(\citealp{PCY07}; cf.{} Fig.~\ref{fig:TbTs} below).  This
very same effect is what limits the maximum flux in the few
days of a magnetar outburst \citep{PonsRea12}. In addition,
for magnetars one must go beyond the plane-parallel
approximation (see Sect.~\ref{sect:2D})

% -----------------------------------
\subsection{The effects of strong magnetic fields}
\label{sect:env-mag}

As seen from Eqs.~(\ref{omegatau}) and (\ref{tau}), the Hall
magnetization parameter is large in the outer neutron-star
envelope at $B\gtrsim10^{11}$~G. Moreover, the magnetic
field can be strongly quantizing in the outermost part of
the envelope. In this case the magnetic field can greatly
affect the heat conduction  and the thermal structure.

\begin{figure}
\includegraphics[width=.5\textwidth]{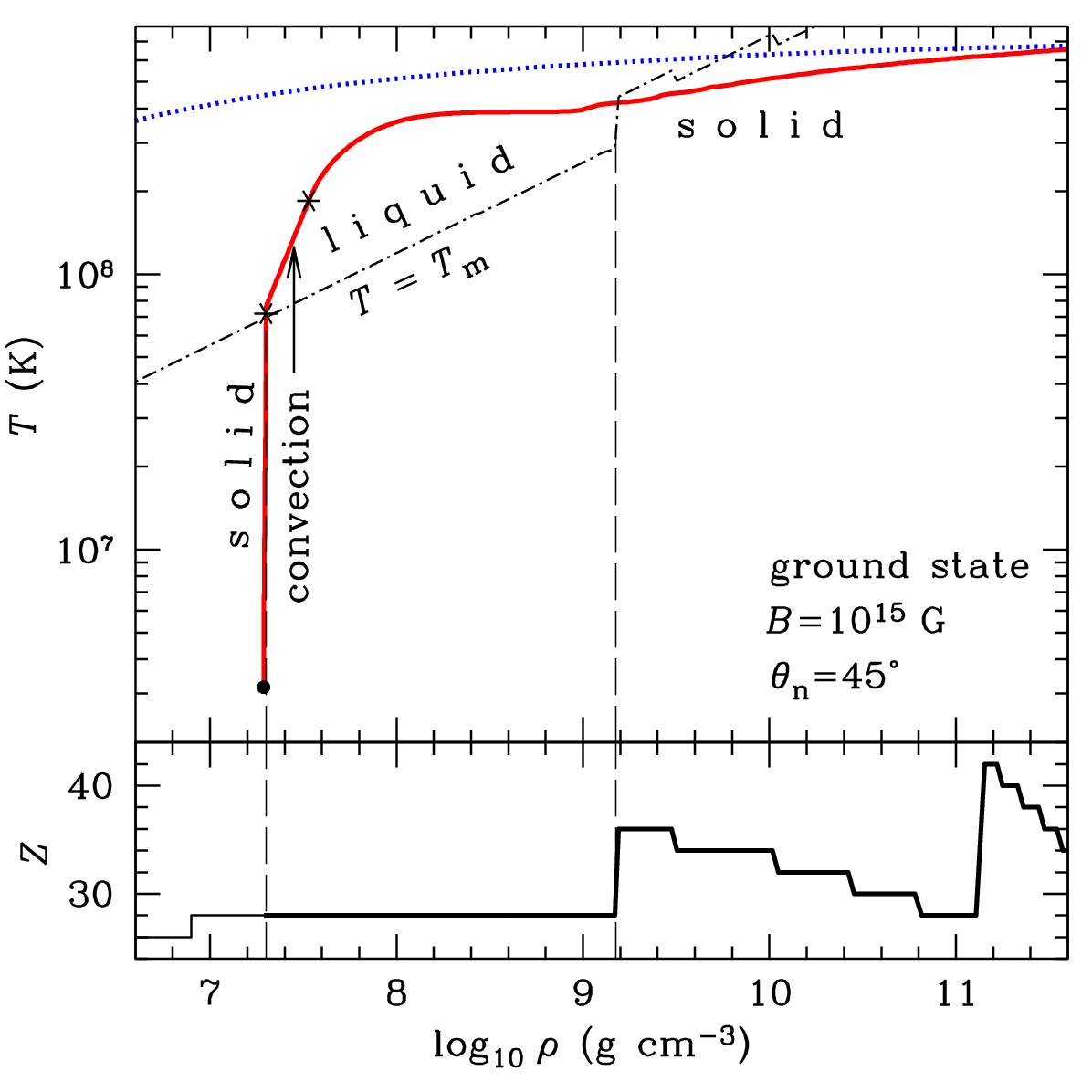}
\includegraphics[width=.5\textwidth]{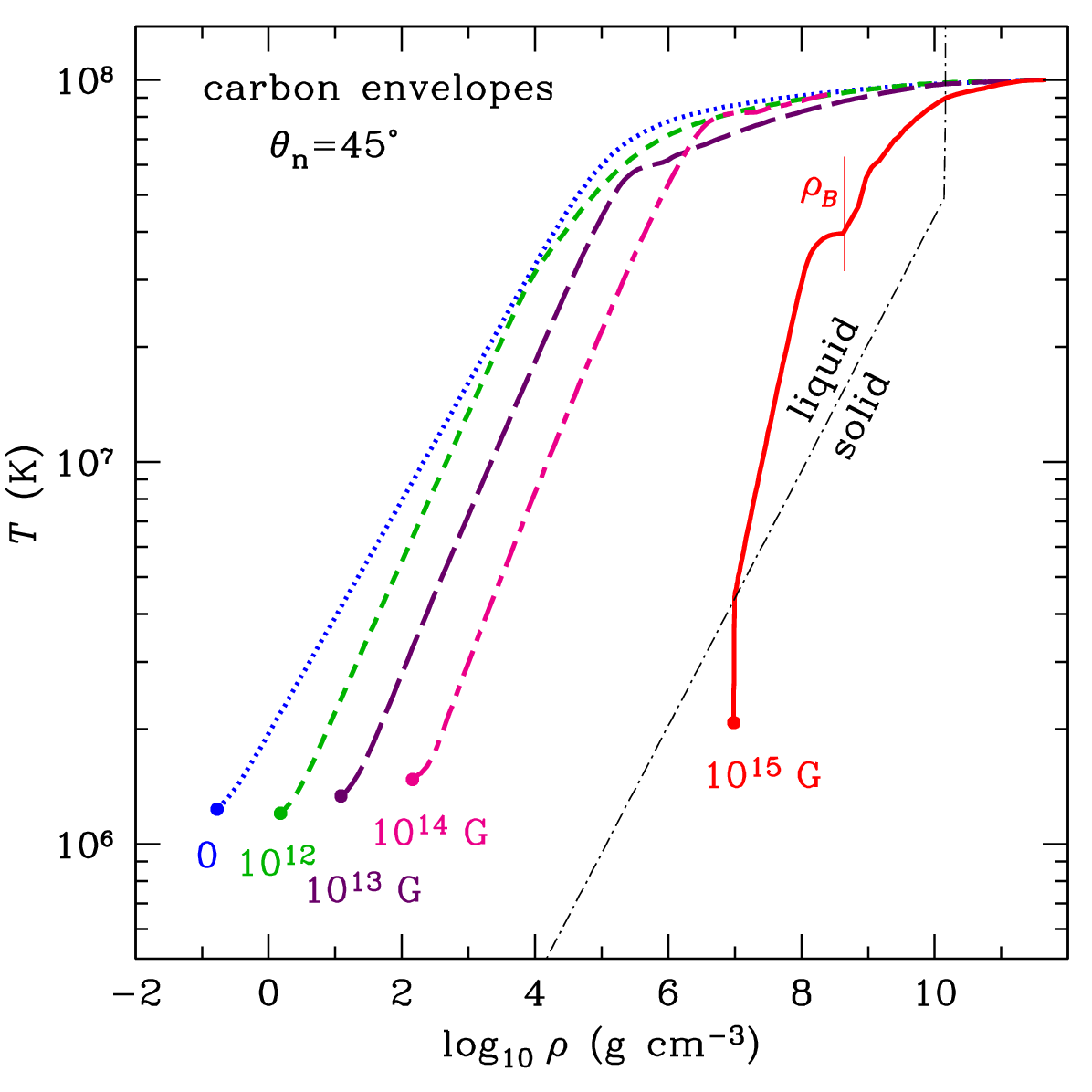}
\caption{Thermal structure of blanketing envelopes with
different magnetic fields. Left panel: temperature profile
(solid line in the left top panel) for an envelope with
ground-state composition, with $Z$ values shown in the
bottom left panel, for a neutron star with surface gravity
$g=1.6\times10^{14}$ cm s$^{-2}$, magnetic field
$B=10^{15}$~G directed at $45^\circ$ to the surface, and
internal temperature $T_\mathrm{b}=6.7\times10^8$~K, which
yields the effective surface temperature
$\Ts=3.16\times10^6$~K. The dot-dashed line is the melting
line. The asterisks confine the part of the profile where
heat is carried by convection. For comparison, the
non-magnetic profile with the same $T_\mathrm{b}$ is shown by
dotted line. Right panel: temperature profiles for carbon
blanketing envelopes for a neutron star with
$g=1.4\times10^{14}$ cm s$^{-2}$,  $T_\mathrm{b}=10^8$~K,
and magnetic fields $B=0$ (dotted line), $10^{12}$~G (short
dashes), $10^{13}$~G (long dashes), $10^{14}$~G (alternating
short and long dashes), and $10^{15}$~G (solid line). The
vertical segment of the dot-dashed melting line corresponds
to the maximum density for carbon, according to the
thermonuclear stability criterion of \citet{PC12}, where carbon
gives way to heavier chemical elements, which form a
crystal. The heavy dots mark the position of the radiative
surface, where $T$ equals the
effective surface temperature $\Ts$.
}
\label{fig:profiles}
\end{figure}

Figure~\ref{fig:profiles} shows examples of the temperature
profiles in the envelopes. The left panel is a recast of
Fig.~8 from \citet{PC13}. Here we show a profile of an
ultra-magnetized neutron star, with $B=10^{15}$~G, and with
relatively high surface temperature, $\log_{10}\Ts{} (K)
=6.5$,  which is similar to the values evaluated for some
magnetars. In this case, thermal photons are radiated from a
solid surface, with high mass density $\rho=2\times10^7$
\gcc{} just below the surface. The temperature quickly
grows  at the solid surface and reaches the melting point at
the depth $z\approx7$~cm. Thus, at the given conditions, the
liquid ocean of a magnetar turns out to be covered by a thin
layer of ``ice'' (solid substance).  We treat the solid
crust as immobile, but the liquid layer below the ``ice'' is
convective up to the depth $z\sim1$~m.  The change of the
heat-transport mechanism from conduction to convection 
causes the break of the temperature profile at the melting
point in \fig{fig:profiles}. We underline that this treatment is only an
approximation. In reality, the superadiabatic growth of
temperature can lead to a hydrostatic instability of the
shell of ``ice'' and eventually to its cracking and
fragmentation into turning-up ``ice floes''. \citet{PC13}
speculated that such events may result in variations of
thermal luminosity of magnetars. The temperature profile
flattens with density increase, and the Coulomb plasma
freezes again at the interface between the layers of
$^{66}$Ni and $^{86}$Kr at $\rho=1.5\times10^9$ \gcc. 

For comparison, we also show the thermal profile without the
magnetic field. It is smooth. There is
neither magnetic condensation nor convection. In this case,
the spectrum is formed in the gaseous atmosphere at much
lower density beyond the
frame of the left panel. 

In the right panel of Fig.~\ref{fig:profiles} we compare
temperature profiles for a neutron star with internal
temperature $10^8$~K and heat blanketing envelopes made of
carbon, endowed with different magnetic fields. For the
field strengths up to $10^{14}$~G, the radiation is formed
in the gaseous atmosphere, whose density gradually becomes
larger with increasing magnetic fields, due to the reduction
of the effective opacities discussed in
Sect.~\ref{sect:radiopa}. The temperature profiles are
rather smooth. The blanketing envelopes are liquid at this
temperature. At the largest field strength $B=10^{15}$~G,
however, the situation is qualitatively different. As well
as in the case of the hotter ultra-magnetized ground-state
envelope in the left panel, the heat is radiated from the
condensed solid surface. Below the surface, at density
$10^7$ \gcc, the temperature quickly grows, which
causes melting of the Coulomb crystal with formation of a
Coulomb liquid beneath the solid surface. With further
density increase, the profile suffers a break at
$\rho_B\approx4.5\times10^8$ \gcc{} [\req{rho_B}], where the
electrons start to populate the first excited Landau level,
which is associated with the peak of the thermal
conductivity around $\rho_B$ (cf.{} Fig.~\ref{fig:cond}).

As we have seen in Sect.~\ref{sect:condmag}, the conduction
is strongly anisotropic in these conditions.
Therefore the effective local surface temperature $\Ts$ is
non-uniform and depends on the magnetic field geometry.
Figure~\ref{fig:TbTs} shows examples of the relations
between $\Ts$ and $T_\mathrm{b}$ deep in the crust for the
magnetic fields $B=10^{12}$~G and $10^{15}$~G perpendicular
and parallel to the radial direction.  The relations
obtained in the 1D approximation \citep{PCY07} with and
without allowance for neutrino emission are plotted by the
solid and dotted lines, respectively. We see that at
$T_\mathrm{b}\lesssim10^8$~K the neutrino emission does not
affect $\Ts$. At higher $T_\mathrm{b}\gtrsim10^9$~K, in
contrast, this emission is crucial: if $Q_\nu=0$, then $\Ts$
continues to grow up with increasing $T_\mathrm{b}$, whereas
with realistic $Q_\nu$ the surface temperature tends to  a
constant limit, which depends on $\bm{B}$. In most cases
this limit is reached when $T_\mathrm{b}\sim10^9$~K.

Since the distribution of $\Ts$ over the neutron-star surface is
non-uniform in strong magnetic fields, it is convenient to
introduce the overall effective temperature of the star,
$\Teff$, defined by
\beq
   4 \pi \sSB R^2 \Teff^4 =
   L_\mathrm{ph}= 
 \int F_\mathrm{ph} \,\mathrm{d}\Sigma =
\sSB \int \Ts^4 \,\mathrm{d}\Sigma ,
\label{L}
\eeq
where $F_\mathrm{ph}$ is the local flux density and
 $\mathrm{d}\Sigma$ is the surface element.
The quantities $\Ts$, $\Teff$, and $L_\mathrm{ph}$
refer to a local reference frame at the neutron-star surface.
The redshifted (``apparent'') quantities
as detected by a distant observer are \citep{Thorne77}:
\beq
R^\infty = R / \sqrt{1-r_g/R},
\qquad
\Teff^\infty = \Teff \, \sqrt{1-r_g/R},
\qquad
L_\mathrm{ph}^\infty = (1-r_g/R)\, L_\mathrm{ph}.
\eeq

The effects of quantizing magnetic fields on the thermal
structure of neutron-star envelopes were first studied by
\citet{Hernquist85} and somewhat later by \citet{VanRiper88}
and \citet{Schaaf90}, using the 1D approximation.
\citet{VanRiper88} considered a neutron star with a constant
radial magnetic field. In this model, the quantum
enhancement of conductivity at $\rho$ near $\rho_B$, seen in
Fig.~\ref{fig:cond}, results
in an overall enhancement of the neutron-star photon
luminosity $L_\mathrm{ph}$ at a fixed $T_\mathrm{b}$.
Consequently, \citet{VanRiper91} found a strong effect of
the magnetic field $B\sim10^{13}$~G on the neutron-star
cooling. However, \citet{ShibanovYakovlev96} showed that, for
the dipole field distribution, the effects of suppression of
the heat conduction across $\bm{B}$ at the loci of nearly
tangential field can compensate or even overpower the effect
of the conductivity increase near the   normal direction of
the field lines. This conclusion confirmed the earlier conjectures  of
\citet{Hernquist85} and \citet{Schaaf90}. In the 2000s,
detailed studies of the \TbTs{} relation in strong magnetic
fields were performed for iron envelopes \citep{PY01} and
accreted envelopes composed of light elements
\citep{Potekhin-ea03}, as well as for the large-scale
(dipole) and small-scale (stochastic) surface magnetic
fields \citep{PUC05}. These studies confirmed the conclusions
of \citet{ShibanovYakovlev96}, but showed that in
superstrong fields $B\gtrsim10^{14}$~G the quantum
enhancement of the conductivity and the corresponding
increase of $\Ts$ at the places where
$\bm{B}$ is nearly radial overpowers the
decrease in the regions of nearly tangential field lines, so
that $\Teff$ at a given $T_\mathrm{b}$ increases. However,
this may not be the case in the configurations where
the field is nearly tangential over a significant portion of
the stellar surface as, e.g., in the case of a superstrong
toroidal field \citep{PerezAMP06,PageGK07}.

% -----------------------------------
\subsection{Non-radial heat transport}
\label{sect:2D}

% ------------------------------------
\begin{figure*}
\includegraphics[height=.53\textwidth]{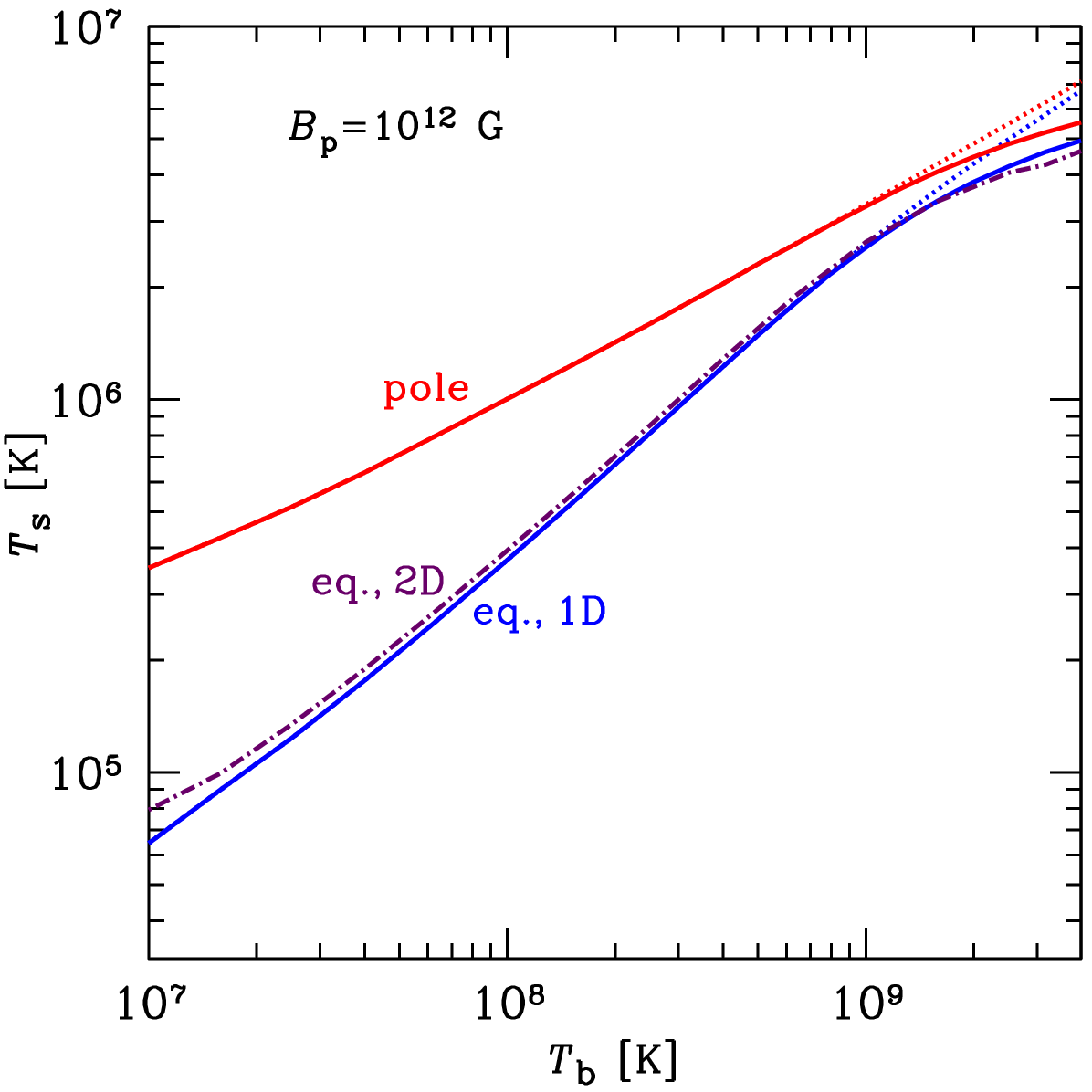}
\includegraphics[height=.53\textwidth]{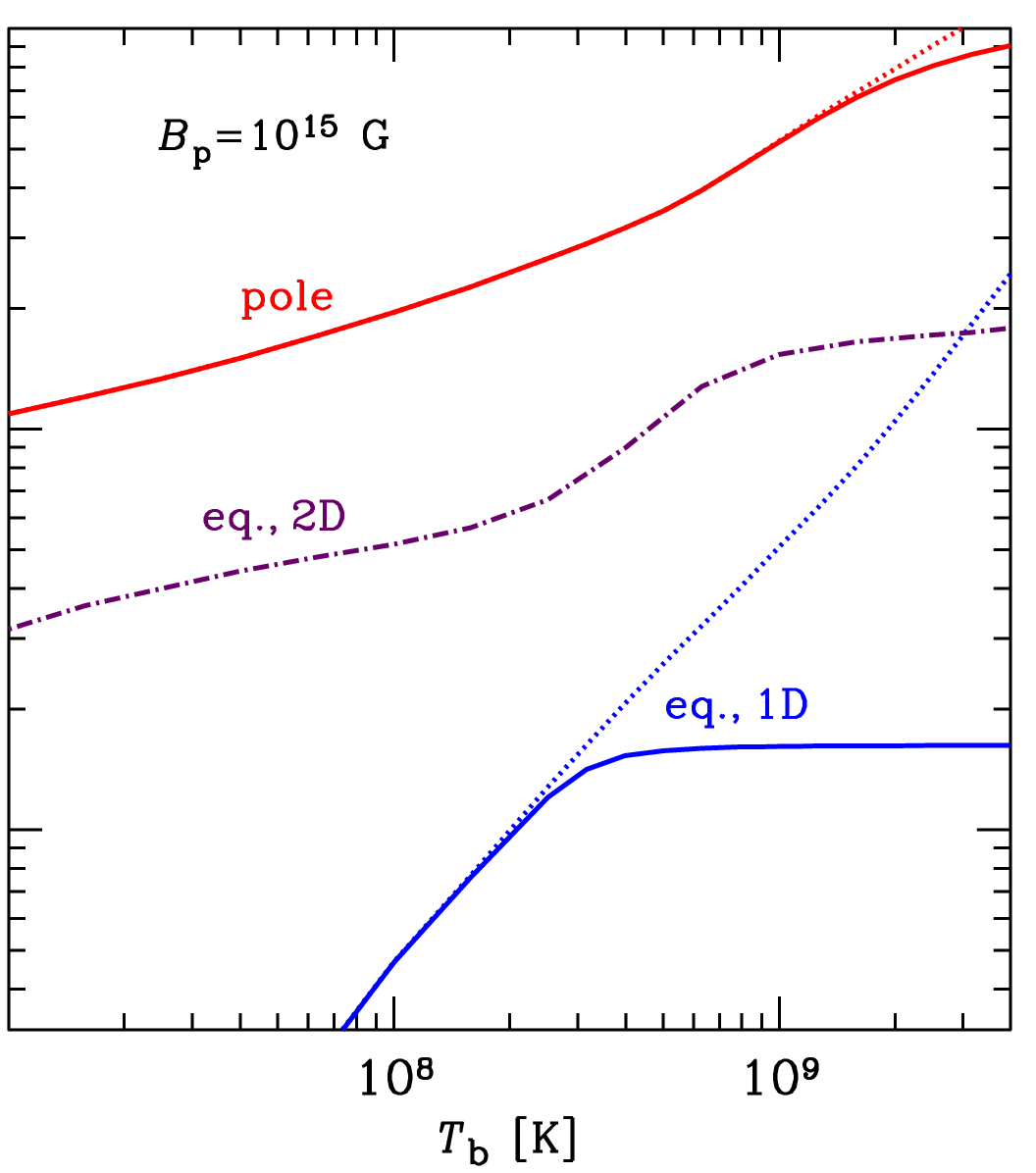}
\caption{Local effective surface temperature $\Ts$ as
function of the temperature $T_\mathrm{b}$ at the bottom of
a non-accreted heat blanketing envelope with $\rhob=10^{10}$
\gcc{} for a neutron star with mass $M=1.4\,M_\odot$, radius
$R=12.6$ km, and the dipole magnetic field with polar
strength $B_\mathrm{p}=10^{12}$~G (left panel) and
$10^{15}$~G (right panel). Solid lines -- 1D calculation
with allowance for neutrino emission from the crust, dotted
lines -- neutrino emission is neglected. The upper dotted or
solid curve shows $\Ts$ at the magnetic pole,  and the lower
curve curve shows $\Ts$ at the equator.  The dot-dashed
curve shows the result of a full 2D calculation for $\Ts$ at
the magnetic equator.
}
\label{fig:TbTs}
\end{figure*}
% ------------------------------------

As we mentioned in Sect.~\ref{sect:blanket}, in the case
where $\bm{B}$ is nearly parallel to the surface, the 1D
approximation fails, because the heat is transported along
the field lines from the hotter surface regions outside the
considered patch of the surface.  Therefore the 1D
approximation overestimates the heat blanketing effect in
regions with  nearly tangential magnetic fields. For a
dipole field geometry it is an equatorial region , whose
width can be estimated as $\sim10$\% of the radius
\citep{PCY07}. Since these regions are also the coldest
ones, their contribution to the total flux is negligible.
Then the 1D approximation well reproduces the integrated
observed flux. However, it is not the case for magnetars,
which may have a complex field geometry. A 2D treatment
shows that the 1D approach is reliable in the regions where
magnetic field lines make a substantial angle to the surface
\citep{Kaminker-ea12, Kaminker-ea14}, but it predicts too
low surface temperatures when the tangential magnetic field
dominates (see Section 3 in \citealp{PonsMG09}). Therefore
for magnetars one must go beyond the 1D approximation.
Complex field configurations which lack cylindrical symmetry
may require the full 3D treatment, which has not been done
yet.

In Fig.~\ref{fig:TbTs} we show results of 2D calculations
in the dipole field geometry, compared with the 1D results. In this case, we see a
substantial increase of $\Ts$ at the magnetic equator. This
effect is especially pronounced for the superstrong field on
the right panel. In Appendix~\ref{sect:TbTs} we give an
analytical approximation to the \TbTs{} relation in the case
of a strong magnetic field, including the effects of
neutrino emission from the crust.

%%%%%%%%%%%%%%%%%%%%%%%%%%%%%%%%%%%%%
\section{Thermal evolution of neutron stars}
\label{sect:cooling}
%%%%%%%%%%%%%%%%%%%%%%%%%%%%%%%%%%%%%%%%%%%%%
\subsection{Cooling scenarios}
\label{sect:scenarios}

Several tens of seconds after birth, the
protoneutron-neutron star has lost its excess lepton 
content, it has finished its residual contraction and
becomes transparent to neutrino emission
\citep{Burrows86,Pons99,Roberts12}.  Soon after that, the
temperature distribution in the highly conductive stellar
core reaches equilibrium, which is preserved thereafter
throughout the star lifetime (except during short periods
after catastrophic phase transitions in the core postulated
by certain hypothetical models). 

In the initial cooling stages, the stellar crust is hotter than the
core, which is rapidly cooled down by the copious neutrino
emission. The cooling wave reaches the surface within
10\,--100 years; thereafter, the star cools down in the
quasistationary regime. Since all currently observed neutron
stars are at least several centuries old, they should be in
the state of quasistationary cooling, except during 
transient events with significant energy release in the
crust or the ocean discussed below.

Cooling in the quasistationary regime goes through two major
stages. The first, \emph{neutrino cooling} stage lasts
$\sim10^5$ years. During this period, the core cools mostly
via neutrino emission. The second, \emph{photon cooling}
stage begins when the low temperature of the core makes the neutrino energy
losses smaller than the losses due to electromagnetic
radiation from the surface (see, e.g.,
\citealp{YakovlevPethick04}, and references therein).
This occurs at the age of $\approx 10^5$ years, depending on the particular
stellar model and local conditions.

A theoretical cooling curve of an isolated neutron star, which shows the photon
luminosity of the star $L_\mathrm{ph}$ or its temperature as
a function of age $t$, depends on the stellar mass $M$, on
the model of superdense matter in the core, which in
particular, determines the intensity of neutrino emission
and the EoS (and hence the stellar radius $R$), and on the
properties of the envelopes. The latter include the thermal
conductivity, which determines $L_\mathrm{ph}$ at a given
internal stellar temperature, the neutrino luminosity
$Q_\nu$ in the stellar crust,  and the intensity of heating
sources $H$. For highly magnetized neutron stars, the
cooling curve also depends on  the magnetic field $\bm{B}$
(on both its strength and configuration), since it affects
the microphysics (conductivities, EoS, specific heat, etc.).
Therefore, in general, the thermal evolution equations
(\ref{Tbalance}) should be supplemented by the equations
that describe evolution of the magnetic field and electric
currents in the star, which leads to the thermomagnetic
evolution scenarios (see the review by
\citealp{Mereghetti-ea15} and references therein).

By comparing theoretical cooling curves with the observed
$L_\mathrm{ph}$ and $t$ of isolated neutron stars, one can
eventually place bounds on the theoretical models of
superdense matter. At contrast, most neutron stars in binary
systems have an additional source of energy (accretion) and
an additional source  of X-ray radiation (accretion disk),
often much more powerful than the surface thermal emission.
For this reason, they cannot be used to test cooling models.

The theoretical cooling scenarios are currently divided into
two main classes: ``minimal cooling'' and ``enhanced
cooling''. The enhanced cooling implies fast neutrino
emission processes, such as Durca reactions
(Sect.~\ref{sect:nu-core}), whereas the minimal cooling does
not include such processes, but may be enhanced at the epoch
of the onset of the baryon superfluidity via the PBF
neutrino emission mechanism (Sect.~\ref{sect:SFnu}), which
helps to explain the variety of the observed surface
temperatures of cooling neutron stars
\citep{Gusakov-ea04,Pageetal2004}. A spectacular example is
the neutron star CXO J232327.9+584842 in the Cassiopeia A
supernova remnant, dubbed Cas A NS, which shows an
unexpectedly appreciable temperature decline during several
years \citep{HeinkeHo10,Elshamouty-ea13} (but see
\citealp{Posselt-ea13} for tentative alternative
interpretations of the observations). This decline can be
comfortably explained by the PBF emission
(\citealp{Page-ea11,Shternin-ea11}; see also \citealp{Ho-ea15}
for a recent analysis including modern observational data).

%%%%%%%%%%%%%%%%%%%%%%%%%%%%%%%%%%%%%%%%%%%%%
\subsection{Heating mechanisms}
\label{sect:heating}

In the course of their evolution,
some regions of neutron stars may not only cool but also heat up by 
different mechanisms. For instance,
the polar cap of a pulsar can be heated by a stream of
electrons or positrons moving along open field lines from
the magnetosphere. The temperature of ``hot spots'' produced
by this additional heat deposited onto the stellar surface may be much higher
than the average temperature of the star \citep[e.g.,][]{GreensteinHartke}.
Non-uniform heating processes occur also during accretion episodes
\cite[e.g.,][]{InogamovSunyaev10}. The hot polar caps emit
much more intense X-rays than the remaining surface; as a
result, such neutron stars become X-ray pulsars. Pulsed
X-ray radiation is also observed from thermonuclear
explosions of accreted matter at the surface of a rotating
neutron star (see, e.g., review by
\citealp{StrohmayerBildsten06}).

On the other hand, a neutron star may also be heated from
inside, for example due to dissipation of a strong magnetic
field \citep[e.g.,][]{MUK98,UrpinKonenkov08,PonsMG09}. It has been
suggested that dissipation of superstrong magnetic fields
may be responsible for the high effective temperatures of
magnetars (\citealp{Thompson01}; see 
\citealp{Mereghetti-ea15}, for a review). Noticeable liberation of
energy in the crust may also occur during starquakes
\citep{HaenselDP90,Franco-ea00,Perna11}. 
Two heating processes related to the  secular spin-down of the star have 
also been proposed: vortex creep, the dissipative motion of superfluid vortices 
through the neutron star crust
\citep{Alpar-ea84}, and rotochemical heating, the energy
released by non-equilibrium beta decays due to the slow
contraction of the neutron star as its centrifugal force
decreases \citep{Reisenegger95}. 
\citet{GonzalezReisenegger10} performed a comparative study
of several heating mechanisms and found that the
rotochemical heating and vortex creep can be most important
for classical and millisecond pulsars. Both processes,
albeit model-dependent, can keep millisecond pulsars at a
surface temperature $\Teff\sim10^5$~K.

Another class of neutron stars undergoing heating episodes are
quasipermanent transients, i.e., those soft X-ray transients
(SXTs) whose active and quiescent periods last a few years
or longer. During high-state accretion episodes, compression
of the crust under the weight of newly accreted matter
results in deep crustal heating, driven by exothermic
nuclear transformations \citep{HZ90,HZ08}. For a given
neutron star model, one can calculate the
heating curve, that is the dependence of the equilibrium
accretion-free $\Teff$ on the accretion rate averaged over a
large period of time. There is a close correspondence
between the theory of thermal states of transiently
accreting neutron stars and the theory of neutron star
cooling \citep{YakLH03}. Comparing the heating curves with a
measured equilibrium $\Teff$ value, one can constrain
parameters connected to properties of dense
matter \citep{Yak-ea04,LevenfishHaensel07,Ho11}.
\citet{WijnandsDP13} discuss prospects of application of
such analysis to various classes of X-ray transients. The
SXTs that have recently turned into quiescence allow one to
probe the state of the neutron-star crust with the observed decline of
$\Teff$. \citet{BBR98} suggested that during this decline
the radiation is fed by the heat that was deposited in the
crust in the preceding active period. Such cooling is
independent of the details of the star structure and 
composition and therefore its analysis directly yields
information on the physics of the crust. Observations of
several sources can be interpreted in terms of this
hypothesis and yield constraints to the heat conductivity in
the neutron-star crust, as, e.g., for KS 1731--260
\citep{Shternin-ea07,BC09}, XTE J1701--462
\citep{Fridriksson-ea11,PageReddy13}, EXO 0748--676
\citep{TurlioneAP13}. The conductivity proves to be rather
high, which means that the crust has a regular crystalline
(not amorphous) structure. On the other hand, there are
similar objects which display variations of thermal flux
that do not conform to the thermal-relaxation scenario,
which may be caused by a residual slow accretion on the
neutron star in quiescence
\citep{Rutledge-ea02a,CotiZelati-ea14,TurlioneAP13}.

%%%%%%%%%%%%%%%%%%%%%%%%%%%%%%%%%%%%%%%%%%%%%
\subsection{Thermal luminosities of isolated neutron stars}

The inferred effective temperature depends on the choice
of the emission model (blackbody vs. atmosphere models, composition,
condensed surface, etc. -- see Paper~I), which typically results in
variation of $\Teff$ by a factor $\approx 2-3$, and it has even larger
theoretical uncertainties in the case of strong magnetic fields.  In addition, photoelectric
absorption in the interstellar medium further constitutes a source of
error in temperature measurements, since the value of the hydrogen
column density $N_\mathrm{H}$ is correlated to the temperature value obtained
in spectral fits. Different choices for the absorption model and the
metal abundances can also yield different results for the temperature.
Last, in the case of data with few photons and/or strong absorption features,
the temperature is poorly constrained by the fit, adding a large
statistical error to the systematic one.

%%%%%%%%%%%%%%%%% TABLE 3 %%%%%%%%%%%%%%%%%%%%%%%%%
\begin{table}
\caption{Cooling neutron stars. $t_\mathrm{c}$ is the
  characteristic age, $t_\mathrm{k}$ is the kinematic age,
  and $f_\mathrm{X}$ is the unabsorbed flux in the
  1--10\,keV band. The range of luminosities $L$ includes
  both statistical and distance errors; for strongly
  absorbed sources (i.e., most magnetars) a minimum
  arbitrary factor of $50\%$ uncertainty is assumed to
  account for systematical model-dependent uncertainties.
  Data have been taken from \cite{vigano13} (see references
  therein and the online catalog in
  \texttt{http://www.neutronstarcooling.info}.) }
\begin{center}
\begin{tabular}{l c c c c c }
\hline
\hline
Source & $\log_{10} (t_\mathrm{c})$ & $\log_{10}
(t_\mathrm{k})$ & $\log_{10}(f_\mathrm{X})$ & $d$  &
$\log_{10}(L)$  \\[0pt]
& [yr] & [yr] & [erg cm$^{-2}$ s$^{-1}$] & [kpc] & [erg/s]
\\[.5ex]
\hline
 CXOU J185238.6+004020 & 8.3 & 3.7--3.9 & $ -12.3$ &        
          7.1 &   33.5--33.7 \rule{0pt}{3ex} \\
        1E 1207.4--5209          &  8.5 & 3.4-4.3 & $-11.8$ &
2.1$^{+ 1.8}_{- 0.8}$ &    33.0--34.0 \\
         RX J0822--4300          &  8.3 & 3.5--3.6 & $-11.3$
&          2.2$\pm 0.3$ &     33.5--33.7 \\
  CXO J232327.9+584842  &   --  & 2.5 & $     -11.8$ &
3.4$^{+ 0.3}_{- 0.1}$ &   33.4--33.6 \\
        PSR J0538+2817       & 5.8 & $\approx$ 4.6 & $-12.1$
&          1.3$\pm 0.2$ &    32.7--32.9 \\
          PSR B1055--52         &  5.7 & -- & $-13.4$ &      
  0.73$\pm0.15$ &    32.2--32.6 \\
        PSR J0633+1746       &  5.5 & -- & $-12.5$
&0.25$^{+0.22}_{-0.08}$ &  31.6--32.5 \\
          PSR B1706--44         &  4.2  & -- & $-12.1$ &
2.6$^{+ 0.5}_{- 0.6}$ &    31.7--32.1 \\
          PSR B0833--45         &  4.1 & 3.7-4.2 & $-10.5$ & 
       0.28$\pm0.02$ &   32.1--32.3 \\
          PSR B0656+14        &  5.0  & $\approx$ 4.9 &
$-12.6$ &         0.28$\pm0.03$ &     32.7--32.8 \\
          PSR B2334+61        &  4.6  & $\approx$ 4.0 &
$-14.0$ & 3.1$^{+ 0.2}_{- 2.4}$ &     30.7--32.1 \\
        PSR J1740+1000       &  3.1  & -- & $ -13.8$ &      
            1.4 &   32.1--32.2 \\
        PSR J1741--2054        & 5.6 & -- & $-12.5$ &   0.8 &
30.4--31.4 \\
        PSR J0726--2612       &  5.3  & -- & $-14.0$ &       
           1.0 &     32.1--32.5 \\
        PSR J1119--6127       &  3.2 & 3.6--3.9 & $-13.0$ &  
       8.4$\pm 0.4$ &    33.1--33.4 \\
        PSR J1819--1458       &  5.1 & -- & $-12.6$ &        
          3.6 &    33.6--33.9 \\
        PSR J1718--3718       & 4.5 & -- & $ -13.2$ & 4.5$^{+
5.5}_{- 0.0}$ &    32.8--33.5 \\
       RX J0420.0--5022        & 6.3 & -- & $ -17.8$ &       
          0.34 &    30.9--31.0 \\
       RX J1856.5--3754       & 6.6 & 5.5-5.7 & $ -14.4$ &   
     0.12$\pm0.01$ &    31.5--31.7 \\
       RX J2143.0+0654      &  6.6 & -- & $-13.1$ &         
        0.43 &    31.8--31.9 \\
       RX J0720.4--3125       & 6.3 & 5.8-6.0 & $ -13.3$
&0.29$^{+0.03}_{-0.02}$ &    32.2--32.4 \\
       RX J0806.4--4123       & 6.5 & -- & $ -13.4$ &        
         0.25 &  31.2--31.4 \\
       RX J1308.6+2127      & 6.2 & 5.9-6.1 & $ -12.1$ &    
             0.50 &   32.1--32.2 \\
       RX J1605.3+3249      &  -- & $ 5.7-6.7 $ & $-13.0$ & 
      0.35$\pm 0.05$ &   30.9--31.0 \\
           1E 2259+586         & 5.4 & 4.0-4.3 & $ -10.3$ & 
        3.2$\pm 0.2$ &    35.0--35.4 \\
           4U 0142+614         & 4.8 & -- & $  -9.8$ &      
   3.6$\pm 0.5$ &    35.4--35.8 \\
  CXO J164710.2--455216 & 5.2 & -- & $ -12.2$ & 4.0$^{+
1.5}_{- 1.0}$ &   33.1--33.6 \\
         XTE J1810--197        & 4.1 & -- & $ -11.7$ &       
  3.6$\pm 0.5$ &   34.0--34.4 \\
        1E 1547.0--5408        &  2.8 & -- & $-11.5$ &       
  4.5$\pm 0.5$ &     34.3--34.7 \\
        1E 1048.1--5937        &  3.7 & -- & $-10.8$ &       
  2.7$\pm 1.0$ &    33.8--34.5 \\
    CXOU J010043.1--721  &  3.8 & -- & $-12.5$ &        
60.6$\pm 3.8$ &   35.2--35.5 \\
 1RXS J170849.0--400910 & 4.0 & -- & $-10.4$ &         
3.8$\pm 0.5$ &  34.8--35.1 \\
 CXOU J171405.7--381031 & 3.0 & $\approx 3.7$ & $-11.4$ &    
    13.2$\pm 0.2$ &    34.9--35.2 \\
           1E 1841--045           & 3.7 & 2.7--3.0 & $-10.4$
& 9.6$^{+ 0.6}_{- 1.4}$ &    35.2--35.5 \\
         SGR 0501+4516       & 4.2 & $\approx 4 $ & $ -11.3$
& 1.5$^{+ 1.0}_{- 0.5}$ &     33.2--34.0 \\
           SGR 1627--41         &  3.3 & $\approx 3.7 $ &
$-11.6$ &         11.0$\pm 0.2$ &   34.4--34.8 \\
           SGR 0526--66         & 3.5 & $\approx 3.7 $ & $
-12.0$ &         49.7$\pm 1.5$ &    35.4--35.8 \\
           SGR 1900+14        &  3.0 & 3.6--3.9 & $-11.1$ & 
       12.5$\pm 1.7$ &    35.0--35.4 \\
           SGR 1806--20        &  2.6 & 2.8--3.0 & $-10.6$
&13.0$^{+ 4.0}_{- 3.0}$ &     35.1--35.5 \\
         SGR 0418+5729{}* & 7.6 & -- & $-14.0$ &            
      2.0 &     30.7--31.1 \\
    Swift J1822.3--1606{}* & 6.2 & -- & $-11.5$ &         
1.6$\pm 0.3$ &    32.9--33.2 \\[.7ex]
\hline
\hline
\end{tabular}
\end{center}
\textit{Notes.}{~}*{\footnotesize The source has been recently discovered in
outburst and it could have not yet reached the quiescence level.}
\label{tab:cooling}
\end{table}
%%%%%%%%%%%%%%%%%%%%%%%%%%%%%%%%%%%%%%%%%%%%%%%%%%%%%%%%%%%%%%

Because of these uncertainties, the luminosity may often be
a better choice to compare data and theoretical cooling
models.  Since it is an integrated quantity, it averages
effects of anisotropy and the choice of spectral model. The
main uncertainty on the luminosity is often due to the
poorly known distance to the source. In many cases, the
distance is known within an error of a few, resulting in up
to one order of magnitude of uncertainty in the luminosity.
In addition, the interstellar absorption acts predominantly
in the energy band in which most of the middle age neutron
stars emit ($E\lesssim 1$ keV). Clearly, hottest (magnetars)
or closest (XINSs) sources are easier to detect (see
\citealp{vigano13} and Paper~I). Similarly to the case of
the temperature, the choice of different models of
absorption and chemical abundances can yield additional
systematic errors on the luminosity. However, for the worst
cases, the relative error is about $30\%$, making it usually
a secondary source of error compared with the distance.

\begin{figure}
 \centering
\includegraphics[width=.85\textwidth]{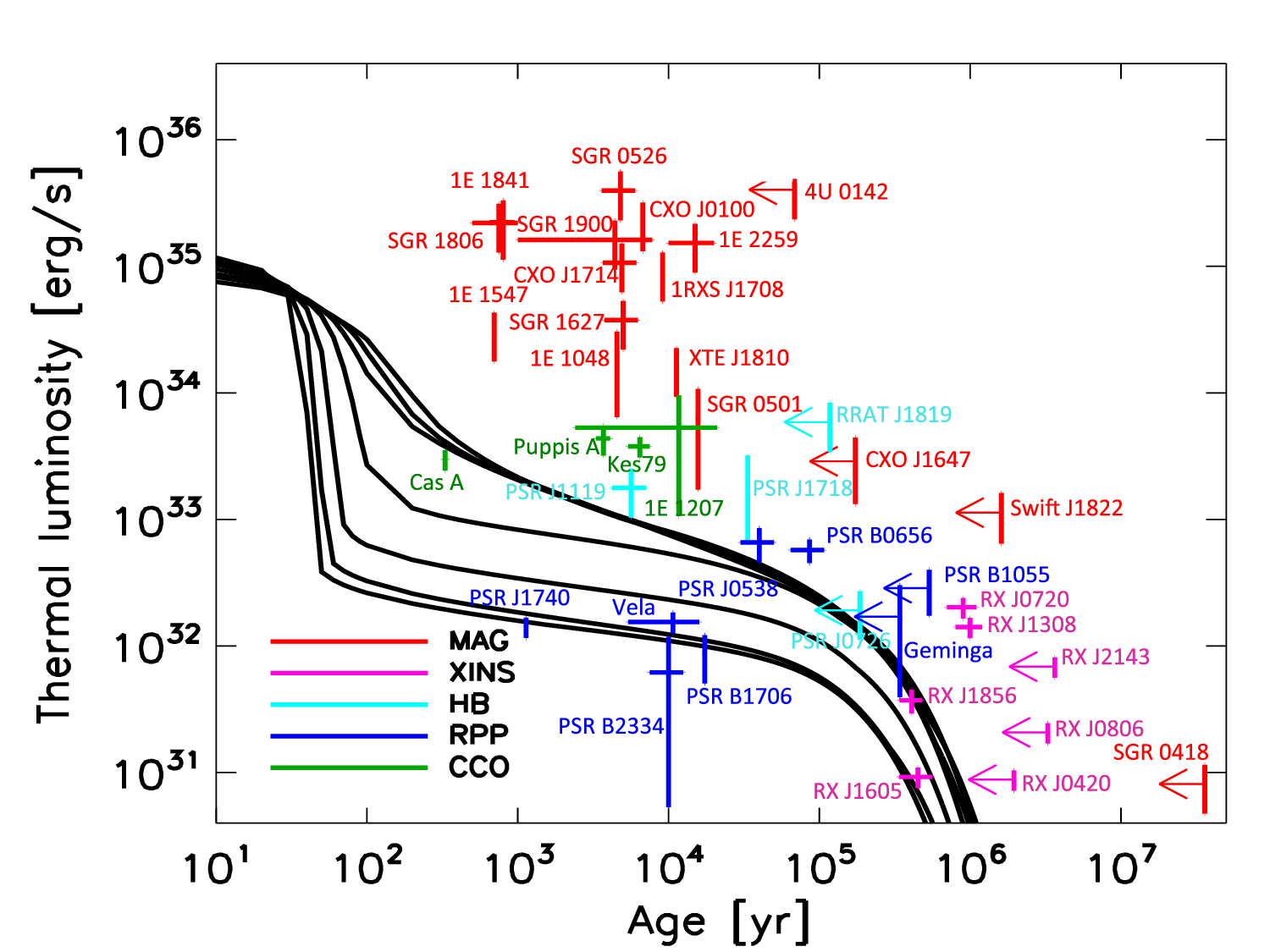}
\includegraphics[width=.85\textwidth]{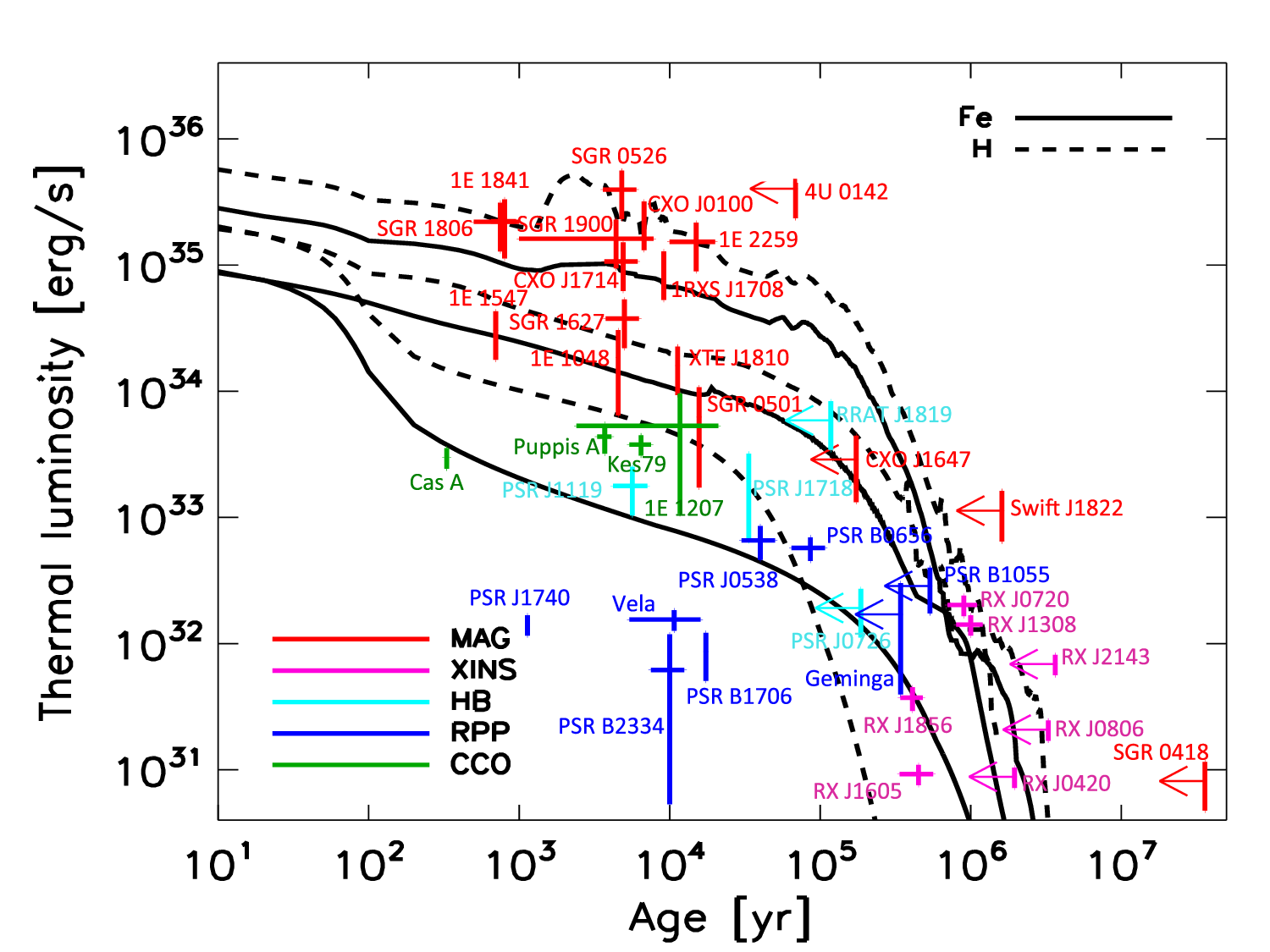}
\caption{Comparison between observational data and
theoretical cooling curves (from \citealp{vigano13}). 
The observational
estimates of (errorbars) or constraints on (arrows) the age
and thermal luminosity correspond to \citet{vigano13} and
Table~\ref{tab:cooling}. The abbreviations in the legend
mark different classes of neutron stars with measured
thermal radiation (MAG -- magnetar candidates, XINS -- X-ray
isolated neutron stars, HB -- high-$B$ radio pulsars, RPP --
rotation powered pulsars, CCO -- central compact objects;
see Paper~I).
Upper panel: non-magnetic neutron stars
with iron envelopes, with $M=(1.10$,
1.25, 1.32, 1.40, 1.48, 1.60, 1.70, 1.76) $M_\odot$ (lines
from top to bottom).
Lower panel: a neutron star
with $M=1.4\,M_\odot$ and  $R=11.6$ km, and three different
cases with initial magnetic field at the pole $B=0$,
$3\times 10^{14}$~G, and $3\times 10^{15}$~G. The magnetic
field topology is that of Model A in \citet{vigano13}
(crustal confined). We show results for iron envelopes
(solid lines) and hydrogen envelopes (dashed lines).}
 \label{fig:cooling}
\end{figure}

In Table \ref{tab:cooling} we summarize the properties of
cooling neutron stars,\footnote{A regularly updated online
catalog can be found at 
\texttt{http://www.neutronstarcooling.info},  with abundant
links to references for each source.} and in
Fig.~\ref{fig:cooling} we compare the observational data to
theoretical cooling curves, from \cite{vigano13}. Here, the
theoretical results are computed by a finite difference
method for 2D (axisymmetric) stellar configurations, using
the SLy EoS model \citep{DouchinHaensel01} at $\rho>\rhod$
and the BPS EoS \citep{BPS} at $\rho < \rhod$. The high
Durca threshold of the SLy EoS has been artificially
lowered for illustrative purpose to $\rho=10^{15}$ \gcc,
corresponding to the central density of a star with
$M=1.42\,M_\odot$ (see \citealp{ViganoPhD} for details). For
superfluid gap energies, the phenomenological model of
\citet{HoGA12} has been adopted. The other microphysics
input is the same as in
Sects.~\ref{sect:phys}\,--\,\ref{sect:magnetic}.

%%%%%%%%%%%%%%%%%%%%%%%%%%%%%%%%%%%%%%%%%%%
%%%%%%%%%%%%%%%%%%%%%%%%%%%%%%%%%%%%%%%%%%%

In the upper panel of Fig.~\ref{fig:cooling} we show cooling
curves for non-magnetized neutron stars with masses ranging
between 1.10 and 1.76 $M_\odot$ (lines from top to bottom).
After $\approx 100$ yr, low mass stars ($M \lesssim 1.4
M_\odot$) are brighter than high mass stars. For the
high-mass family, $M\gtrsim 1.4 M_\odot$, the Durca
processes in the central part of the star result in fast
cooling before one hundred years. Within the low-mass
family, cooling curves are similar at early ages ($<100$
yr). The differences at $t\sim 10^2-10^3$ yr are due to the
delayed transition of neutrons in the core to a superfluid
state, which activates the PBF neutrino emission. After the
effect of the transition to a superfluid core is finished,
at $t\gtrsim 10^3$ yr, cooling curves for low-mass neutron
stars tend to converge again, following the same curve
independently of the mass. 

We see that luminosities of some objects in the upper panel
of Fig.~\ref{fig:cooling} are systematically above the
theoretical cooling curves. For the CCOs this discrepancy
can be eliminated by considering accreted (more
heat-transparent) blanketing envelopes, as the lowest dashed
line in the lower panel of Fig.~\ref{fig:cooling}
demonstrates. However, the high-$B$ objects still remain
systematically hotter than what the theory can explain at
$B=0$. This provides strong evidence in favor of the
scenario in which magnetic field decay powers their larger
luminosity. In the lower panel we compare the observational
data to theoretical cooling curves for different values of
the initial magnetic field up to $3\times 10^{15}$~G. The
most relevant effect of the inclusion of the magnetic field
is that it allows to explain objects with high luminosities.
Magnetic fields $B\gtrsim 10^{14}$~G are strong enough to
noticeably  heat up the crust and power the observed X-ray
radiation. Another important difference is that the cooling
timescale for strongly magnetized objects is several times
larger than for the weakly magnetized neutron stars.

%%%%%%%%%%%%%%%%%%%%%%%%%%%%%%%%%%%%%
\section{Conclusions}
\label{sect:concl}
We have considered the basic physical ingredients needed for
theoretical modeling neutron-star thermal evolution and
briefly reviewed some recent results on cooling of
magnetized neutron stars. The physics behind such thermal
evolution is extremely rich. Clearly, we could not consider
it in depth in a single review paper. However, the
information that we have given, together with the references
to the formulae and online resources elsewhere, should be
sufficient to build a neutron-star cooling model involving
only the simplest assumptions. We considered the basic
equations that govern the mechanical and thermal structure
of a neutron star and its thermal evolution, the main
contributions to the physical quantities that enter these
equations -- namely, EoS and heat capacity, thermal
conductivity, neutrino emissivity, the effects of baryon
superfluidity and proton superconductivity and of strong
magnetic fields. In addition, we present a novel fit to
the relation between the internal and external temperatures
and heat fluxes
in the blanketing envelope, which includes the effects
of neutrino emission from the crust and the effects of
non-radial heat transport.

In this paper we have restricted ourselves by the $npe\mu$
matter, without either hyperons or ``exotic'' models that
involve hyperon condensates, quark phases, mixed phases, or
phase transitions. We hope that an interested reader should
be able to study these issues in depth following the
literature references that we have provided. We have not
considered also the equations of magnetic-field evolution,
coupled to the thermal evolution, which is especially
important in magnetars. These equations are given, for
instance, in the paper by \citet{Mereghetti-ea15} in this
volume, where origin, evolution, and observational
manifestations of magnetars are reviewed in depth.

%%%%%%%%%%%%%%%%%%%%%%%%%%%%%%%%%%%%%%%%%%%%%%%%%%%%%%%%%%
\begin{acknowledgements}
The authors acknowledge hospitality of organizers and useful
discussions with participants at the ISSI Workshop ``The
Strongest Magnetic Fields in the Universe'' (Bern,
Switzerland, 3\,--7 February 2014), where this joint review
was initiated.
A.P.{} is grateful to D.G.\,Yakovlev for useful discussions.
The work of A.P.{} on the effects of strong magnetic fields
on blanketing envelopes (Sect.~\ref{sect:env-mag} and
Appendix~\ref{sect:TbTs}) has been
supported by the Russian Science Foundation (grant
14-12-00316). 

Conflict of Interest: 
The authors declare that they have no conflict of interest.
\end{acknowledgements}
%%%%%%%%%%%%%%%%%%%%%%%%%%%%%%%%%%%%%%%%%%%%%%%%%%%%%%%%%%

\begin{appendix}
\section*{Appendices}

\renewcommand{\thesection}{\Alph{section}.}
\renewcommand{\thesubsection}{\Alph{section}.\arabic{subsection}}

\section{Electron thermal conductivities}
\label{sect:conduct}
\renewcommand{\theequation}{A.\arabic{equation}}
\setcounter{equation}{0}
In this Appendix, we briefly overview the physics of
electron heat conduction in the neutron-star envelopes,
which is the most important heat conduction mechanism as
regards the neutron-star thermal evolution, in the case of
$B=0$.  The magnetic field effects on the heat conduction
are considered in Sect.~\ref{sect:condmag}.

% -----------------------------------
\subsection{Weakly degenerate electron gas}
\label{sect:nondeg}

In the case of non-degenerate and non-relativistic electrons
\citep{SpitzerHarm,Braginskii58,Spitzer},
the effective energy-averaged electron-ion collision
frequency is
\begin{equation}
 \nu_{e\mathrm{i}} = \frac43 \sqrt{\frac{2\pi}{\mel}}
       \,\frac{Z^2 e^4}{T^{3/2}}
 \,\nion \Lambda_{e\mathrm{i}},
\label{nu_ei_nondeg}
\end{equation}
where  $\Lambda_{e\mathrm{i}}$ is the Coulomb logarithm. In
the considered case $\Lambda_{e\mathrm{i}}$ is a slowly
varying function of density and temperature. Its
precise value depends on the approximations used to solve
the Boltzmann equation, but its order of magnitude is given
by the elementary theory, where the Coulomb collision
integral is truncated at small and large impact parameters
of the electrons. Then
$\Lambda_{e\mathrm{i}}\sim\ln(r_{\mathrm{max}}/r_{\mathrm{min}})$,
where $r_{\mathrm{max}}$ and $r_{\mathrm{min}}$ are the
maximum and minimum electron impact parameters. The
parameter $r_{\mathrm{max}}$ can be set equal to the Debye
screening length, $r_{\mathrm{max}}^{-2}=4\pi (\nel +Z^2
\nion) e^2/T$. The second parameter can be estimated as
$r_{\mathrm{min}} = \max(\lambde,\,r_{\mathrm{cl}})$,
where $\lambde$ (defined in Sect.~\ref{sect:param}) limits
$r_{\mathrm{min}}$ in the high-temperature regime (where the
Born approximation holds), and $r_{\mathrm{cl}} = Ze^2/T$
is the classical closest-approach distance of a thermal
electron, which limits $r_{\mathrm{min}}$ in the
low-temperature, quasiclassical regime.

A similar effective frequency
\begin{equation}
 \nu_{ee} = \frac83 \sqrt{\frac{\pi}{\mel}}\,
\frac{e^4}{T^{3/2}}
 \,\nel \Lambda_{ee}
\label{nu_ee_nondeg}
\end{equation}
characterizes the efficiency of the $ee$ collisions.
If $\Lambda_{ee}\sim\Lambda_{e\mathrm{i}}$, then $\nu_{e\mathrm{i}}/\nu_{ee}\sim Z$,
therefore for large $Z$ the $e\mathrm{i}$ collisions are much more
efficient than the $ee$ collisions.

% -----------------------------------
\subsection{Strongly degenerate electron gas}
\subsubsection{Electron-ion
scattering}
\label{sect:deg}

The thermal conductivity of strongly degenerate electrons in
a fully ionized plasma is given by \req{elementary} with
$a=\pi^2/3$. In order to determine the effective collision
frequency that enters this equation, we use the
Matthiessen rule $\nu=\nu_{e\mathrm{i}}+\nu_{ee}$.

The effective electron-ion collision frequency
can be written in the form \citep{Lee50,YakovlevUrpin80}
\begin{equation}
   \nu_{e\mathrm{i}} = \frac{ 
   4 Z \mel^\ast e^4 \Lambda_{e\mathrm{i}}}{ 3 \pi \hbar^3 }
   =\frac{Z\Lambda_{e\mathrm{i}}
                   \,\sqrt{1+\xr^2}}{5.7\times10^{-17}\mbox{~s}}.
\label{tau}
\end{equation}
\citet{Lee50} gave an estimate of the Coulomb logarithm
$\Lambda_{e\mathrm{i}} = \ln 
(r_\mathrm{max}/r_\mathrm{min})$, with the minimum impact
parameter $r_\mathrm{min} = \hbar/2 \pF$ and the maximum
impact parameter $r_\mathrm{max} = \aion$.
\citet{YakovlevUrpin80} calculated the conductivities for
relativistic degenerate electrons, neglecting electron
screening, and obtained a more accurate estimate
$r_\mathrm{max} \approx 0.4 \aion$ in the liquid regime. In
the solid regime, where the electrons scatter on phonons
(collective ion excitations), \citet{YakovlevUrpin80}
obtained different approximations for the two distinct
cases, $\Theta_\mathrm{D} < T < \Tm$ and $T<
\Theta_\mathrm{D}$.

\citet{Potekhin-ea99} derived a unified treatment of the
electron conductivities in the Coulomb liquid and solid and
described both regimes by \req{tau}. Then qualitatively, by
order of magnitude, $\Lambda_{e\mathrm{i}}\sim1$ in the ion
liquid, and $\Lambda_{e\mathrm{i}}\sim T/T_\mathrm{m}$ in
the Coulomb solid with a melting temperature $T_\mathrm{m}$.
The effects of multiphonon scattering, electron screening,
and non-Born corrections, have been taken into account, and
the Coulomb logarithms in both liquid and solid phases have
been fitted by a single analytical formula. A Fortran code 
and a table of thermal conductivities, based on this
formalism, are available
online.\footnote{\texttt{http://www.ioffe.ru/astro/conduct/}\label{conduct}}

At the conditions typical for the envelopes of neutron
stars, the electron-phonon scattering proceeds mainly via
the Umklapp processes, where the wave vector corresponding
to the change of electron momentum lies outside the first
Brillouin zone. \citet{RaikhYakovlev82} noticed that if
$T\lesssim T_\mathrm{U} = \Tp Z^{1/3} \alphaf
\sqrt{1+\xr^2}/3\xr$, then the Umklapp processes occur less
often (``freeze out''). Then the scattering rate decreases.
\citet{RaikhYakovlev82} assumed an extremely strong
(exponential) decrease. This implied that at $T <
T_\mathrm{U}$ the conductivity would be in practice
determined by impurities and structure defects of the
lattice, rather than by the electron-phonon scattering
\citep{GYP01}. However, \citet{Chugunov12} showed that
distortion of electron wave functions due to interaction
with the Coulomb lattice destroys this picture and strongly
slows down the increase of the conductivity. As a result,
the conductivities in neutron star envelopes can be treated
neglecting the ``freezing-out'' of the Umklapp processes.

% -----------------------------------
\subsubsection{Electron-electron scattering}
\label{sect:ee}

Although the electron-ion scattering is usually most
important for degenerate plasmas, the electron-electron
scattering still can be non-negligible for relatively light
elements ($Z\lesssim10$) \citep{Lampe68b}. The expression of
$\nu_{ee}$ for the relativistic degenerate electrons
at $T\ll T_\mathrm{p}$ was obtained by
\citet{FlowersItoh76}. \citet{UrpinYakovlev80} extended it
to higher temperatures, where $\Tp\lesssim T\ll \EF $.

\citet{ShterninYakovlev06} reconsidered
the problem including the Landau damping of transverse
plasmons, neglected by the previous authors. This effect 
is due to the difference of the components of the
polarizability tensor, responsible for screening the
charge-charge and current-current interactions: the
transverse current-current interactions undergo  ``dynamical
screening.'' \citet{ShterninYakovlev06} showed that the
Landau damping of transverse plasmons strongly increases
$\nu_{ee}$ in the domain of $\xr \gtrsim 1$
and $T\ll \Tp$ and presented a new fit to $\nu_{ee}$
(also implemented in the code referenced in footnote
\ref{conduct}).

% ---------------------------------------------------
\subsection{The case of intermediate degeneracy}
\label{sect:interpolate}

In the case where the electron gas is partially degenerate,
that is $T\sim\EF$, the thermal and electrical
conductivities determined by the electron-ion scattering are
satisfactorily evaluated by the thermal averaging procedure
[\req{sigma-tau} in Sect.~\ref{sect:condmag-e}]. For conductivities
determined by the electron-electron collisions, there is no
such averaging procedure, but we can use an
interpolation between the two limiting cases,
\beq
 \nu_{ee} = \nu_{ee}^{\mathrm{deg}}
    \frac{1+625\, (T/\EF)^2}{1+25\, T/\EF+ 271\, (T/\EF)^{5/2}}.
\label{nu-ee}
\eeq
A satisfactory accuracy of this interpolation has been
verified by \citet{Cassisi-ea07}.

% ---------------------------------------------------
\subsection{Impurities and mixtures}
\label{sect:imp}

If the plasma in an envelope is not a pure substance of a single
chemical element,
then the
effective collision frequency $\nu_{e\mathrm{i}}$ should 
be modified. The required modification can be different,
depending on the state of the plasma and on the amount of
impurities. For example, \citet{FlowersItoh76},
\citet{YakovlevUrpin80}, and
\citet{ItohKohyama93} considered electron scattering by charged
impurities in a Coulomb crystal. If the fraction of
impurities is small and they are randomly distributed, then
electron-impurity scattering can be treated as scattering by
charge fluctuations, controlled by the impurity parameter
$
Q = \langle (Z-\langle Z\rangle)^2\rangle,
$
where $\langle Z\rangle \equiv \sum_j Y_j Z_j$,
$Y_j=n_j/\sum_j n_j$ is the number fraction of ions of the
$j$th kind, and $Z_j$ is their charge number. Then, using
the Matthiessen rule, one can obtain  $\nu_{e\mathrm{i}}$ as
a sum of the terms corresponding to the electron-phonon
scattering in a homogeneous lattice and to the electron
scattering by charge fluctuations. The effective relaxation
time for the latter term is given by \req{tau} with
$Z\Lambda_{e\mathrm{i}}$ replaced by $\sum_j Y_j(Z_j-\langle
Z\rangle)^2 \Lambda_j/\langle Z\rangle$, where the Coulomb
logarithm $\Lambda_j$ depends generally on $j$. Neglecting the
differences between the Coulomb logarithms, one can thus simply
replace $Z$ by $Q/\langle Z\rangle$ in \req{tau} to estimate
the conductivity due to electron scattering by charged
impurities.

An alternative approach is relevant when there is no
dominant ion species  which forms a crystal (e.g., in a
liquid, a gas, or a glassy alloy). In this case, one can use
\req{tau} with $Z^2 \nion \Lambda_{e\mathrm{i}}$ replaced by
$\sum_j Z_j^2 n_j \Lambda_j$. An approximation to
$\Lambda_j$ based on the plasma  ``additivity rule'' has
been suggested by \citet{Potekhin-ea99}. Neglecting the
differences between the Coulomb logarithms, one arrives at
\req{tau} with $Z$ replaced by $\sqrt{\langle Z^2\rangle}$. If tabulated
conductivities $\kappa_j$ for pure substances are used, then
the best agreement with calculations based on the
``additivity rule'' is usually given by the estimate
\beq
   \kappa \approx \frac{\sum_j Y_j Z_j \kappa_j }{
        \sum_j Y_j Z_j}
 \equiv \frac{\langle \kappa Z \rangle }{ \langle Z \rangle}.
\eeq

%%%%%%%%%%%%%%%%%%%%%%%%%%%%%%%%%%%%%%%%%%%%

%%%%%%%%%%%%%%%%%%%%%%%%%%%%%%%%%%%%%%
\section{Temperature relations for envelopes
of neutron stars with magnetic fields}
\label{sect:TbTs}
\renewcommand{\theequation}{B.\arabic{equation}}
\setcounter{equation}{0}
Here we present an analytical fit to the temperature
distribution over a surface of a neutron star with a
non-accreted envelope and a dipole magnetic field. We have
chosen $\rhob=10^{10}$ \gcc{} and used the BSk21 EoS
\citep{Pearson-ea12} in the parametrized form
\citep{Potekhin-ea13}. The numerical data have been produced
with the 2D code of \citet{vigano13}  for 5 values of
internal temperature $\Tb$ from $10^7$~K to $10^9$~K, 5
values of the magnetic field at the pole $B_\mathrm{p}$ from
$10^{11}$~G to $10^{15}$~G, and 20 values of magnetic
colatitude $\theta$ at the surface of the neutron star from
0 to $\pi/2$. The use of the 2D code corrects the
temperature distribution near the magnetic equator, because
the non-radial heat flow increases the equatorial $\Ts$ as
compared to the 1D model that was employed previously (see
Fig.~\ref{fig:TbTs} in Sect.~\ref{sect:env-mag}). These data
have been supplemented with more detailed calculations at
the magnetic pole ($\theta=0$) using the 1D code of
\citet{PCY07} for 36 values of $\Tb$ from $10^{6.5}$~K to
$10^{10}$~K and 9 values of $B_\mathrm{p}$ from $10^{11}$~G
to $10^{15}$~G. An important difference from the old results
is the inclusion of the neutrino emission from the crust,
which is especially important for the magnetars (see
Sect.~\ref{sect:th_str}). Because of the 2D treatment and
the allowance for neutrino emission, the new fit supersedes
the previous one \citep{Potekhin-ea03}, whenever
$B>10^{12}$~G or $\Tb\gtrsim10^8$~K. We stress that
its use is restricted by non-accreted (i.e., composed of
heavy chemical elements) envelopes in the range of
$10^{6.5}\mbox{~K}\lesssim\Tb\lesssim10^{10}$~K and
$B_\mathrm{p}\lesssim10^{15}$~G, which is covered by the
underlying numerical data. For envelopes with
$B\lesssim10^{12}$~G (either non-accreted or accreted), the
previous fit can be used, however the surface temperature
$\Ts$ (but not the flux at the inner boundary,
$F_\mathrm{b}$ -- see item 4 below) should be limited for
hot stars according to \req{nu-corr} below.

The fit consists of 3 stages: 
(1) an expression for the
surface temperature at the magnetic pole, $T_\mathrm{p}$,
as function of $\Tb$, $g$, and $B_\mathrm{p}$;
(2) an expression for the
\emph{ratio} of the polar to the equatorial surface
temperatures, $T_\mathrm{p}/T_\mathrm{eq}$;
(3) an expression for 
the dependence of $\Ts$ on the magnetic colatitude $\theta$.
Since the thermal conductivities for quantizing magnetic
fields (Sect.~\ref{sect:condmag-e}) are known for the
electron-ion but not electron-electron collision mechanism,
we multiplied $\Ts$ by a correction factor,
obtained numerically from a comparison of the results of
thermal-structure calculations with and without the $ee$
collisions at $B=0$. At the end of this Appendix we suggest
a recipe for relating the flux $F_\mathrm{b}$ at the bottom
of the heat-blanketing envelope to temperature $\Ts$ and thereby to
$\Tb$.

\textbf{1.}
At the magnetic pole, the effective surface temperature,
neglecting neutrino emission from the crust, is
approximately given by the expression
\beq
     T_\mathrm{p}^{(0)}=\left[g_{14}(T_1^4+
(1+0.15\sqrt{B_{12}})\,T_0^4)\right]^{1/4}
\times10^6\mbox{~K},
\eeq
where
\beq
T_0 = (15.7 T_9^{3/2}+1.36
T_9)^{0.3796},
\quad
      T_1 = 1.13\,B_{12}^{\,0.119}T_9^{a},
\quad
      a=0.337/(1+0.02\sqrt{B_{12}}), 
\eeq
$T_9 = \Tb/10^9$~K, and
$B_{12}=B_\mathrm{p}/10^{12}$~G.
The limiting temperature, at which $T_\mathrm{p}(\Tb)$
levels off due to the neutrino emission from the crust is
approximately given by
\beq
      T_\mathrm{p}^\mathrm{(max)}= (5.2 g_{14}^{0.65}+
0.093\sqrt{g_{14}\,B_{12}}) \times10^6\mbox{~K}.
\eeq
The corrected surface temperature at the pole, which takes
this limit into account, is reproduced by the
expression
\beq
   T_\mathrm{p} = T_\mathrm{p}^{(0)} \left[ 1+
(T_\mathrm{p}^{(0)}/T_\mathrm{p}^\mathrm{(max)})^4
\right]^{-1/4}
\label{nu-corr}
\eeq

\textbf{2.}
The ratio of the polar to equatorial surface
temperatures can be roughly evaluated as
\beq
\frac{T_\mathrm{p}}{T_\mathrm{eq}} =
1+ \frac{(1230\,T_9)^{3.35}\,B_{12}\,
     \sqrt{1+2B_{12}^{\,2}} }{ (B_{12}+450\,T_9+
119\,B_{12}\,T_9)^4}
     +  \frac{0.0066\,B_{12}^{\,5/2} }{
         T_9^{1/2}+0.00258\,B_{12}^{\,5/2}}.
\label{TpTeq}
\eeq
The numerically calculated $T_\mathrm{p}/T_\mathrm{eq}$
ratio has a complex dependence on $\Tb$ and $B$ at $B >
10^{13}$~G. In order to keep our fitting formulae relatively
simple, we do not reproduce these oscillations, but instead
force the ratio (\ref{TpTeq}) to converge to some average
value at $B \gg 10^{13}$~G. The numerical data oscillate in
a complicated manner around this average, with deviations
reaching up to 35\%. For smaller fields,
$B\lesssim3\times10^{12}$~G, \req{TpTeq} reproduces the
numerical data with typical errors of several percent (up to
10\%). Note that these significant deviations affect only
nearly tangential field case, viz.\ the equatorial region,
which is substantially colder than the rest of the surface.
Therefore its contribution to the observed flux is usually
not very important.

\textbf{3.}
Finally, the dependence of the surface temperature on the magnetic
colatitude $\theta$ is approximately described by the
expression
\beq
      \frac{\Ts(\theta)-T_\mathrm{eq}}{T_\mathrm{p}-T_\mathrm{eq}}
         = \frac{(1+a_1+a_2) \cos^2\theta}{
            1+a_1\cos\theta+a_2\cos^2\theta},
\quad\mbox{where~~}
      a_1=\frac{a_2 T_9^{1/2}}{3},
\quad
   a_2  =
     \frac{ 10\,B_{12}}{
      T_9^{1/2} +
0.1\,B_{12}\,T_9^{-1/4}}.
\eeq

\textbf{4.}
Note that the outer boundary condition to the thermal evolution
equations (\ref{Tbalance}) involves the
relation between the heat flux density $F_\mathrm{b}$ through the
boundary at $\rho=\rhob$ and the temperature $\Tb$ at this
boundary. In the absence of the neutrino emission from the
crust, this boundary condition is directly provided by the
\TbTs{} relation, because in this case (in the
plane-parallel approximation) $F_\mathrm{b}=\sSB\Ts^4$. It
is not so if a significant part of the energy is
carried from the outer crust by neutrinos. In this case we
suggest to evaluate the flux through the boundary by the
relation $F_\mathrm{b}=\sSB T_\ast^4$, where
$T_\ast$ is given by the above approximations for $\Ts$,
but without the correction (\ref{nu-corr}).

\end{appendix}

% =============== JOURNAL ABBREVIATIONS =============
\newcommand{\artref}[4]{{#1} {#2}, #3 (#4)}
\newcommand{\AandA}[3]{\artref{Astron.\ Astrophys.}{#1}{#2}{#3}}
\newcommand{\AIPC}[3]{\artref{AIP Conf.\ Proc.}{#1}{#2}{#3}}
\newcommand{\AnnPhysNY}[3]{\artref{Ann.\ Phys. (N.Y.)}{#1}{#2}{#3}}
\newcommand{\ApJ}[3]{\artref{Astrophys.\ J.}{#1}{#2}{#3}}
\newcommand{\ApJS}[3]{\artref{Astrophys.\ J.\ Suppl.\ Ser.}{#1}{#2}{#3}}
\newcommand{\ApSS}[3]{\artref{Astrophys.\ Space Sci.}{#1}{#2}{#3}}
\newcommand{\ARAA}[3]{\artref{Annu.\ Rev.\ Astron.\ Astrophys.}{#1}{#2}{#3}}
\newcommand{\AL}[3]{\artref{Astron.\ Lett.}{#1}{#2}{#3}}
\newcommand{\JPB}[3]{\artref{J.\ Phys.\ B: At.\ Mol.\ Opt.\ Phys.}{#1}{#2}{#3}}
\newcommand{\jpb}[3]{\artref{J.\ Phys.\ B: At.\ Mol.\ Phys.}{#1}{#2}{#3}}
\newcommand{\MNRAS}[3]{\artref{Mon.\ Not.\ R.\ Astron.\ Soc.}{#1}{#2}{#3}}
\newcommand{\NPA}[3]{\artref{Nucl.\ Phys.\ A}{#1}{#2}{#3}}
\newcommand{\PL}[4]{\artref{Phys.\ Lett. #1}{#2}{#3}{#4}}
\newcommand{\PR}[4]{\artref{Phys.\ Rev. #1}{#2}{#3}{#4}}
\newcommand{\PRL}[3]{\artref{Phys.\ Rev.\ Lett.}{#1}{#2}{#3}}
\newcommand{\RMP}[3]{\artref{Rev.\ Mod.\ Phys.}{#1}{#2}{#3}}
\newcommand{\SSRv}[3]{\artref{Space Sci.\ Rev.}{#1}{#2}{#3}}
\newcommand{\SvAL}[3]{\artref{Sov.\ Astron.\ Lett.}{#1}{#2}{#3}}
\newcommand{\SvA}[3]{\artref{Sov.\ Astron.}{#1}{#2}{#3}}
% ===================================================


\begin{thebibliography}{}
\addcontentsline{toc}{section}{References}
\small
\bibitem[Aguilera et al.(2008))Aguilera, Pons and Miralles]{AguileraPM08}
D.N.~Aguilera, J.A.~Pons, J.A.~Miralles,
% 2D cooling of magnetized neutron stars,
\AandA{486}{255}{2008}

\bibitem[Aguilera et al.(2009)]{aguilera09}
D.N.~Aguilera, V.~Cirigliano, J.A.~Pons, S.~Reddy, R.~Sharma,
\PRL{102}{091101}{2009}

\bibitem[Akmal et al.(1998)Akmal, Pandharipande, and Ravenhall]{APR}
A.~Akmal, V.R.~Pandharipande, D.G.~Ravenhall,
% Equation of state of nucleon matter and neutron star structure
\PR{C}{58}{1804}{1998}

\bibitem[Ainsworth et al.(1989)Ainsworth, Wambach, and Pines]{AinsworthWP89}
T.L.~Ainsworth, J.~Wambach, D.~Pines,
\artref{Phys.\ Lett. B}{222}{173}{1989}

\bibitem[Alpar et al.(1984)]{Alpar-ea84}
M.A.~Alpar, D.~Pines, P.W.~Anderson, J.~Shaham,
%Vortex creep and the internal temperature of neutron stars. I - General theory
\ApJ{276}{325}{1984}

\bibitem[Amundsen and {\O}stgaard(1985a)]{AmundsenOstgaard85a}
L.~Amundsen, E.~{\O}stgaard,
% Superfluidity of neutron matter (I). Singlet pairing,
\NPA{437}{487}{1985a}

\bibitem[Amundsen and {\O}stgaard(1985b)]{AmundsenOstgaard85b}
L.~Amundsen, E.~{\O}stgaard,
% Superfluidity of neutron matter (I). Singlet pairing,
\NPA{442}{163}{1985b}

\bibitem[Bahcall and Wolf(1965a)]{BahcallWolf65a}
J.N.~Bahcall, R.A.~Wolf,
% ``An observational test of theories of neutron-star cooling,''
\ApJ{142}{1254}{1965a}

\bibitem[Bahcall and Wolf(1965b)]{BahcallWolf65b}
J.N.~Bahcall, R.A.~Wolf,
% ``Neutron stars. II. Neutrino-cooling and observability,''
 \PR{}{140}{B1452}{1965b}

\bibitem[Baiko(2009)]{Baiko09}
D.A.~Baiko,
\PR{E}{80}{046405}{2009}

\bibitem[Baiko and Yakovlev(1999)]{BaikoYakovlev99}
D.A.~Baiko, D.G.~Yakovlev,
% Direct URCA process in strong magnetic fields and neutron star cooling
\AandA{342}{192}{1999}

\bibitem[Baiko et al.(1998)]{Baiko-ea98}
D.A.~Baiko, A.D.~Kaminker, A.Y.~Potekhin, D.G.~Yakovlev,
% Ion structure factors and electron transport
% in dense Coulomb plasmas,
\PRL{81}{5556}{1998}

\bibitem[Baiko et al.(2001a)Baiko, Haensel, and Yakovlev]{BaikoHY01}
D.A.~Baiko, P.~Haensel, D.G.~Yakovlev,
\AandA{374}{151}{2001a}

\bibitem[Baiko et al.(2001b)Baiko, Potekhin, and Yakovlev]{BPY01}
D.A.~Baiko, A.Y.~Potekhin, D.G.~Yakovlev,
%Thermodynamic functions of harmonic Coulomb crystals
\PR{E}{64}{057402}{2001b}

\bibitem[Balberg and Barnea(1998)]{BalbergBarnea98}
S.~Balberg, N.~Barnea,
\PR{C}{57}{409}{1998}

\bibitem[Baldo and Schulze(2007)]{BaldoSchulze07}
M.~Baldo, H.-J.~Schulze,
% Proton pairing in neutron stars
\PR{C}{75}{025802}{2007}

\bibitem[Baldo et al.(1992)]{Baldo-ea92}
M.~Baldo, J.~Cugnon, A.~Lejeune, U.~Lombardo, 
\NPA{536}{349}{1992}

\bibitem[Baldo et al.(1998)]{Baldo-ea98}
M.~Baldo, {\O}.~Elgar{\o}y, L.~Engvik, M.~Hjorth-Jensen, H.-J.~Schulze,
% 3P2-3F2Proton pairing in neutron stars pairing in neutron matter with modern nucleon-nucleon potentials
\PR{C}{58}{1921}{1998}

\bibitem[Bardeen et al.(1957)Bardeen, Cooper, and Schrieffer]{BCS57}
J.~Bardeen, L.N.~Cooper, J.R.~Schrieffer,
\PR{}{108}{1175}{1957}

\bibitem[Baym and Pethick(1991)]{BaymPethick}
G.~Baym, C.~Pethick,
\textit{Landau Fermi-Liquid Theory: Concepts and
Applications}.
John Wiley \& Sons, New York (1991)

\bibitem[Baym et al.(1971)Baym, Pethick, \& Sutherland]{BPS}
G.~Baym, C.~Pethick, P.~Sutherland,
% ``The ground state of matter at high densities: Equation of state and stellar models,''
\ApJ{170}{299}{1971}

\bibitem[Baym et al.(1969)Baym, Pethick, \& Pines]{BPP69a}
G.~Baym, C.~Pethick, D.~Pines,
% Superfluidity in neutron stars
\artref{Nature}{224}{673}{1969}

\bibitem[Bezchastnov et al.(1997)]{Bezchastnov-ea97}
V.G.~Bezchastnov, P.~Haensel, A.D.~Kaminker, D.G.~Yakovlev,
% Neutrino synchrotron emission from dense magnetized electron gas of neutron stars.
\AandA{328}{409}{1997}

\bibitem[Bisnovatyi-Kogan and Romanova(1982)]{BisnoRomanova82}
G.S.~Bisnovatyi-Kogan, M.M.~Romanova,
% Diffusion and heat transfer of neutrons in the envelopes neutron stars
% \ZhETF{83}, 449--459.
\artref{Sov.\ Phys. JETP}{56}{243}{1982}

\bibitem[Blaschke et al.(1995)]{Blaschke-ea95}
D.~Blaschke, G.~R\"opke, H.~Schulz, A.D.~Sedrakian, D.N.~Voskresensky,
\MNRAS{273}{596}{1995}

\bibitem[Bohr et al.(1958)Bohr, Mottelson, and Pines]{BohrMP}
A.~Bohr, B.R.~Mottelson, D.~Pines,
% Possible analog between the excitation spectra of nuclei and those of superconducting metal state,
\PR{}{110}{936}{1958}

\bibitem[Bowyer et al.(1964a)]{Bowyer-ea64a}
S.~Bowyer, E.T.~Byram, T.A.~Chubb, H.~Friedman,
% X-ray Sources in the Galaxy,
\artref{Nature}{201}{1307}{1964a}

\bibitem[Bowyer et al.(1964b)]{Bowyer-ea64b}
S.~Bowyer, E.T.~Byram, T.A.~Chubb, H.~Friedman,
% Lunar occultation of X-ray emission from the Crab nebula,
\artref{Science}{146}{912}{1964b}

\bibitem[Braginski{\u\i}(1958)]{Braginskii58}
S.I.~Braginski{\u\i},
% Transport phenomena in a completely ionized two-temperature
% plasma,
% Zh.\ Eksp.\ Teor. Fiz.1957 {33}, 459
% (Engl.\ transl.: 
\artref{Sov.\ Phys. JETP}{6}{358}{1958}

\bibitem[Brown et al.(1998)Brown, Bildsten, and Rutledge]{BBR98}
E.F.~Brown, L.~Bildsten, R.E.~Rutledge,
\ApJ{504}{L95}{1998}

\bibitem[Broderick et al.(2000)Broderick, Prakash, and Lattimer]{Broderick00}
A.~Broderick, M.~Prakash, J.M.~Lattimer,
% The equation of state of neutron star matter in strong magnetic fields,
\ApJ{537}{351}{2000}

\bibitem[Brown and Cumming(2009)]{BC09}
E.F.~Brown, A.~Cumming,
%Mapping Crustal Heating with the Cooling Light Curves of Quasi-Persistent Transients
\ApJ{698}{1020}{2009}

\bibitem[Brown et al.(1988)]{Brown-ea88}
G.E.~Brown, K.~Kudobera, D.~Page, P.M.~Pizzochero,
% Strangeness condensation and cooling of neutron stars,
\PR{D}{37}{2042}{1988}

\bibitem[Burrows and Lattimer(1986)]{Burrows86}
A.~Burrows, J.M.~Lattimer, 
% Neutron starquake models for gamma-ray bursts,
\ApJ{307}{178}{1986}

\bibitem[Cameron(1967)]{CameronScoX1}
A.G.W.~Cameron,
% ``Stellar Accretion and X-ray Emission,''
\artref{Nature}{215}{464}{1967}

\bibitem[Carr(1961)]{Carr61}
W.J.~Carr,
% Energy, Specific Heat, and Magnetic Properties of the Low-Density Electron Gas
\PR{}{122}{1437}{1961}

\bibitem[Cassisi et al.(2007)]{Cassisi-ea07}
S.~Cassisi, A.Y.~Potekhin, A.~Pietrinferni, M.~Catelan,
M.~Salaris,
% Updated electron-conduction opacities: the impact on low-mass stellar models,
\ApJ{661}{1094}{2007}

\bibitem[Chamel(2005)]{Chamel2005}
N. Chamel,
\NPA{747}{109}{2005}

\bibitem[Chamel(2012)]{Chamel12}
N.~Chamel,
\PR{C}{85}{035801}{2012}

\bibitem[Chamel et al.(2012)]{Chamel-ea12}
N.~Chamel, R.L.~Pavlov, L.M.~Mihailov, et al.,
% Properties of the outer crust of strongly magnetized neutron stars from Hartree-Fock-Bogoliubov atomic mass models
\PR{C}{86}{055804}{2012}

\bibitem[Chamel et al.(2013)Chamel, Page and Reddy]{Chamel13}
N.~Chamel, D.~Page, S.K.~Reddy, 
\PR{C}{87}{035803}{2013}

\bibitem[Chang et al.(2010)Chang, Bildsten, and Arras]{ChangBA10}
P.~Chang, L.~Bildsten, P.~Arras,
\ApJ{723}{719}{2010}

\bibitem[Chatterjee et al.(2015)Chatterjee, Elghozi, Novak, and Oertel]{Chatt15}
D.~Chatterjee, T.~Elghozi, J.~Novak, M.~Oertel,
\MNRAS{447}{3785}{2015}

\bibitem[Chen et al.(1993)]{Chen-ea93}
J.M.C.~Chen, J.W.~Clark, R.D.~Dav\'e, V.V.~Khodel,
% Pairing gaps in nucleonic superfluids
\NPA{555}{59}{1993}

\bibitem[Chiu and Salpeter(1964)]{ChiuSalpeter64}
H.-Y.~Chiu, E.E.~Salpeter,
% Surface X-ray emission from neutron stars,
\PRL{12}{413}{1964}

\bibitem[Chugunov(2012)]{Chugunov12}
A.I.~Chugunov,
% Electrical conductivity of the neutron star crust at low temperatures,
\AL{38}{25}{2012}

\bibitem[Chugunov and Haensel(2007)]{ChugunovHaensel07}
A.I.~Chugunov, P.~Haensel,
% Thermal conductivity of ions in a neutron star envelope.
\MNRAS{381}{1143}{2007}

\bibitem[Cooper(1956)]{Cooper56}
L.N.~Cooper,
\PR{}{104}{1189}{1956}

\bibitem[Coti Zelati et al.(2014)]{CotiZelati-ea14}
F.~Coti Zelati, S.~Campana, P.~D'Avanzo, A.~Melandri,
\MNRAS{438}{2634}{2014}

\bibitem[De Blasio(2000)]{DeBlasio00}
F.V.~De Blasio,
% A dense two-component plasma in a strong gravity field and thermal conductivity of neutron stars
\AandA{353}{1129}{2000}

\bibitem[de Freitas Pacheco et al.(1977)de Freitas Pacheco, Steiner, \& Neto]{deFreitas77}
J.A. de Freitas Pacheco, J.E.~Steiner, A.D.~Neto,
\AandA{55}{111}{1977}

\bibitem[DeWitt et al.(1993)]{DeWittSY93}
 H.E.~DeWitt, W.L.~Slattery, J.~Yang,
% ``Monte Carlo simulations of the OCP freezing transition,''
in 
\textit{Strongly Coupled Plasma Physics}, 
ed.~by H.M.~Van Horn, S.~Ichimaru.
Univ.~Rochester, Rochester (1993), p.~425

\bibitem[Douchin and Haensel(2001)]{DouchinHaensel01}
F.~Douchin, P.~Haensel,
\AandA{380}{151}{2001}

\bibitem[Elshamouty et al.(2013)]{Elshamouty-ea13}
K.G.~Elshamouty, C.O.~Heinke, G.R.~Sivakoff, et al.,
% W.C.G.~Ho, P.S.~Shternin, D.G.~Yakovlev, D. J. Patnaude, and L. David, 
\ApJ{777}{22}{2013}

\bibitem[Farouki and Hamaguchi(1993)]{FaroukiHamaguchi}
R.T.~Farouki, S.~Hamaguchi,
\PR{E}{47}{4330}{1993}

\bibitem[Flowers and Itoh(1976)]{FlowersItoh76}
E.~Flowers, N.~Itoh,
% ``Transport properties of dense matter,''
\ApJ{206}{218}{1976}

\bibitem[Flowers and Itoh(1979)]{FlowersItoh79}
 E.~Flowers, N.~Itoh,
%  ``Transport properties of dense matter. II,''
 \ApJ{230}{847}{1979}

\bibitem[Flowers and Itoh(1981)]{FlowersItoh81}
E.~Flowers, N.~Itoh,
% ``Transport properties of dense matter. 
% III. Analytic formulae for thermal conductivity,''
\ApJ{250}{750}{1981}

\bibitem[Flowers et al.(1976)Flowers, Ruderman, and Sutherland]{FlowersRS76}
E.~Flowers, M.~Ruderman, P.~Sutherland,
\ApJ{205}{541}{1976}

\bibitem[Franco et al.(2000)Franco, Bennett, and Epstein]{Franco-ea00}
L.M.~Franco, L.~Bennett, R.I.~Epstein,
% Quaking neutron stars,
\ApJ{543}{987}{2004}

\bibitem[Fridriksson et al.(2011)]{Fridriksson-ea11}
J.K.~Fridriksson, J.~Homan, R.~Wijnands, et al.,
\ApJ{736}{162}{2011}

\bibitem[Frieben and Rezzolla(2012)]{FriebenRezzolla12}
J.~Frieben, L.~Rezzolla,
\MNRAS{427}{3406}{2012}

\bibitem[Gamow and Schoenberg(1941)]{GamowSchoenberg41}
G.~Gamow, M.~Schoenberg,
% Neutrino theory of stellar collapse,
\PR{}{59}{539}{1941}

\bibitem[Gandolfi et al.(2008)]{Gandolfi-ea08}
S.~Gandolfi, A.Yu.~Illarionov, F.~Pederiva, K.E.~Schmidt, S.~Fantoni,
% Equation of state of low-density neutron matter, and the 1S0 pairing gap
\PR{C}{80}{045802}{2008}

\bibitem[Geppert et al.(2004)Geppert, K\"uker, and Page]{GeppertKP04}
U.~Geppert, M.~K\"uker, D.~Page,
% Temperature distribution in magnetized neutron star crusts,
\AandA{426}{267}{2004}

\bibitem[Geppert et al.(2006)Geppert, K\"uker, and Page]{GeppertKP06}
U.~Geppert, M.~K\"uker, D.~Page,
% Temperature distribution in magnetized neutron star crusts.
% II. The effect of a strong toroidal component,
\AandA{457}{937}{2006}

\bibitem[Giacconi et al.(1962)]{Giacconi-ea62}
R.~Giacconi, H.~Gursky, F.R.~Paolini, B.B.~Rossi,
% Evidence for X-rays from sources outside the solar system,
\PRL{9}{439}{1962}

\bibitem[Ginzburg(1970)]{Ginzburg70}
V.L.~Ginzburg,
\artref{Sov.\ Phys.\ Usp.}{12}{241}{1970}

\bibitem[Ginzburg(1971)]{Ginzburg71}
V.L.~Ginzburg,
\artref{Sov.\ Phys.\ Usp.}{14}{83}{1971}

\bibitem[Ginzburg and Kirzhnits(1965)]{GinzburgKirzhnits}
V.L.~Ginzburg, D.A.~Kirzhnits,
\artref{Sov.~Phys. JETP}{20}{1346}{1965}

\bibitem[Glen and Sutherland(1980)]{GlenSutherland80}
G.~Glen, P.~Sutherland,
% On the cooling of neutron stars,
\ApJ{239}{671}{1980}

\bibitem[Gnedin and Yakovlev(1995)]{GnedinYakovlev95}
O.Y.~Gnedin, D.G.~Yakovlev,
% Thermal conductivity of electrons and muons in neutron star cores
\NPA{582}{697}{1995}

\bibitem[Gnedin et al.(2001)Gnedin, Yakovlev, and Potekhin]{GYP01}
O.Y.~Gnedin, D.G.~Yakovlev, A.Y.~Potekhin, 
% ``Thermal relaxation in young neutron stars,''
\MNRAS{324}{725}{2001}

\bibitem[Gold(1968)]{Gold68}
T.~Gold,
% `Rotating neutron stars as the origin of the pulsating radio sources,''
\artref{Nature}{218}{731}{1968}

\bibitem[Gonzalez and Reisenegger(2010)]{GonzalezReisenegger10}
D.~Gonzalez, A.~Reisenegger,
% Internal heating of old neutron stars: contrasting different mechanisms
\AandA{522}{A16}{2010}

\bibitem[Goriely et al.(2010)]{Goriely-ea10}
S.~Goriely, N.~Chamel, J.M.~Pearson,
\PR{C}{82}{035804}{2010}

\bibitem[Greenstein and Hartke(1983)]{GreensteinHartke}
G.~Greenstein, G.J.~Hartke,
% Pulselike character of blackbody radiation from neutron stars
\ApJ{271}{283}{1983}

\bibitem[Gudmunds\-son et al.(1983)Gudmundsson, Pethick, and Epstein]{GPE83}
E.Y.~Gudmundsson, C.J.~Pethick, R.I.~Epstein,
% Structure of neutron star envelopes,
\ApJ{272}{286}{1983}

\bibitem[Gusakov(2002)]{Gusakov02}
M.E.~Gusakov,
% Neutrino emission from superfluid neutron-star cores: Various types of neutron pairing
\AandA{389}{702}{2002}

\bibitem[Gusakov et al.(2004)]{Gusakov-ea04}
M.E.~Gusakov, A.D.~Kaminker, D.G.~Yakovlev, O.Y.~Gnedin,
% Enhanced cooling of neutron stars via Cooper-pairing neutrino emission
\AandA{423}{1063}{2004}

\bibitem[Haensel and Pichon(1994)]{HP94}
P.~Haensel, B.~Pichon,
\AandA{283}{313}{1994}

\bibitem[Haensel and Zdunik(1990)]{HZ90}
P.~Haensel, J.L.~Zdunik,
\AandA{227}{431}{1990}

\bibitem[Haensel and Zdunik(2008)]{HZ08}
P.~Haensel, J.L.~Zdunik,
\AandA{480}{459}{2008}

\bibitem[Haensel et al.(1990)Haensel, Denissov, and Popov(1990)]{HaenselDP90}
P.~Haensel, A.~Denissov, S.~Popov,
% Neutron star corequake implied by pion condensation.
% Dynamic, neutrino and thermal effects,
\AandA{240}{78}{1990}

\bibitem[Haensel et al.(2007)Haensel, Potekhin, and Yakovlev]{HPY07}
P.~Haensel, A.Y.~Potekhin, D.G.~Yakovlev,
\textit{Neutron Stars 1: Equation of State and Structure.}
Springer, New York (2007)

\bibitem[Hansen et al.(1977)Hansen, Torrie, and Vieillefosse]{HTV77}
J.P.~Hansen, G.M.~Torrie, P.~Vieillefosse,
\PR{A}{16}{2153}{1977}

\bibitem[Heinke and Ho(2010)]{HeinkeHo10}
C.O.~Heinke, W.C.G.~Ho,
\ApJ{719}{L167}{2010}

\bibitem[Heiselberg and Pethick(1993)]{HeiselbergPethick93}
H.~Heiselberg, C.J.~Pethick,
\PR{D}{48}{2916}{1993}

\bibitem[Hernquist(1984)]{Hernquist84}
L.~Hernquist,
% Relativistic electron transport in a quantizing magnetic field,
\ApJS{56}{325}{1984}

\bibitem[Hernquist(1985)]{Hernquist85}
L.~Hernquist,
% Thermal structure of magnetized neutron-star envelopes,
\MNRAS{213}{313}{1985}

\bibitem[Hewish and Okoye(1965)]{HewishOkoye65}
A.~Hewish, S.E.~Okoye,
% ``Evidence for an unusual source of high radio brightness temperature in the Crab nebula,''
\artref{Nature}{207}{59}{1968}

\bibitem[Hewish et al.(1968)]{Hewish-ea68}
A.~Hewish, S.J.~Bell, J.D.H.~Pilkington, P.F.~Scott, R.F.~Collins,
% ``Observation of a rapidly rotating radio source,''
\artref{Nature}{217}{709}{1968}

\bibitem[Ho(2011)]{Ho11}
W.C.G.~Ho,
\MNRAS{418}{L99}{2011}

\bibitem[Ho et al.(2012)]{HoGA12}
W.C.G.~Ho, K.~Glampedakis, N.~Andersson,
\MNRAS{422}{2632}{2012}

\bibitem[Ho et al.(2015)]{Ho-ea15}
W.C.G.~Ho, K.G.~Elshamouty, C.O. Heinke, A.Y.~Potekhin,
\PR{C}{91}{015806}{2015}

\bibitem[Hubbard and Lampe(1969)]{HubbardLampe}
W.~Hubbard, M.~Lampe,
% Thermal conduction by electrons in stellar matter,
\ApJS{18}{297}{1969}

\bibitem[Hughto et al.(2011)]{Hughto-ea11}
J.~Hughto, A.S.~Schneider, C.J.~Horowitz, D.K.~Berry,
\PR{E}{84}{016401}{2011}

\bibitem[Hughto et al.(2012)]{Hughto-ea12}
J.~Hughto, C.J.~Horowitz, A.S.~Schneider, et al.,
% Z.~Medin, A.~Cumming, D.K.~Berry,
\PR{E}{86}{066413}{2012}

\bibitem[Inogamov and Sunyaev(2010)]{InogamovSunyaev10}
N.A.~Inogamov, R.A.~Sunyaev,
\artref{Astron.\ Lett.}{36}{848}{2010}

\bibitem[Itoh and Kohyama(1983)]{ItohKohyama83}
N.~Itoh, Y.~Kohyama,
% Neutrino-pair bremsstrahlung in dense stars. I. Liquid metal case,''
\ApJ{275}{858}{1983}

\bibitem[Itoh and Kohyama(1993)]{ItohKohyama93}
N.~Itoh, Y.~Kohyama,
% Electrical and thermal conductivities of dense matter in the crystalline lattice phase. II - Impurity scattering
\ApJ{404}{268}{1993}; 
erratum: \artref{\textit{ibid.}}{420}{943}{1994}

\bibitem[Kaminker and Yakovlev(1981)]{KaminkerYakovlev81}
A.D.~Kaminker, D.G.~Yakovlev,
% Description of a relativistic electron in a quantizing magnetic field. Transverse transport coefficients of an electron gas,
\artref{Theor.\ Math.\ Phys.}{49}{1012}{1981}

\bibitem[Kaminker and Yakovlev(1994)]{KaminkerYakovlev94}
A.D.~Kaminker, D.G.~Yakovlev,
% Annihilation and synchrotron emission of neutrino pairs by electrons and positrons in the neutron star envelopes.
% Astron.\ Zh., 71, 910-915, 1994
\artref{Astron.\ Rep.}{38}{809}{1994}

\bibitem[Kaminker et al.(1992)]{Kaminker-ea92}
A.D.~Kaminker, K.P.~Levenfish, D.G.~Yakovlev, P.~Amsterdamski, P.~Haensel,
% Neutrino-pair bremsstrahlung by electrons in neutron star crusts
\PR{D}{46}{3256}{1992}

\bibitem[Kaminker et al.(1997)Kaminker, Yakovlev, and Haensel]{KaminkerYH97}
A.D.~Kaminker, D.G.~Yakovlev, P.~Haensel,
% Neutrino pair emission due to scattering of electrons off fluxoids in superfluid neutron star cores
\AandA{325}{391}{1997}

\bibitem[Kaminker et al.(1999)]{Kaminker-ea99}
A.D.~Kaminker, C.J.~Pethick, A.Y.~Potekhin, V.~Thorsson, D.G.~Yakovlev,
% Neutrino-pair bremsstrahlung by electrons in neutron star crusts
\AandA{343}{1009}{1999}

\bibitem[Kaminker et al.(2001)Kaminker, Haensel, and Yakovlev]{KaminkerHY01}
A.D.~Kaminker, P.~Haensel, D.G.~Yakovlev,
% Nucleon superfluidity vs. observations of cooling neutron stars
\AandA{373}{L17}{2001}

\bibitem[Kaminker et al.(2009)]{Kaminker-ea09}
A.D.~Kaminker, A.Y.~Potekhin, D.G.~Yakovlev, G.~Chabrier,
% Heating and cooling of magnetars with accreted envelopes,
\MNRAS{395}{2257}{2009}

\bibitem[Kaminker et al.(2012)]{Kaminker-ea12}
A.D.~Kaminker, A.A.~Kaurov, A.Y.~Potekhin, D.G.~Yakovlev,
% Heating Magnetar Surface from the Crust
in \textit{Electromagnetic Radiation from Pulsars and
Magnetars},
ed.~by W.~Lewandowski, J.~Kijak, A.~Slowikowska, O.~Maron,
\artref{Astron.\ Soc.\ Pacific Conf. Ser.}{466}{237}{2012}

\bibitem[Kaminker et al.(2014)]{Kaminker-ea14}
A.D.~Kaminker, A.A.~Kaurov, A.Y.~Potekhin, D.G.~Yakovlev,
% Thermal emission of neutron stars with internal heaters,
\MNRAS{442}{3484}{2014}

\bibitem[Kantor and Gusakov(2007)]{KantorGusakov07}
E.M.~Kantor, M.E.~Gusakov,
%The neutrino emission due to plasmon decay and neutrino luminosity of white dwarfs
\MNRAS{381}{1702}{2007}

\bibitem[Kardashev(1964)]{Kardashev64}
N.S.~Kardashev,
% ``Magnetic collapse and the nature of powerful sources of cosmic radio emission,''
\artref{Astron.\ Zh.}{41}{807}{1964}

\bibitem[Kittel(1963)]{Kittel63}
C.~Kittel,
\textit{Quantum Theory of Solids}.
Wiley, New York (1963)

\bibitem[Kondratyev et al.(2001)Kondratyev, Maruyama, Chiba]{KondratyevMC01}
V.N.~Kondratyev, T.~Maruyama, S.~Chiba,
% Magnetic Field Effect on Masses of Atomic Nuclei,
\ApJ{546}{1137}{2001}

\bibitem[Lai and Shapiro(1991)]{LaiShapiro91}
D.~Lai, S.L.~Shapiro,
% Cold equation of state in a strong magnetic field -- Effects of inverse beta-decay,
\ApJ{383}{745}{1991}

\bibitem[Lampe(1968)]{Lampe68b}
M.~Lampe,
% Transport theory of a partially degenerate plasma,
\PR{}{174}{276}{1968}

\bibitem[Lattimer et al.(1991)]{Lattimer-ea91}
J.M.~Lattimer, C.J.~Pethick, M.~Prakash, P.~Haensel,
% Direct URCA process in neutron stars,
\PRL{66}{2701}{1991}

\bibitem[Lee(1950)]{Lee50}
T.D.~Lee, 
% Hydrogen content and energy-production mechanism of white dwarfs,
\ApJ{111}{625}{1950}

\bibitem[Leinson(2009)]{Leinson09}
L.B.~Leinson,
% Superfluid response and the neutrino emissivity of baryon matter: Fermi-liquid effects
\PR{C}{79}{045502}{2009}

\bibitem[Leinson(2010)]{Leinson10}
L.B.~Leinson,
% Neutrino emission from triplet pairing of neutrons in neutron stars
\PR{C}{81}{025501}{2010}

\bibitem[Leinson and P\'erez(1998)]{LeinsonPerez98}
L.B.~Leinson, A.~P\'erez,
% Direct Urca process in neutron stars with strong magnetic fields,
\artref{J.\ High Energy Phys.}{1998}{issue 9, id.\,020}{1998}

\bibitem[Levenfish and Haensel(2007)]{LevenfishHaensel07}
K.P.~Levenfish, P.~Haensel,
\ApSS{308}{457}{2007}

\bibitem[Levenfish and Yakovlev(1994)]{LY94}
K.P.~Levenfish, D.G.~Yakovlev,
\artref{Astron.\ Rep.}{38}{247}{1994}

\bibitem[Lifshitz and Pitaevski{\u\i}(2002)]{LaLi-SP2}
E.M.~Lifshitz, L.P.~Pitaevski{\u\i},
\textit{Statistical Physics, Part~2: Theory of the Condensed State}.
Butterworth-Heinemann, Oxford (2002)

\bibitem[Lorenz et al.(1993)]{LorenzRP93}
C.P.~Lorenz, D.G.~Ravenhall, C.J.~Pethick
% ``Neutron star crusts,''
\PRL{70}{379}{1993}

\bibitem[Machleidt et al.(1987)Machleidt, Holinde, and Elster]{Machleidt-ea87}
R.~Machleidt, K.~Holinde, Ch.~Elster,
\artref{Phys.\ Rep.}{149}{1}{1987}

\bibitem[Margueron et al.(2008)]{MargueronSH08}
J.~Margueron, H.~Sagawa, K.~Hagino,
\PR{C}{77}{054309}{2008}

\bibitem[Medin and Cumming(2010)]{MedinCumming10}
Z.~Medin, A.~Cumming,
\PR{E}{81}{036107}{2010}

\bibitem[Mereghetti et al.(2015)Mereghetti, Pons, and Melatos]{Mereghetti-ea15}
S.~Mereghetti, J.A.~Pons, A.~Melatos,
% Magnetars: Properties, Origin and Evolution,
\artref{Space Sci.\ Rev.}{191}{315}{2015}
[DOI: 10.1007/s11214-015-0146-y]

\bibitem[Migdal(1959)]{Migdal59}
A.B.~Migdal,
\artref{Sov.\ Phys. JETP}{10}{176}{1960}

\bibitem[Miralles, Urpin and Konenkov(1998)]{MUK98}
J.A.~Miralles, V.~Urpin, D.~Konenkov,
\ApJ{503}{368}{1998}

\bibitem[Misner et al.(1973)Misner, Thorn, \& Wheeler]{MisnerTW}
C.W.~Misner, K.S.~Thorne, J.A.~Wheeler,
\textit{Gravitation}.
W.H.~Freeman and Co., San Francisco (1973)
	
\bibitem[Muzikar et al.(1980)Muzikar, Sauls, and Serene]{MuzikarSS80}
P.~Muzikar, J.A.~Sauls, J.W.~Serene,
\PR{D}{21}{1494}{1980}

\bibitem[Morton(1964)]{Morton64}
D.C.~Morton,
% Neutron stars and X-ray sources,
\artref{Nature}{201}{1308}{1964}

\bibitem[Negele and Vautherin(1973)]{NV73}
J.W.~Negele, D.~Vautherin,
\NPA{207}{298}{1973}

\bibitem[Nomoto and Tsuruta(1981a)]{NomotoTsuruta81a}
K.~Nomoto, S.~Tsuruta,
\SSRv{30}{123}{1981a}

\bibitem[Nomoto and Tsuruta(1981b)]{NomotoTsuruta81b}
K.~Nomoto, S.~Tsuruta,
\ApJ{250}{L19}{1981b}

\bibitem[Nomoto and Tsuruta(1986)]{NomotoTsuruta86}
K.~Nomoto, S.~Tsuruta,
\ApJ{305}{L19}{1986}

\bibitem[Nomoto and Tsuruta(1987)]{NomotoTsuruta87}
K.~Nomoto, S.~Tsuruta,
\ApJ{312}{711}{1987}

\bibitem[Ofengeim et al.(2014)Ofengeim, Kaminker, and Yakovlev]{OfengeimKY14}
D.D.~Ofengeim, A.D.~Kaminker, D.G.~Yakovlev,
\artref{Europhys.\ Lett.}{108}{31002}{2014}

\bibitem[Pacini(1967)]{Pacini67}
F.~Pacini,
% ``Energy emission from a neutron star,''
\artref{Nature}{216}{567}{1967}

\bibitem[Page(2009)]{Page09}
D.~Page,
% Neutron Star Cooling: I,
in \textit{Neutron Stars and Pulsars},
ed.~by W.~Becker,
\artref{Astrophys.\ Space Sci.\ Library}{357}{247}{2009}

\bibitem[Page and Applegate(1992)]{PageApplegate92}
D.~Page, J.H.~Applegate,
% The cooling of neutron stars by the direct Urca process,
\ApJ{394}{L17}{1992}

\bibitem[Page and Baron(1990)]{PageBaron90}
D.~Page, E.~Baron,
% Strangeness  condensation, nucleon superfluidity and
% cooling of neutron stars,
\ApJ{354}{L17}{1990};
erratum: \artref{\textit{ibid.}}{382}{L111--L112}{1991}

\bibitem[Page and Reddy(2013)]{PageReddy13}
D.~Page, S.~Reddy,
\PRL{111}{241102}{2013}

\bibitem[Page et al.(2004)]{Pageetal2004}
 D.~Page, J.M.~Lattimer, M.~Prakash, A.W.~Steiner,
 \ApJS{155}{623}{2004}

\bibitem[Page et al.(2007)Page, Geppert, and K\"uker]{PageGK07}
D.~Page, U.~Geppert, M.~K\"uker,
\ApSS{308}{403}{2007}

\bibitem[Page et al.(2011)]{Page-ea11}
D.~Page, M.~Prakash, J.M.~Lattimer, A.W.~Steiner,
\PRL{106}{081101}{2011}

\bibitem[Page et al.(2014)]{Page-ea14}
D.~Page, J.M.~Lattimer, M.~Prakash, A.W.~Steiner,
% Stellar Superfluids
in \textit{Novel Superfluids},
ed.~K.H.~Bennemann, J.B.~Ketterson.
Oxford Univ.\ Press, Oxford (2014), p.~550%--579

\bibitem[Pandharipande and Ravenhall(1989)]{Pandharipande89}
V.R.~Pandharipande, D.G.~Ravenhall,
in \textit{Nuclear Matter and Heavy Ion Collisions},
NATO ADS Ser., vol.~B205, 
ed.~M.~Soyeur, H.~Flocard, B.~Tamain, and M.~Porneuf.
Reidel, Dordrecht (1989), p.~103

\bibitem[Pastore et al.(2015)Pastore, Chamel, \& Margueron]{PastoreCM15}
A.~Pastore, N.~Chamel, J.~Margueron,
\MNRAS{448}{1887}{2015}

\bibitem[Pearson et al.(2011)]{PearsonGC11}
J.M.~Pearson, S.~Goriely, N.~Chamel,
\PR{C}{83}{065810}{2011}

\bibitem[Pearson et al.(2012)]{Pearson-ea12}
J.M.~Pearson, N.~Chamel, S.~Goriely, C.~Ducoin,
\PR{C}{85}{065803}{2012}

\bibitem[P\'erez-Azorin et al.(2006)P\'erez-Azorin, Miralles, and Pons]{PerezAMP06}
J.F.~P\'erez-Azorin, J.A.~Miralles, J.A.~Pons,
% Anisotropic thermal emission from magnetized neutron stars,
\AandA{451}{1009}{2006}

\bibitem[Perna and Pons(2011)]{Perna11}
R.~Perna, J.A.~Pons,
\ApJ{727}{L51}{2011}

\bibitem[Pethick and Potekhin(1998)]{PethickPotekhin96}
C.J.~Pethick, A.Y.~Potekhin,
% Liquid crystals in the mantles of neutron stars,''
\artref{Phys.\ Lett. B}{427}{7}{1996}

\bibitem[Pethick and Ravenhall(1995)]{PethickRavenhall95}
C.J.~Pethick, D.G.~Ravenhall,
% Matter at large neutron excess and the physics of neutron-star crusts,
\ARAA{45}{429}{1995}

\bibitem[Pinaev(1964)]{Pinaev}
V.S.~Pinaev,
\artref{Sov.\ Phys. JETP}{18}{377}{1964}

\bibitem[Pons et al.(1999)]{Pons99}
J.A.~Pons, S.~Reddy, M.~Prakash, J.M.~Lattimer, J.A.~Miralles,
\ApJ{513}{780}{1999}

\bibitem[Pons et al.(2009)Pons, Miralles, and Geppert]{PonsMG09}
J.A.~Pons, J.A.~Miralles, U.~Geppert,
% Magneto-thermal evolution of neutron stars,
\AandA{496}{207}{2014}

\bibitem[Pons and Perna(2011)]{PonsPerna11}
J.A.~Pons, R.~Perna,
% Magneto-thermal evolution of neutron stars,
\ApJ{741}{123}{2011}

\bibitem[Pons and Rea(2012)]{PonsRea12}
J.A.~Pons, N.~Rea,
% Magneto-thermal evolution of neutron stars,
\ApJ{750}{L6}{2012}

\bibitem[Posselt et al.(2013)]{Posselt-ea13}
B. Posselt, G.G.~Pavlov, V.~Suleimanov, O.~Kargaltsev,
\ApJ{779}{186}{2013}

\bibitem[Potekhin(1996)]{P96}
A.Y.~Potekhin,
% Electron conduction along quantizing magnetic fields in neutron star crusts. I. Theory,
\AandA{306}{999}{1996}; erratum: \artref{\textit{ibid.}}{327}{441}{1997}

\bibitem[Potekhin(1999)]{P99}
A.Y.~Potekhin,
% Electron conduction in magnetized neutron star envelopes,
\AandA{351}{787}{1999}

\bibitem[Potekhin(2014)]{P14}
A.Y.~Potekhin,
% Atmospheres and radiating surfaces of neutron stars
\artref{Phys.\ Usp.}{57}{735}{2014}

\bibitem[Potekhin and Chabrier(2000)]{PC00}
A.Y.~Potekhin, G.~Chabrier,
\PR{E}{62}{8554}{2000}

\bibitem[Potekhin and Chabrier(2010)]{PC10}
A.Y.~Potekhin, G.~Chabrier,
%Thermodynamic Functions of Dense Plasmas: 
%Analytic Approximations for Astrophysical Applications
\artref{Contr.\ Plasma Phys.}{50}{82}{2010}

\bibitem[Potekhin and Chabrier(2012)]{PC12}
A.Y.~Potekhin, G.~Chabrier,
\AandA{538}{A115}{2012}

\bibitem[Potekhin and Chabrier(2013)]{PC13}
A.Y.~Potekhin, G.~Chabrier,
\AandA{550}{A43}{2013}

\bibitem[Potekhin and Yakovlev(2001)]{PY01}
A.Y.~Potekhin, D.G.~Yakovlev,
% Thermal structure and cooling of neutron stars with magnetized envelopes,
\AandA{374}{213}{2001}

\bibitem[Potekhin et al.(1997)Potekhin, Chabrier, and Yakovlev]{PCY97}
A.Y.~Potekhin, G.~Chabrier, D.G.~Yakovlev, 
% Internal temperatures and cooling of neutron stars
% with accreted envelopes,
\AandA{323}{415}{1997}

\bibitem[Potekhin et al.(1999)]{Potekhin-ea99}
A.Y.~Potekhin, D.A.~Baiko, P.~Haensel, D.G.~Yakovlev,
% Transport properties of degenerate electrons in 
%       neutron star envelopes and white dwarf cores,
\AandA{346}{345}{1999}

\bibitem[Potekhin et al.(2003)]{Potekhin-ea03}
A.Y.~Potekhin, D.G.~Yakovlev, G.~Chabrier, Gnedin O.Y.,
% Thermal structure and cooling of superfluid neutron stars
% with accreted magnetized envelopes,
\ApJ{594}{404}{2003}

\bibitem[Potekhin et al.(2005)Potekhin, Urpin, and Chabrier]{PUC05}
A.Y.~Potekhin, V.A.~Urpin, G.~Chabrier,
% The magnetic structure of neutron stars and their surface-to-core temperature relation
\AandA{443}{1025}{2005}

\bibitem[Potekhin et al.(2007)]{PCY07}
A.Y.~Potekhin, G.~Chabrier, D.G.~Yakovlev,
% ``Heat blanketing envelopes and thermal radiation of strongly magnetized neutron stars,''
\ApSS{308}{353}{2007}

\bibitem[Potekhin et al.(2009)]{Pmix09}
A.Y.~Potekhin, G.~Chabrier, A.I.~Chugunov, H.E.~DeWitt, F.J.~Rogers,
\PR{E}{80}{047401}{2009}

\bibitem[Potekhin et al.(2013)]{Potekhin-ea13}
A.Y.~Potekhin, A.F.~Fantina, N.~Chamel, J.M.~Pearson, S.~Goriely,
% Analytical representations of unified equations of state for neutron-star matter,
\AandA{560}{A48}{2013}

\bibitem[Potekhin et al.(2015)Potekhin, De Luca, and Pons]{NSemitters}
A.Y.~Potekhin, A.~De Luca, J.A.~Pons,
% Neutron stars -- Thermal emitters,
\artref{Space Sci.\ Rev.}{191}{176}{2015}
[DOI: 10.1007/s11214-014-0102-2] (Paper~I)

\bibitem[Raikh and Yakovlev(1982)]{RaikhYakovlev82}
M.E.~Raikh, D.G.~Yakovlev,
% Thermal and electrical conductivities of crystals in neutron stars and degenerate dwarfs,
\ApSS{87}{193}{1982}

\bibitem[Reisenegger(1995)]{Reisenegger95}
A.~Reisenegger,
% Deviations from chemical equilibrium due to spin-down as an internal heat source in neutron stars
\ApJ{442}{749}{1995}

\bibitem[Richarson et al.(1982)]{Richardson-ea82}
M.B.~Richardson, H.M.~Van Horn, K.F.~Ratcliff, R.C.~Malone,
% Neutron star evolutionary sequences,
\ApJ{255}{624}{1982}

\bibitem[Roberts(2012)]{Roberts12}
L.F.~Roberts,
\ApJ{755}{126}{2012}

\bibitem[R\"uster et al.(2006)]{RuesterHSB06}
S.B.~R\"uster, M.~Hempel, J.~Schaffner-Bielich,
% ``Outer crust of nonaccreting cold neutron stars,''
\PR{C}{73}{035804}{2006}

\bibitem[Rutledge et al.(2002)]{Rutledge-ea02a}
R.E.~Rutledge, L.~Bildsten, E.F.~Brown, G.G.~Pavlov, V.E.~Zavlin,
\ApJ{577}{346}{2002}

\bibitem[Sandage et al.(1966)]{Sandage-ea66}
A.~Sandage, P.~Osmer, R.~Giacconi et al.,
% ``On the optical identification of Sco X-1,''
\ApJ{146}{316}{1966}

\bibitem[Schaab et al.(1997)]{Schaab-ea97}
Ch.~Schaab, D.~Voskresensky, A.D.~Sedrakian, F.~Weber, M.K.~Weigel,
\AandA{321}{591}{1997}

\bibitem[Schaaf(1990)]{Schaaf90}
M.E.~Schaaf,
% Surface-to-core temperature variation of homogeneously
% magnetized neutron stars,
\AandA{227}{61}{1990}

\bibitem[Schwenk et al.(2003)Schwenk, Friman, and Brown]{SchwenkFB03}
A.~Schwenk, B.~Friman, G.E.~Brown,
\NPA{713}{191}{2003}

\bibitem[Schwinger(1988)]{Schwinger}
J.~Schwinger,
\textit{Particles, Sources, and Fields}.
Addison-Wesley, Redwood City (1988)

\bibitem[Shibanov and Yakovlev(1996)]{ShibanovYakovlev96}
Yu.A.~Shibanov, D.G.~Yakovlev,
% On cooling of magnetized neutron stars,
\AandA{309}{171}{1996}

\bibitem[Shklovsky(1967)]{Shklovsky67}
I.S.~Shklovsky,
% ``On the nature of the source of X-ray emission of Sco XR-1,''
\ApJ{148}{L1}{1967}


\bibitem[Shternin and Yakovlev(2006)]{ShterninYakovlev06}
P.S.~Shternin, D.G.~Yakovlev,
% Electron thermal conductivity owing to collisions between degenerate electrons,
\PR{D}{74}{043004}{2006}

\bibitem[Shternin and Yakovlev(2007)]{ShterninYakovlev07}
P.S.~Shternin, D.G.~Yakovlev,
% Electron-muon heat conduction in neutron star cores via the exchange of transverse plasmons
\PR{D}{75}{103004}{2007}

\bibitem[Shternin et al.(2007)]{Shternin-ea07}
P.S.~Shternin, D.G.~Yakovlev, P.~Haensel, A.Y.~Potekhin,
% Neutron star cooling after deep crustal heating in the X-ray
% transient KS 1731-260,
\MNRAS{382}{L43}{2007}

\bibitem[Shternin et al.(2011)]{Shternin-ea11}
P.S.~Shternin, D.G.~Yakovlev, C.O.~Heinke, W.C.G.~Ho, D.J.~Patnaude,
% Cooling neutron star in the Cassiopeia A supernova remnant: evidence for superfluidity in the core
\MNRAS{412}{L108}{2011}

\bibitem[Shternin et al.(2013)]{ShterninBH13}
P.S.~Shternin, M.~Baldo, P.~Haensel,
% Transport coefficients of nuclear matter in neutron star cores
\PR{C}{88}{065803}{2013}

\bibitem[Silant'ev and Yakovlev(1980)]{SilYak}
N.A.~Silant'ev, D.G.~Yakovlev,
% Radiative heat transfer in surface layers of neutron stars with a magnetic field,
\ApSS{71}{45}{1980}

\bibitem[Sokolov and Ternov(1986)]{SokTer}
A.A.~Sokolov, I.M.~Ternov,
\textit{Radiation from Relativistic Electrons}
(New York: Am.\ Inst.\ Phys., 1986)

\bibitem[Spitzer(1962)]{Spitzer}
L.~Spitzer, Jr., 
\textit{Physics of Fully Ionized Gases}.
Wiley, New York (1962)

\bibitem[Spitzer and H\"arm(1953)]{SpitzerHarm}
L.~Spitzer, Jr., R.~H\"arm,
% Transport phenomena in a completely ionized gas
\artref{Phys.\ Rev.}{89}{977}{1953}

\bibitem[Stabler(1960)]{Stabler60}
R.~Stabler,
Ph.D.\ Thesis. Cornell Univ., Ithaca, NY (1960)

\bibitem[Stejner et al.(2009)Stejner, Weber, and Madsen]{StejnerWM09}
M.~Stejner, F.~Weber, J.~Madsen,
% Signature of Deconfinement with Spin-Down Compression in Cooling Hybrid Stars
\ApJ{694}{1019}{2009}

\bibitem[Strohmayer and Bildsten(2006)]{StrohmayerBildsten06}
T.~Strohmayer, L.~Bildsten,
in \textit{Compact Stellar X-Ray Sources},
ed.~by W.H.G.~Lewin and M.~van der Klis.
Cambridge Univ.\ Press, Cambridge (2006), p.~113

\bibitem[Suh and Mathews(2001)]{SuhMathews}
I.-S.~Suh, G.J.~Mathews,
% Cold ideal equation of state for strongly magnetized neutron star
% matter: Effects on muon production and pion condensation,
\ApJ{546}{1126}{2001}

\bibitem[Takatsuka and Tamagaki(1995)]{TakatsukaTamagaki95}
T.~Takatsuka, R.~Tamagaki,
\artref{Prog.\ Theor.\ Phys.}{94}{457}{1995}

\bibitem[Takatsuka and Tamagaki(2004)]{TakatsukaTamagaki04}
T.~Takatsuka, R.~Tamagaki,
% Baryon Superfluidity and Neutrino Emissivity of Neutron Stars
\artref{Prog.\ Theor.\ Phys.}{112}{37}{2004}

\bibitem[Thompson(2001)]{Thompson01}
C.~Thompson,
% ``Astrophysics of the soft gamma repeaters and the anomalous X-ray pulsars,''
 in \textit{The Neutron Star -- Black Hole Connection},
ed.~by C.~Kouveliotou, J.~Ventura, E.~Van den Heuvel.
Kluwer, Dordrecht (2001), p.~369

\bibitem[Thorne(1977)]{Thorne77}
K.S.~Thorne,
% The relativistic equations of stellar structure and evolution,
\ApJ{212}{825}{1977}

\bibitem[Tilley and Tilley(1990)]{Tilley90}
D.R.~Tilley, J.~Tilley,
\textit{Superfluidity and Superconductivity}.
IOP Publishing, Bristol (1990)

\bibitem[Tsuruta(1964)]{Tsuruta64}
S.Tsuruta,
\textit{Neutron Star Models},
Ph.D.\ Thesis. Columbia Univ., New York (1964)

\bibitem[Tsuruta(1979)]{Tsuruta79}
S.~Tsuruta,
% Thermal  properties and detectability of neutron stars.
% I. Cooling and heating of neutron stars,
\artref{Phys.\ Rep.}{56}{237}{1979}

\bibitem[Tsuruta(1986)]{Tsuruta86}
S.~Tsuruta,
\artref{Comments on Astrophysics}{11}{151}{1986}

\bibitem[Tsuruta(1998)]{Tsuruta98}
S.~Tsuruta,
% Thermal properties and detectability of neutron stars. II.
% Thermal evolution of rotation-powered neutron stars,
\artref{Phys.\ Rep.}{292}{1}{1998}

\bibitem[Tsuruta(2009)]{Tsuruta09}
S.~Tsuruta,
% Neutron Star Cooling: II,
in \textit{Neutron Stars and Pulsars},
ed.~by W.~Becker,
\artref{Astrophys.\ Space Sci.\ Library}{357}{289}{2009}

\bibitem[Tsuruta and Cameron(1966)]{TsurutaCameron66}
S.~Tsuruta, A.G.W.~Cameron,
% Cooling and detectability of neutron stars,
\artref{Can.\ J.\ Phys.}{44}{1863}{1966}

\bibitem[Turlione et al.(2015)Turlione, Aguilera, and Pons]{TurlioneAP13}
A.~Turlione, D.~Aguilera, J.~Pons,
\AandA{577}{A5}{2015}

\bibitem[Urpin and Konenkov(2008)]{UrpinKonenkov08}
V.~Urpin, D.~Konenkov,
% ``Joule heating in high magnetic field pulsars,''
\AandA{483}{223}{2008}

\bibitem[Urpin and Yakovlev(1980)]{UrpinYakovlev80}
V.A.~Urpin, D.G.~Yakovlev,
% Thermal conductivity due to collisions between electrons
% in a degenerate relativistic electron gas,
% \textit{Astron.\ Zh.} \textbf{57}, 213--215
\SvA{24}{126}{1980}

\bibitem[Van Riper(1988)]{VanRiper88}
K.A.~Van Riper,
% Magnetic neutron star atmospheres,
\ApJ{329}{339}{1988}

\bibitem[Van Riper(1991)]{VanRiper91}
K.A.~Van Riper,
% Neutron star thermal evolution,
\ApJS{75}{449}{1991}

\bibitem[Van Riper and Lamb(1981)]{VanRiperLamb81}
K.A.~Van Riper, F.~Lamb,
% Neutron star evolution and results from the EINSTEIN X-ray
% observatory,
\ApJ{244}{L13}{1981}

\bibitem[Ventura and Potekhin(2001)]{VP01}
J.~Ventura, A.Y.~Potekhin,
% Neutron Star Envelopes and Thermal Radiation from the
% Magnetic Surface,
in
\textit{The neutron star -- black hole connection},
Proceedings of the NATO \textit{ASI}, Ser.~C, vol.~567,
ed.~by C.~Kouveliotou, J.~Ventura, and E.~Van den Heuvel.
Kluwer Academic Publishers, Dordrecht (2001), p.~393

\bibitem[Vidaurre et al.(1995)]{Vidaurre95}
A.~Vidaurre, A.~Perez, H.~Sivak, J.~Bernabeu, J.M.~Ibanez, 
\ApJ{448}{264}{2013}

\bibitem[Vigan{\`o}(2013)]{ViganoPhD}
D.~Vigan{\`o},
\textit{Magnetic Fields in Neutron Stars},
Ph.D.\ Thesis. Univ.\ of Alicante, Alicante (2013)

\bibitem[Vigan{\`o} et al.(2013)]{vigano13}
D.~Vigan{\`o}, N.~Rea, J.A.~Pons, D.N.~Aguilera, 
J.A.~Miralles,
\MNRAS{434}{123}{2013}

\bibitem[Voskresensky(2001)]{Voskresensky01}
D.N.~Voskresensky,
% Neutrino Cooling of Neutron Stars: Medium Effects
in
\textit{Physics of Neutron Star Interiors},
ed.~by D.~Blaschke, N.K.~Glendenning, and A.~Sedrakian, 
\artref{Lecture Notes in Physics}{578}{467}{2001}

\bibitem[Voskresensky and Senatorov(1986)]{VoskresenskySenatorov86}
D.N.~Voskresensky, A.V.~Senatorov,
\artref{Sov.\ Phys. JETP}{63}{885}{1986}

\bibitem[Voskresensky and Senatorov(1987)]{VoskresenskySenatorov87}
D.N.~Voskresensky, A.V.~Senatorov,
\artref{Sov.\ J.\ Nucl.\ Phys.}{45}{411}{1987}

\bibitem[Wambach et al.(1993)Wambach, Ainsworth, and Pines]{WambachAP93}
J.~Wambach, T.L.~Ainsworth, D.~Pines,
\NPA{555}{128}{1993}

\bibitem[Wijnands et al.(2013)Wijnands, Degenaar, and Page]{WijnandsDP13}
R.~Wijnands, N.~Degenaar, D.~Page,
\MNRAS{432}{2366}{2013}

\bibitem[Yakovlev(1984)]{Yakovlev84}
D.G.~Yakovlev,
% Transport properties of the degenerate electron gas of neutron stars along the quantizing magnetic field,
\ApSS{98}{37 }{1984}

\bibitem[Yakovlev and Pethick(2004)]{YakovlevPethick04}
D.G.~Yakovlev, C.J.~Pethick,
% Neutron Star Cooling,
\ARAA{42}{169}{2004}

\bibitem[Yakovlev and Urpin(1980)]{YakovlevUrpin80}
D.G.~Yakovlev, V.A.~Urpin,
% Thermal and electrical conductivity in white dwarfs and neutron stars,
\SvA{24}{303}{1980}

\bibitem[Yakovlev et al.(1999a)Yakovlev, Kaminker, and Levenfish]{YKL99}
D.G.~Yakovlev, A.D.~Kaminker, K.P.~Levenfish,
% Neutrino emission due to Cooper pairing of nucleons in cooling neutron stars
\AandA{343}{650}{1999a}

\bibitem[Yakovlev et al.(1999b)]{YLS99}
D.G.~Yakovlev, K.P.~Levenfish, Yu.A.~Shibanov,
% Cooling of neutron stars and superfluidity in their cores
\artref{Phys. Usp.}{42}{737}{1999b}

\bibitem[Yakovlev et al.(2001)]{YKGH01}
D.G.~Yakovlev, A.D.~Kaminker, O.Y.~Gnedin, P.~Haensel,
% Neutrino emission from neutron stars,
\artref{Phys.\ Rep.}{354}{1}{2001}

\bibitem[Yakovlev et al.(2003)Yakovlev, Levenfish, and Haensel]{YakLH03}
D.G.~Yakovlev, K.P.~Levenfish, P.~Haensel,
% Thermal state of transiently accreting neutron stars
\AandA{407}{265}{2003}

\bibitem[Yakovlev et al.(2004)]{Yak-ea04}
D.G.~Yakovlev, K.P.~Levenfish, A.Y.~Potekhin, O.Y.~Gnedin, G.~Chabrier,
% Thermal states of coldest and hottest neutron stars in soft X-ray transients
\AandA{417}{169}{2004}

\bibitem[Yakovlev et al.(2008)]{Yakovlev-ea08}
D.G.~Yakovlev, O.Y.~Gnedin, A.D.~Kaminker, A.Y.~Potekhin,
% Theory of cooling neutron stars versus observations,
in 
\textit{40 Years of Pulsars: Millisecond Pulsars, Magnetars and More},
ed.~by C.~Bassa, Z.~Wang, A.~Cumming and V.M.~Kaspi,
\AIPC{983}{379}{2008}

\bibitem[Ziman(1960)]{Ziman}
J.M.~Ziman,
\textit{Electrons and Phonons}.
Oxford Univ.\ Press, Oxford (1960)

\end{thebibliography}
\end{document}